%% file: 0-main.tex
\documentclass[11pt]{article}
\usepackage{amsmath, amsthm, amssymb, amsthm, dcolumn}

\usepackage{graphicx} 
\usepackage{color}
\usepackage{float}
\usepackage[utf8]{inputenc}
\usepackage{subcaption}
\usepackage[titletoc, title]{appendix}
\usepackage{apacite}
\usepackage{tabu,multirow}
\usepackage[titletoc]{appendix}

\usepackage{import}
\usepackage{rotating}
\usepackage[flushleft]{threeparttable}
\usepackage{caption}
\usepackage{soul}
\usepackage{ragged2e}
\usepackage{authblk}

\begin{document}

\begin{titlepage}
\title{Predicting crashes in oil prices during the COVID-19 pandemic with mixed causal-noncausal models}
\author[ ]{Alain Hecq\footnote{Corresponding author : Alain Hecq, Maastricht University, Department of Quantitative Economics, School of Business and Economics, P.O.box 616, 6200 MD, Maastricht, The Netherlands. Email: a.hecq@maastrichtuniversity.nl}}
\author[ ]{Elisa Voisin}

\affil[ ]{Maastricht University}
\date{December 2021}

\maketitle

\begin{abstract}
    This paper aims at shedding light upon how transforming or detrending a series can substantially impact predictions of mixed causal-noncausal (\textit{MAR}) models, namely dynamic processes that depend not only on their lags but also on their leads. \textit{MAR} models have been successfully implemented on commodity prices as they allow to generate nonlinear features such as locally explosive episodes (denoted here as bubbles) in a strictly stationary setting. We consider multiple detrending methods and investigate, using Monte Carlo simulations, to what extent they preserve the bubble patterns observed in the raw data.  \textit{MAR} models relies on the dynamics observed in the series alone and does not require economical background to construct a structural model, which can sometimes be intricate to specify or which may lack parsimony. We investigate oil prices and estimate probabilities of crashes before and during the first 2020 wave of the COVID-19 pandemic. We consider three different mechanical detrending methods and compare them to a detrending performed using the level of strategic petroleum reserves.\\
    
    \textit{Keywords:}  {\small  noncausal models, detrending, forecasting, predictive densities, bubbles, crashes, simulations-based forecasts, Hodrick-Prescott filter, COVID-19 pandemic}
\end{abstract}
\end{titlepage}

\newpage

\graphicspath{ {Pictures/} }

\import{./}{1-Introduction.tex}

\import{./}{2-Discussion.tex}

\import{./}{3-Montecarlo_MAR.tex}

\import{./}{4-Empirics.tex}

\section{Conclusion}\label{sec:Conclusion}
This paper aims at shedding light upon how transforming or detrending a series can substantially impact predictions of mixed causal-noncausal models. Assuming a polynomial trend of order 4 for WTI and Brent series probably alters the dynamics in the remaining cycle. The HP filter (with penalizing parameter $\lambda=129\,600$) does not require any further assumptions with respect to the trend and can therefore be an adequate filter in cases where the trend is unknown. Knowing the actual trend or using exogenous variables for it is also not straightforward. We use US crude oil strategic petroleum reserves (SPR) to detrend oil price series to illustrate this option. We show that by detrending with SPR we obtain similar results to the $HP$ and polynomial trend of order 6 detrending. However, detrending with a variable that has seasonality or dynamics will alter the dynamics left in the cycle. Overall, caution is needed when detrending a series, and some filtering such as polynomial trends may require additional understanding regarding the deviations of the series from its fundamental trend. Nonetheless, once the series is detrended, resulting in a stationary series, using \textit{MAR} models is a straightforward approach to model nonlinear time series. They capture the locally explosive episodes observed in oil prices in a strictly stationary setting. While the bi-modality of the predictive density would not be detected with standard Gaussian \textit{ARMA} models, it could be detected with complex nonlinear models, but such model lacks the parsimonious characteristic of \textit{MAR} models. The data-driven prediction methods may lack theoretical grounds but provide valuable information based on the estimated model and on past behaviors of the series in a parsimonious way. This paper focuses on one-step ahead predictions of decrease in crude oil prices during the first wave of the COVID-19 pandemic. \\

\newpage
\section*{Acknowledgments}
The authors would like to thank Francesco Giancaterini, an anonymous referee and the editors for valuable comments and suggestions. All remaining errors are ours.
\bibliography{references.bib}
\newpage
\import{./}{Appendix.tex}

\end{document}

%% file: 1-Introduction.tex
\section{Introduction}\label{sec:Introduction}
This paper aims at forecasting Brent and WTI oil price series during the first wave of the COVID-19 pandemic outbreak in 2020 using the recent literature on mixed causal-noncausal autoregressive models (hereafter \textit{MAR}). Namely, time series processes with lags but also leads components and non-Gaussian errors. This new specification can, in a parsimonious way, model locally explosive episodes in a strictly stationary setting. It can therefore capture nonlinear features such as bubbles (which is defined here as a persistent increase followed by a sudden crash), often observed in commodities prices, while standard linear autoregressive models (e.g. \textit{ARMA} models) cannot do so. \textit{MAR} models have successfully been implemented on several commodity price series (see inter alia \citeNP{voisin2019forecasting}, \citeNP{hecq2020mixed}, \citeNP{fries2019mixed}, \citeNP{gourieroux2017local}, \citeNP{cubadda2019detecting}, \citeNP{lof2017noncausality}, \citeNP{karapanagiotidis2014dynamic}).\footnote{An alternative strategy to ours is to consider autoregressive processes with breaks in coefficients. Indeed, autoregressive processes with successively unit roots, explosive and stable stationary episodes are also able to capture locally explosive episodes. See among many others \citeA{phillips2011explosive} and the survey papers by \citeA{homm2012testing} or \citeA{bertelsen2019comparing}. Yet, for the purpose of forecasting, we argue for the choice of a model with constant coefficients as more adequate.} Similarly to \citeA{gourieroux2013explosive}, our goal when introducing a lead component in oil prices is not to provide an economic justification for the existence of a rational bubble. However, the link with a present value model between prices and dividends \cite{campbell1987cointegration} can enrich the discussion and it also explains the difficulties to find economic fundamentals for oil prices. This motivates our choice to use proxies such as technical methods to extract the bubble component. Let us indeed consider a general model (see \citeNP{diba1988explosive}) in which the real current stock price $P_{t}$ is linked to the present value of next period's expected stock price $P_{t+1}$, dividend payments $D_{t+1}$ and an unobserved variable $u_{t+1},$
\begin{equation}
P_{t}=\frac{1}{1+r}\mathbb{E}_{t}\big[P_{t+1}+\alpha D_{t+1}+u_{t+1}\big],  \label{PV1}
\end{equation}
with $\mathbb{E}_{t}$ the conditional expectation given the information set known at time \textit{t}. The discount factor
is $\frac{1}{1+r}$ with $r$ being a time-invariant interest rate. The general
solution of \eqref{PV1} is (e.g. \citeNP{diba1988explosive}) 
\begin{eqnarray}
P_{t}\;=\;\sum\limits_{i=1}^{\infty }\Bigg(\frac{1}{1+r}\Bigg)^{i}\mathbb{E}_{t}\big[\alpha D_{t+i}+u_{t+i}\big]+B_{t}\;=\;P_{t}^{F}+B_{t},  \label{pv3}
\end{eqnarray}
where the actual price deviates from its fundamental value $P_{t}^{F}$ by
the amount of the rational bubble $B_{t}.$ As shown by \citeA{gourieroux2020stationary}, \textit{MAR} processes provide stationary solutions for the modeling of the bubbles component in \eqref{pv3} (see also \citeNP{fries2021conditional}). \\

However, oil prices are challenging time series to forecast and model (see \citeNP{baumeister2016forty} and for a survey on oil prices forecasting see \citeNP{alquist2013forecasting}). Unlike for equity prices, measuring commodities fundamentals might not be as straightforward \cite{brooks2015booms}. \citeA{pindyck1992present} and \citeA{alquist2010we} consider the convenience yield, that is, a premium associated with holding an underlying product instead of derivative securities or contracts. It typically increases when costs associated with physical storage are low. Yet, not only is the convenience yield not easy to measure but there also are other factors driving each of the demand and supply side of crude oil: the level of stocks, economic activity, geopolitical considerations, shifts in expectations regarding the oil market, etc. While there is a large literature on modeling and forecasting the price of oil using structural models that incorporate economic fundamentals (see \citeNP{kilian2020econometrics}), our model is parsimonious and exploits the statistical properties of oil prices only.  \\

As can be seen in Figure \ref{fig:oilprices} in Section \ref{sec:Empirics}, oil prices series do not appear to be stationary over time. Consequently, before estimating \textit{MAR} models we intend to extract a smooth time-varying trend to render the series stationary without affecting the dynamics. By extracting a trend from the series we do not claim to identify the fundamental values of oil prices but instead detrend the series while preserving the dynamics of the prices in the remaining cycle and more specifically the noncausal component. As such, we obtain stationary series that retain their forward-looking aspect and which can be modeled as \textit{MAR} processes. Obviously, a wrong detrending can give misleading results if it alters the dynamics of the cycle. Consequently, investigating the impact of different technical detrending filters on the identification of \textit{MAR} models is the first contribution of this paper. Similarly to what \citeA{canova1998detrending} does for business cycles, we investigate the extent to which the identification of causal and noncausal dynamics are sensitive to different filters. We then study the consequences on the predictive densities of oil prices after applying different detrending methods. Inspired by the work of \citeA{kilian2014role}, who constructed a structural VAR model of the global market for crude oil, we make use of US crude oil Strategic Petroleum Reserve (SPR), a sub-part of total petroleum stocks, for a potential trend in oil prices in Section \ref{sec:Empirics}. Hence, the second contribution of this paper is to compare the \textit{MAR} estimations and predictions of oil price series after using technical detrending with the results obtained after detrending with the SPR levels. \\

The rest of this paper is as follows. Section \ref{sec:Discussion} describes mixed causal-noncausal models and explains the different technical detrending methods employed in this analysis, leaving the locally explosive components in the cycle. In Section \ref{sec:Montecarlo_MAR}, the impact of the different detrending filters on model identifications is investigated using a Monte Carlo study, based on trends estimated in oil prices series. We investigate the identification of the models but also the magnitude of the coefficients estimated as they are the main drivers of the predictions. Section \ref{sec:Empirics} analyzes the impact of these filters on the WTI and the Brent crude oil price series for ex-post and real-time analyses. We compare the results with those obtained after detrending with US SPR levels. We show how each detrending approach affects probabilities that oil price crashes in the period capturing the first 2020 wave of the COVID-19 pandemic. Section \ref{sec:Conclusion} concludes.

%% file: 2-Discussion.tex
\section{Mixed causal-noncausal models and filtering}\label{sec:Discussion}

\subsection{The model}
\textit{MAR}$(r,s)$ denotes dynamic processes that depend on their $r$ lags as for
usual autoregressive processes but also on their $s$ leads in the following
multiplicative form%
\begin{equation}
\Phi(L)\Psi(L^{-1})y_{t}=\varepsilon_{t},\label{eq:MAR2}%
\end{equation}
with $L$ the backward operator, i.e., $Ly_{t}=y_{t-1}$ gives lags and
$L^{-1}y_{t}=y_{t+1}$ produces leads. When $\Psi(L^{-1})=(1-\psi
_{1}L^{-1}-...-\psi{s}L^{-s})=1,$ namely when $\psi_{1}=...=\psi
_{s}=0,$ the univariate process $y_{t}$ is a purely causal autoregressive process,
denoted \textit{MAR}($r$,0) or simply \textit{AR}($r$) model, $\Phi(L)y_{t}=\varepsilon_{t}$. Reciprocally, the
process is a purely noncausal \textit{MAR}($0,s$) model $\Psi(L^{-1})y_{t}%
=\varepsilon_{t},$ when $\phi_{1}=...=\phi_{r}=0$ in $\Phi(L)=(1-\phi
_{1}L-...-\phi_{r}L^{r}).$ The roots of both the causal and noncausal
polynomials are assumed to lie outside the unit circle, that is $\Phi(z)=0$
and $\Psi(z)=0$ for $|z|>1$ respectively. These conditions imply that the
series $y_{t}$ admits a two-sided moving average representation
$y_{t}=\sum_{j=-\infty}^{\infty}\gamma_{j}\varepsilon_{t-j},$ such that
$\gamma_{j}=0$ for all $j<0$ implies a purely causal process $y_{t}$ (with
respect to $\varepsilon_{t}$) and a purely noncausal model when $\gamma_{j}=0$
for all $j>0$ \cite{lanne2011noncausal}. Error terms $\varepsilon_{t}$ are
assumed $iid$ (and not only weak white noise) non-Gaussian (with potentially infinite variance) to ensure the identifiability of the causal and the noncausal parts (\citeNP{breid1991maximum}, \citeNP{gourieroux2015uniqueness}). While noncausal models are strictly stationary, their conditional moments are time-varying. A purely stationary noncausal \textit{MAR}(0,1) Cauchy-distributed process, has a unit root in its conditional mean and exhibit ARCH-type effects (see \citeNP{gourieroux2017local} and \citeNP{cavaliere2018bootstrapping}). \\

\begin{figure}[h!]
    \centering
    \begin{subfigure}{0.45\textwidth}
      \centering
      \caption{\footnotesize \textit{MAR}(1,0) with $\phi=0.8$}
      \includegraphics[width=\linewidth,trim={0cm 0cm 0cm 0cm},clip]{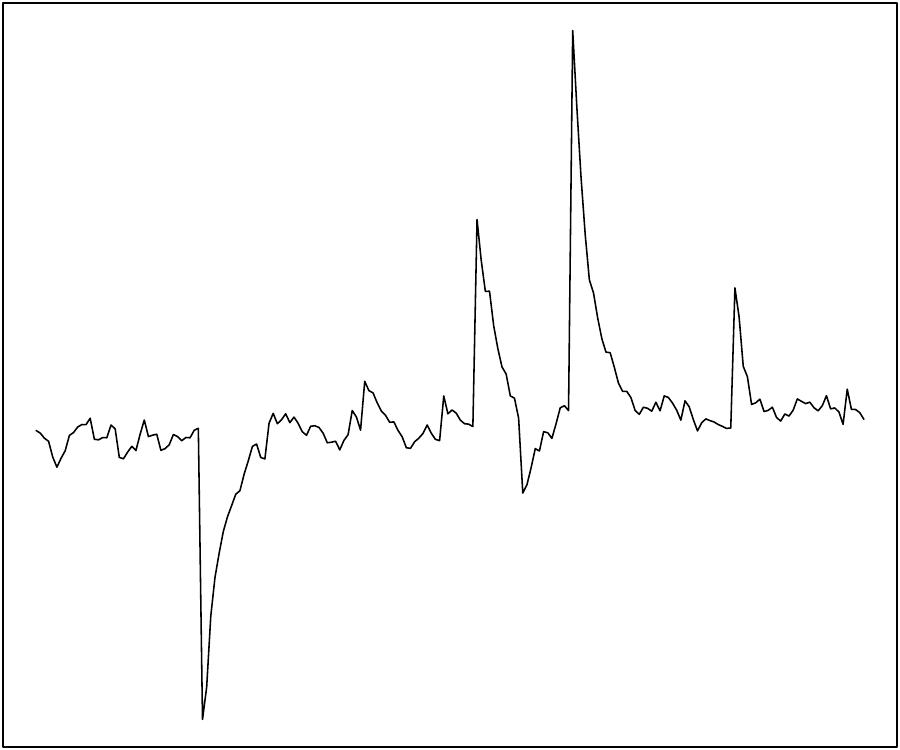}
    \end{subfigure}%
    \hspace{0.8cm}
    \begin{subfigure}{0.45\textwidth}
      \centering
        \caption{\footnotesize \textit{MAR}(0,1) with $\psi=0.8$}
      \includegraphics[width=\linewidth,trim={0cm 0cm 0cm 0cm},clip]{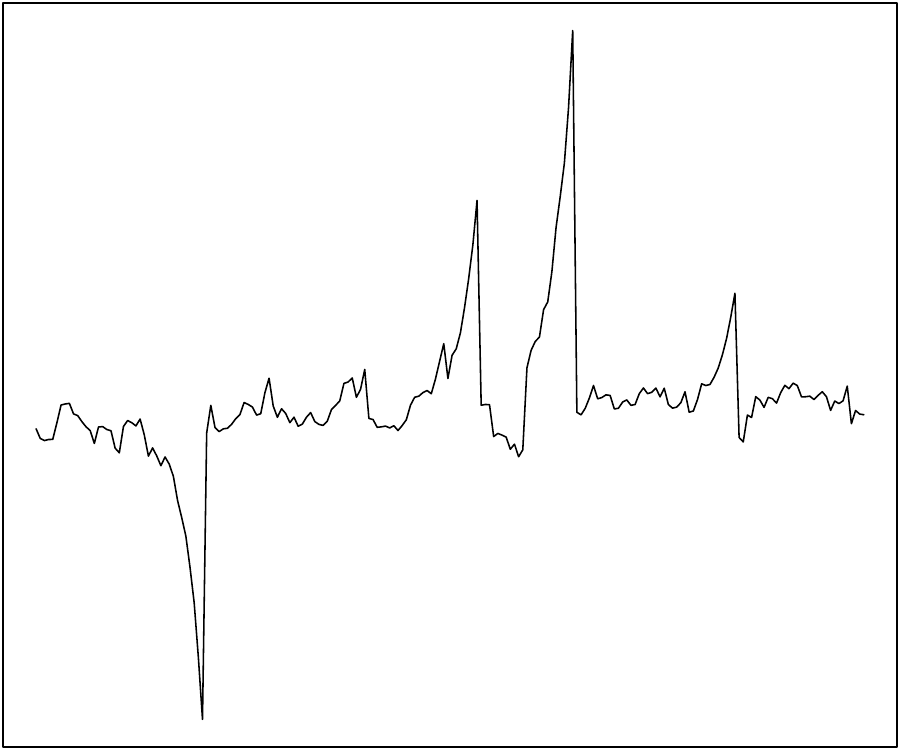}
    \end{subfigure}
    \caption{Purely causal (a) and noncausal (b) trajectories}
    \label{fig:causalvsnoncausal}
\end{figure}

Figure \ref{fig:causalvsnoncausal} shows a purely causal (a) and a purely noncausal (b) trajectories induced by the same Student's \textit{t}(2)-distributed errors, both with coefficient 0.8 and 200 observations. For the purely causal process, a shock is unforeseeable and affects the series only once it happened, inducing a large jump in the series. On the other hand, for purely noncausal processes, a shock impacts the process ahead of time, mirroring the purely causal trajectory. Indeed, we see that the series already reacts to a positive shock by increasing until a sudden crash, creating bubble patterns. This anticipative aspect is widely observed in financial and economics time series. The detrended Brent crude oil prices as shown in Figure \ref{fig:detrended_series} noticeably exhibit such features, the most apparent episode being the 2008 financial crisis. A combination of causal and noncausal dynamics consequently creates some asymmetry around a shock, varying with the magnitude of the respective coefficients. \\

The advantage with oil prices is that they already underwent bubbles in the past, and those previous locally explosive episodes will help identifying \textit{MAR} models. In the case where series are for the first time following a long and abnormal increase, an explosive process is difficult to distinguish from a stationary locally explosive one. \\

The focus of this paper is on the probabilities of crashes. Predictions are performed using the approximation methods of \citeA{filtering} and \citeA{lanne2012optimal} since no closed-form of the predictive density exists when the errors of the process follow a Student'\textit{t} distribution. For a detailed analysis of the two approximation methods see \citeA{voisin2019forecasting}.\footnote{A description of how the methods are used in this analysis can be found in Appendix A in the online material.}

\subsection{Filtering the data}
The requirement of $y_{t}$ being stationarity for both lag and lead
polynomials gave rise to different strategies to transform nonstationary
series to stationary ones. \citeA{hecq2020mixed} and \citeA{cubadda2019detecting} assume\footnote{The locally explosive features of the data make unit root tests
doubtful.} that their commodity price series are \textit{I}(1) and work with the returns
$\Delta y_{t}$. However, this operation eliminates most of the locally explosive behaviors
and the transformed series consist of many spikes instead.\\

In this paper, we capture the trending behavior of the observed
series denoted $\tilde{y}_{t}$ in different ways using the general form%
\begin{equation*}
    \tilde{y}_{t}  = f_t +y_{t},
\end{equation*}
where
\begin{equation*}
    \Phi(L)\Psi(L^{-1})y_{t}  =\varepsilon_{t}.
\end{equation*}

In this framework, $\tilde{y}_{t}$ is the (potentially nonstationary)
observed series and $f_t$ a generic trend function. The deviation of
$\tilde{y}_{t}$ from its trend is an \textit{MAR}($r,s$) process. Several authors, although
sometimes not explicitly, use this decomposition. \citeA{cavaliere2018bootstrapping} opt for the choice of a particular time period with no trend and hence
use only an intercept $f_t=\mu$. \citeA{hencic2015noncausal} detrend $\tilde{y}_{t}$ using a polynomial trend
function of order three. In summary, we could consider several choices among the following deterministic trends,
\begin{align*}
f_t^{(1)}  & =\mu,\\
f_t^{(2)}  & =\mu+\beta D_{t}, \quad \text{ with }D_{t}=1\text{ when
}t\geq t_{break\text{ }}\text{ and }0\text{ otherwise,}\\
f_t^{(3)}  & =\alpha_{0}+\alpha_{1}t+...+\alpha_{k}t^{k}, \quad \text{ with } k \text{ some positive integer and } t=1,2,\dots,T.
\end{align*}
Note (see Section \ref{sec:Empirics}) that since a larger order of polynomial allows for more flexibility, we consider polynomial
trends of order four and six for the trending pattern of the monthly oil prices series considered in this analysis. More complex trends, constructed as a combination of the aforementioned examples could also be considered, such as (multiple) breaks in trends for instance.\\

\citeA{voisin2019forecasting} use the Hodrick-Prescott filter (HP) before detecting bubbles in Nickel monthly prices. The HP filter, as opposed to the aforementioned deterministic trends, extracts the trend process $f^{(4)}_t$ via a minimization that relies on a penalizing parameter denoted $\lambda$.
\begin{equation*}
    \min_{\{f^{(4)}_t\}^T_{t=1}}\Bigg\{\sum_{t=1}^T\big(\tilde{y}_t-f^{(4)}_t\big)^2+\lambda\sum^T_{t=3}\Big[\big(f^{(4)}_t-f^{(4)}_{t-1}\big)-\big(f^{(4)}_{t-1}-f^{(4)}_{t-2}\big)\Big]^2\Bigg\}.
\end{equation*}
The larger this parameter, the smoother the trend component is (that is, with $\lambda$ approaching infinity, the extracting trend becomes linear). For details about the HP filter see \citeA{hodrick1997postwar}. It is now commonly accepted to use $\lambda=1\,600$ for quarterly data. For other frequencies, the rule of thumb consists in adjusting the parameter to the frequency relative to quarterly data,
$$\lambda=\Bigg(\frac{\text{number of observations per year}}{4}\Bigg)^i\times1\,600,$$
with either $i=2$ \cite{backus1992international} or $i=4$ \cite{ravn2002adjusting}, yielding respectively a penalizing parameter of 14\,400 and 129\,600 for monthly series. Most criticisms of the HP filter concern its application on series with complicated stochastic and deterministic trends. \citeA{phillips2019boosting} propose an adaptation of the filter improving its accuracy for such series.\footnote{In our case, the proposed boosting algorithm absorbs too much dynamics and captures the bubble in the trend component.} We investigate in Section \ref{sec:Montecarlo_MAR} the potential dynamic distortions that can be induced by HP filtering (see among others \citeNP{hamilton2018you}) but find no significant distortions of the mixed causal-noncausal dynamics. \\

Note that we are not interested in the exact value of a forecast but rather in its direction and potential magnitude. This is why we extract smooth trends to preserve the dynamics in the series. This allows to estimate predictive densities of oil prices based on the statistical properties of the data alone in a parsimonious way, and not from the construction of complicated structural models. However, wrongly detrending the series could have a significant impact on the estimation of the noncausal dynamics of the process, which could in turn strongly under- or over-estimate the longevity of explosive episodes and therefore of the probabilities of crashes and of turning points.

%% file: 3-Montecarlo_MAR.tex
\newcolumntype{d}[1]{D{.}{.}{#1} }
\section{Monte Carlo analysis - Effects of detrending}\label{sec:Montecarlo_MAR}
The aim of this section is to analyze the effect of wrongly detrending a series, both on the identification of the \textit{MAR} model and on the subsequent predictions performed with the resulting model. We base this analysis on stylized facts observed in oil prices series. \\

\subsection{Accuracy of detrending}
\begin{figure}[h!]
    \centering
    \includegraphics[width=0.8\linewidth]{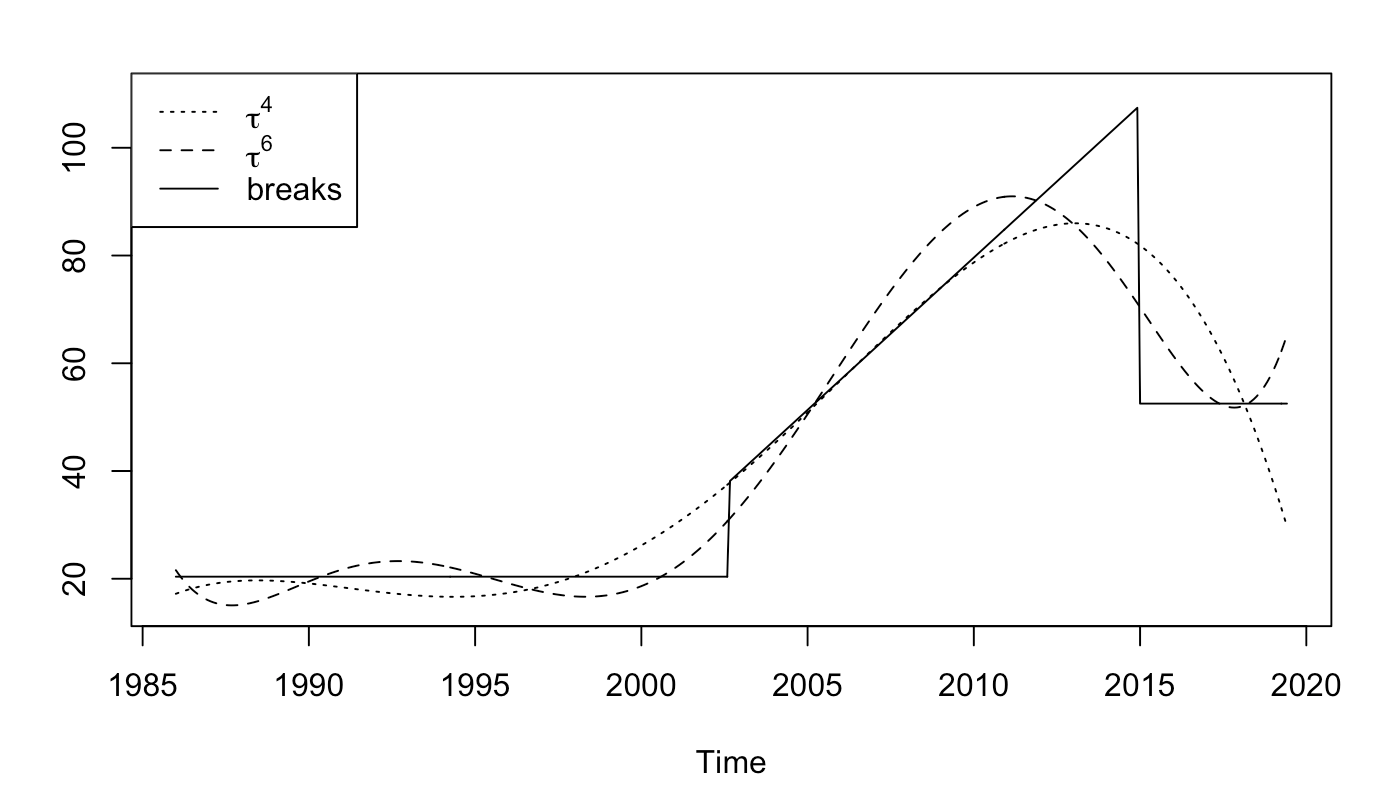}
    \vspace{-0.1cm}
    \caption{Trends estimated on WTI oil prices series}
    \label{fig:MC_trends}
\end{figure}

We simulate 5\,000 trajectories for 12 distinct data generating processes (hereafter \textit{dgp}), composed of a trend and a stationary dynamic process denoted as cycle. All \textit{dgp}s are generated by Student's \textit{t}-distributed errors with 2 degrees of freedom, a value frequently observed in financial time series, and with 400 observations. For the cycles, we consider purely noncausal processes with a lead coefficient of 0.8, purely causal processes with a lag coefficient of 0.6 and mixed causal-noncausal processes with a lag coefficient of 0.6 and a lead coefficient of 0.8. The heavy-tailed distribution generates extreme values, inducing bubble-like phenomena in processes with noncausal components. We are interested in mostly forward looking processes characterized by long lasting bubbles hence the choice of coefficients. We consider three different deterministic trends: a linear trend with breaks (denoted \textit{breaks}) and two trend polynomials up to orders 4 and 6 (denoted respectively $\tau^4$ and $\tau^6$ for simplicity). The coefficients of the trends were estimated on the monthly WTI crude oil prices series between 1986 and 2019. Figure \ref{fig:MC_trends} depicts the three mentioned trends to which purely causal, noncausal and mixed causal-noncausal trajectories are added. Additionally, we consider processes with an intercept only. This results overall in 12 sets of 5\,000 trajectories of the form $\tilde{y}_t=f_t+y_t$. \\

Four detrending methods are employed for each trajectories, with the general form $\tilde{y}_t=\hat{f}_t+\hat{y}_t$. Estimated polynomial trends of orders 4 and 6 and HP filters with $\lambda=14\,000$ and $\lambda=129\,600$ are applied (respectively denoted $t^4$, $t^6$, $HP_1$ and $HP_{2}$).\footnote{The estimated trend polynomials are denoted $t^4$ and $t^6$ to distinguish them from the trend polynomials part of the \textit{dgp}s $\tau^4$ and $\tau^6$.} To gauge and compare the accuracy of the detrending methods, Table \ref{tab:MSEs} shows the average mean square errors (MSE) between the true cycle of $\tilde{y}_t$ ($y_t$) and the one obtained after detrending ($\hat{y}_t$). The average MSEs are computed over the 5\,000 replications of each \textit{dgp} and for the four detrending approaches,
\begin{equation*}
    MSE_{k,d}=\frac{1}{5\,000}\sum_{i=1}^{5\,000}\frac{1}{400}\sum_{t=1}^{400}(y^{(k,i)}_t-\hat{y}^{(k,i,d)}_t)^2,
\end{equation*}
where \textit{k} indicates the \textit{dgp}, \textit{d} the detrending method used, and \textit{i} the \textit{i}-th replication with $1\leq i\leq5\,000$. \\

The MSEs are minimized when the correct polynomial trend is employed or when the lower order is employed (4 in this case) in the absence of trend in the \textit{dgp}. However, underestimating the order of the polynomial trend leads to significantly larger discrepancies.  Distortions between the true cycle and the detrended series are larger for mixed causal-noncausal processes than for purely causal or noncausal processes. Furthermore,  in the presence of noncausal dynamics the HP filter with $\lambda=14\,400$ ($HP_1$) distorts more the series than $HP_2$. Hence, we can expect that a low penalizing parameter in the HP filter mostly captures some of the noncausal dynamics. However, $HP_1$ distorts the least the cycles to which the linear trend with breaks was added. It is the method that best manages to mimic this non-smooth trend due to this flexibility induced by its low penalizing parameter.  \\

\begin{table}[h]\small
    \caption{Average Mean Squared Errors between true cycles and detrended series}
	\centering
	\begin{threeparttable}
    	\begin{tabular}{lld{2}d{2}d{2}d{2}}
            \hline \hline 
            \multirow{2}{*}{\textit{DGP}} && \multicolumn{4}{c}{Detrended with} \\
              && \multicolumn{1}{c}{$t^4$} & \multicolumn{1}{c}{$t^6$} & \multicolumn{1}{c}{$HP_1$} & \multicolumn{1}{c}{$HP_2$}\\\hline
            \textit{MAR}(0,1) + no trend    &&5.23    &7.61   &11.44  &7.15\\
            \textit{MAR}(0,1) + $\tau^4$       &&4.55    &6.03   &9.62   &7.50\\
            \textit{MAR}(0,1) + $\tau^6$       &&62.42   &6.38   &11.35  &11.26\\
            \textit{MAR}(0,1) + breaks      &&79.02   &55.78  &31.84  &47.65\\
            &\\
            \textit{MAR}(1,1) + no trend    &&22.69   &31.05  &48.58  &30.81\\
            \textit{MAR}(1,1) + $\tau^4$       &&42.74   &65.18  &91.42  &57.60\\
            \textit{MAR}(1,1) + $\tau^6$       &&85.91   &39.57  &61.21  &43.02\\
            \textit{MAR}(1,1) + breaks      &&101.48  &86.93  &78.36  &77.18\\
            &\\
            \textit{MAR}(1,0) + no trend    &&1.20    &1.64   &2.55   &1.58\\
            \textit{MAR}(1,0) + $\tau^4$       &&0.96    &1.34   &2.14   &2.70\\
            \textit{MAR}(1,0) + $\tau^6$       &&59.24   &2.45   &4.21   &6.73\\
            \textit{MAR}(1,0) + breaks      &&76.42   &52.19  &26.30  &44.10\\\hline
        \end{tabular}
        \begin{tablenotes}
    	    \item \scriptsize Notes: Are reported the average MSEs over 5\,000 trajectories with sample size $T=400$. $HP_1$ corresponds to the HP filter with $\lambda=14\,400$ and $HP_2$ to the HP filter with $\lambda=129\,600$.
    	\end{tablenotes}
    \end{threeparttable}	
    \label{tab:MSEs}
\end{table}

\subsection{Effects of detrending on model identification}
To investigate the impact of detrending on dynamic processes, we perform \textit{MAR} estimations on the raw and detrended series from each \textit{dgp}. The estimation of \textit{MAR} models first consists in estimating the pseudo causal lag order. Since the autocorrelation structure of mixed or purely causal and noncausal processes are identical, we can estimate the order of autocorrelation (\textit{p}) with information criteria by OLS. Once this order \textit{p} is estimated, the identification of the lag and lead orders (\textit{r} and \textit{s} respectively) is performed by maximum likelihood among all \textit{MAR}(\textit{r},\textit{s}) models such that $r+s=p$ \cite{lanne2011noncausal}. We do so using the MARX package in R \cite{MARX}. \\

Table \ref{tab:estimated_MARs_pmax} presents the frequencies of identifying wrong models in each of the 12 \textit{dgp}, based on the detrending methods, with a maximum pseudo causal lag order of 4.\footnote{Results when the pseudo lag order is fixed to the correct one ($p=1$ or $p=2$ for mixed models) are available upon request.} Proportions of a wrongly identified the pseudo lag order in the first step of the estimation using BIC are reported ($p\neq1$ and $p\neq2$), as well as the proportions of wrongly identified \textit{MAR} models, namely when at least one of the lag or lead order mis-identified. We also report the frequency with which no noncausal dynamics is identified ($s=0$). For the purely causal processes we only report in the last column ($s>0$), i.e. the frequency with which spurious noncausal dynamics is detected. \\

Let us first focus on the models with noncausal dynamics (the \textit{MAR}(0,1) and \textit{MAR}(1,1) \textit{dgp}s) for which we report the frequencies with which we over- or underestimate the pseudo causal lag order in the first step of the estimation. We can see that $HP_1$ under-performs relative to the other approaches. Indeed, around twice as many lag orders are wrongly estimated in the first step on average, with a maximum of 22.84\% for the \textit{MAR}(1,1) processes with breaks in the linear trend. However, this non-smooth trend seems to be difficult to capture by the filters considered in this analysis. We can see from the last five rows of Table \ref{tab:estimated_MARs_pmax} that detrending this type of processes with breaks -- with the four methods employed here -- does not improve the correct identification of the orders of the model, and can even make it worse for \textit{MAR}(1,1). This can be explained by the construction of the trend, mimicking somehow a bubble pattern, with a long and persistent expansion when the linear trend is present and followed by a sudden crash when the series returns to a stationary process. This might be mistaken for noncausal dynamics, ensuring a non zero lead order identification when the series is not detrended. This claim is supported by the results in the last column, indicating large proportions of wrongly detected noncausal dynamics for each detrending approaches, with 7.54\% for $HP_1$ and more than 28\% for the others. For the \textit{dgp}s with other trends (or only intercept) $HP_1$ wrongly estimates the pseudo causal lag order at most 10.78\% of the time. For the three other detrending methods the pseudo lag order is wrongly identified in less than 7.3\% of the cases. Note that when the lag order is wrongly identified, it is almost always due to over-identification. The discrepancy between the two HP filters is explained by the low penalizing parameter in $HP_1$ allowing the trend to mimic the series too much. By that, some of the dynamics of the \textit{MAR} process are absorbed by the trend. \\

It is notably more harmful not to detrend when necessary than the contrary. As can be seen on the upper rows of Table \ref{tab:estimated_MARs_pmax}, applying polynomial trends or $HP_2$ do not increase the proportions of wrongly identified models by more than 1.6\% compared to estimations on the raw series. However, when the existing trend is ignored, the pseudo lag order is wrongly estimated twice as much on the raw series than for the detrended series, and the \textit{MAR} models are wrongly identified up to 6 times more than the best performing detrending method. Furthermore, the incorrect identification of the pseudo lag order \textit{p} accounts for most of the proportion of wrongly identified \textit{MAR} models. If \textit{p} is correctly estimated, the model is also correctly identified in more than 99\% of the cases. Note that the pseudo causal lag order identified is never zero, meaning that no detrending completely absorbs all dynamics. Besides, in no more than 0.62\% the detrending methods killed the noncausal dynamics, as is indicated by the columns $s=0$. \\

Let us now consider the last column, displaying the results for purely causal processes. We here investigate whether detrending can create spurious noncausal dynamics ($s>0$). We find that (ignoring the \textit{dgp} composed of the trend with breaks) as long as the polynomial trend order is not underestimated, in less than 3.46\% of the cases noncausal dynamics was wrongly detected. For the processes with a polynomial trend of order 6, detrending with a polynomial trend of order 4 creates spurious noncausal dynamics in 60.02\% of the cases. \\

\begin{table}[h!]\small
    \caption{Percentages of mis-identified \textit{MAR} models}
	\centering
	\resizebox{\textwidth}{!}{%
	\begin{threeparttable}
    	\begin{tabular}{cc d{2}d{2}d{2}d{2}d{2}d{2}d{2}cc}
            \hline \hline 
            Detrending &&  
            \multicolumn{1}{c}{\multirow{2}{*}{$p\neq 1$}} & \multicolumn{1}{c}{\multirow{1}{*}{wrong}} & \multicolumn{1}{c}{\multirow{2}{*}{$s=0$}} && 
            \multicolumn{1}{c}{\multirow{2}{*}{$p\neq 2$}} & \multicolumn{1}{c}{\multirow{1}{*}{wrong}} & \multicolumn{1}{c}{\multirow{2}{*}{$s=0$}} &&
            \multicolumn{1}{c}{\multirow{2}{*}{$s>0$}} \\
            method &&&\multicolumn{1}{c}{\multirow{1}{*}{\textit{MAR}}}&&&&\multicolumn{1}{c}{\multirow{1}{*}{\textit{MAR}}}&&&\\\hline
            &\\
            &&\multicolumn{3}{c}{\textit{MAR}(0,1) + no trend}    &&\multicolumn{3}{c}{\textit{MAR}(1,1) + no trend} &&\multicolumn{1}{c}{\textit{MAR}(1,0) + no trend} \\\hline
            raw         && 5.50   & 5.50   & 0.00     && 4.74  & 4.74  & 0.00     && {\color{white}0}0.52 \\
            $t^4$       && 5.46   & 5.58   & 0.14  && 4.64  & 4.68  & 0.04  && {\color{white}0}0.72 \\
            $t^6$       && 5.70   & 5.86   & 0.22  && 4.94  & 5.04  & 0.06  && {\color{white}0}0.88 \\
            $HP_1$          && 10.78  & 11.24  & 0.52  && 8.18  & 8.44  & 0.22  && {\color{white}0}1.86 \\
            $HP_2$ && 6.84   & 7.10   & 0.28  && 5.56  & 5.66  & 0.06  && {\color{white}0}1.02\\\hline
            &\\
            &&\multicolumn{3}{c}{\textit{MAR}(0,1) + $\tau^4$}    &&\multicolumn{3}{c}{\textit{MAR}(1,1) + $\tau^4$} &&\multicolumn{1}{c}{\textit{MAR}(1,0) + $\tau^4$}\\\hline
            raw         && 12.10  & 43.84  & 35.04  && 9.70  & 16.38  & 7.22  && 32.76 \\
            $t^4$       && 6.44   & 6.72   & 0.28   && 4.70  & 4.74   & 0.04  && {\color{white}0}0.78\\
            $t^6$       && 6.76   & 6.96   & 0.20   && 4.86  & 4.94   & 0.08  && {\color{white}0}0.76 \\
            $HP_1$          && 10.28  & 10.88  & 0.60   && 7.64  & 7.90   & 0.22  && {\color{white}0}2.52 \\
            $HP_2$ && 6.24   & 6.56   & 0.32   && 5.36  & 5.56   & 0.14  && {\color{white}0}1.92 \\\hline
            &\\
            &&\multicolumn{3}{c}{\textit{MAR}(0,1) + $\tau^6$}    &&\multicolumn{3}{c}{\textit{MAR}(1,1) + $\tau^6$} &&\multicolumn{1}{c}{\textit{MAR}(1,0) + $\tau^6$}\\\hline
            raw         && 13.18  & 36.14  & 26.04  && 9.04  & 15.56  & 7.12  && 35.44 \\
            $t^4$       && 7.30   & 7.36   & 0.08   && 4.00  & 4.14   & 0.04  && 60.02 \\
            $t^6$       && 6.54   & 6.68   & 0.14   && 4.48  & 4.68   & 0.04  && {\color{white}0}0.92 \\
            $HP_1$          && 9.40   & 9.84   & 0.44   && 7.90  & 8.24   & 0.22  && {\color{white}0}2.86 \\
            $HP_2$ && 5.94   & 6.12  & 0.18    && 4.86  & 5.12   & 0.06  && {\color{white}0}3.46 \\\hline
            &\\
            &&\multicolumn{3}{c}{\textit{MAR}(0,1) + breaks}    &&\multicolumn{3}{c}{\textit{MAR}(1,1) + breaks} &&\multicolumn{1}{c}{\textit{MAR}(1,0) + breaks}\\\hline
            raw         && 4.54  & 4.92  & 0.68  && 6.34   & 7.68   & 1.38  && 94.60\\
            $t^4$       && 3.44  & 4.00  & 0.60  && 8.68   & 8.74   & 0.22  && 38.24 \\
            $t^6$       && 3.40  & 3.86  & 0.58  && 10.70  & 10.86  & 0.24  && 28.24 \\
            $HP_1$          && 4.00  & 4.58  & 0.62  && 22.84  & 23.24  & 0.26  && {\color{white}0}7.54 \\
            $HP_2$ && 3.38  & 3.70  & 0.40  && 12.70  & 12.82  & 0.18  && 28.92 \\\hline
        \end{tabular}
        \begin{tablenotes}
    	    \item \scriptsize Notes: During the first stage of the model identification, the maximum number of lags in the pseudo lag model is set to 4. Results are in percentages of the 5\,000 trajectories. $T=400$. $HP_1$ corresponds to the HP filter with $\lambda=14\,400$ and $HP_2$ to the HP filter with $\lambda=129\,600$.
    	\end{tablenotes}
    \end{threeparttable}%
        }	
    \label{tab:estimated_MARs_pmax}
\end{table}

Overall, for a \textit{dgp} with noncausal dynamics, the impact of ignoring a trend is quite significant while detrending when not necessary has negligible effects on model identification. Both the polynomial trends and the HP filter with $\lambda=129\,600$ ($HP_2$) perform equally well with respect to identifying the correct orders of the model. Choosing a penalizing parameter $\lambda$ too low alters the dynamics of the process as shown by the results from $HP_1$. All of the approaches almost always retain the noncausal dynamics, but rarely create spurious noncausal dynamics when nonexistent in the \textit{dgp} (except when the polynomial trend order is underestimated). The lead order is not always the correct one but in less than 0.62\% for all cases no noncausal dynamics is found. The presented results only report identification of the model lag and lead orders. To have a better understanding of the impact of the detrending methods on the dynamics, focus needs to be put on the impact on the estimated coefficients and parameters of the models identified. \\

Detailed results on the impact on estimated coefficients are available in Appendix B in the online material. Overall, we find that due to low penalization, $HP_1$ absorbs too much of the dynamics (mostly the noncausal ones) in the resulting trend. Hence, for monthly data, we advise to use the HP filter with penalization parameter 129\,600. It is also rather harmful to underestimate the order of the polynomial trend, which results in a significantly larger lead coefficient. When the fundamental trend consists of breaks (mimicking bubbles), the smooth detrending methods do not succeed in capturing the trend and this translates in much more persistent noncausal dynamics. We also investigate the effect of detrending white noise series; while for the raw series, 6.82\% of the models were identified with dynamics, 7.34\% were identified with dynamics for the HP filtered series with penalizing parameters 129\,600. Hence we find no significant creation of dynamics when applying the HP filter to a white noise.

%% file: 4-Empirics.tex
\section{Predicting crashes in oil prices}\label{sec:Empirics}
\begin{figure}[h!]
    \centering
    \includegraphics[width=0.9\linewidth]{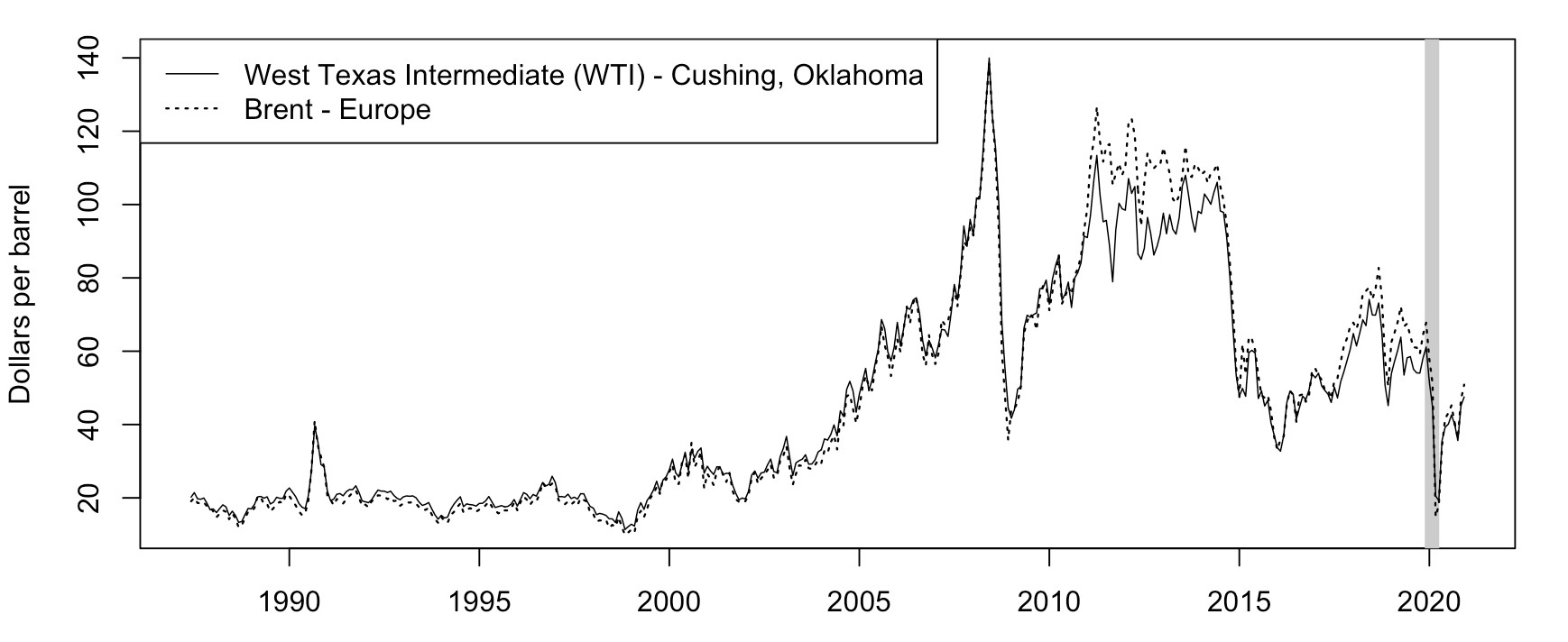}
    \vspace{-0.5cm}
    \caption{Monthly crude oil prices}
    \label{fig:oilprices}
\end{figure}

This section investigates the impact of detrending both for in-sample and real-time analyses. WTI and Brent crude oil monthly prices series are employed, ranging from June 1987 to December 2020. The series consist of end-of-period prices, which enables us to adequately time our analysis based on the outbreak of the COVID-19 pandemic and the appearances of worldwide regulations and lock-downs to counter its spread. Figure \ref{fig:oilprices} shows that both series are characterized by bubble episodes, which we define in this paper as rapidly increasing episodes followed by a sharp decline, the main one being during the financial crises in 2008. The series are also characterized by various sudden crashes. The highlighted gray bar represents the period of interest in this analysis. The earliest point of the period is December 2019; at this point almost no information was available on the coronavirus and no worldwide outbreak had already taken place. Then, we can see that as the outbreak started and regulations were increasingly being imposed worldwide, the price of crude oil significantly dropped. Brent crude oil prices fell from around \$68 at the end of December 2019 to around \$15 by the end of March 2020, point at which most European countries imposed national lock-downs. The restrictions of movement within and between countries thus induced a sharp and sudden decrease in the demand for crude oil. \\

As shown in Figure \ref{fig:oilprices}, the series are probably nonstationary but considering their growth rate would eliminate the locally explosive episodes that are interesting to exploit. The two series appear almost identical until the 2008 financial crisis, period from which we can observe more apparent discrepancies. The last part of the samples is rather noisy and volatile, and estimating a trend on such a part is not straightforward.\footnote{A Figure of the prices deflated with the consumer price index can be found in Appendix C in the online material.} We seek to extract a smooth trend without affecting the dynamics of the series. Based on the findings of Section \ref{sec:Montecarlo_MAR}, we consider the deterministic polynomial trends of orders 4 and 6 as well as the HP filter with $\lambda=129\,600$ (denoted $t^4$, $t^6$ and $HP$ respectively). We furthermore employ an economic variable -- described in the following section -- as another trend to compare economically motivated detrending with mechanical detrendings. The analysis focuses on the probabilities for oil prices to drop and investigates the potential magnitude of such decrease. We first consider an in-sample analysis, that is, the trends and the \textit{MAR} models are estimated over the whole sample, from June 1987 to December 2020. Then, we fix the estimated parameters and use this information to perform one-month ahead density forecasts for the months of January, February, March and April 2020. The in-sample analysis includes as much information as possible and therefore reduces estimation uncertainty. We then compare the in-sample analysis to a real-time forecast exercise. In the real time analysis, we re-estimate the trends and the \textit{MAR} models at each point of the period of interest. That is, we consider an expanding sample and perform one-month ahead density forecasts for points that are out-of-sample.  

\subsection{Economic variables to detrend series}
\begin{figure}[h!]
    \centering
      \includegraphics[width=\linewidth,trim={0cm 1cm 0cm 0cm},clip]{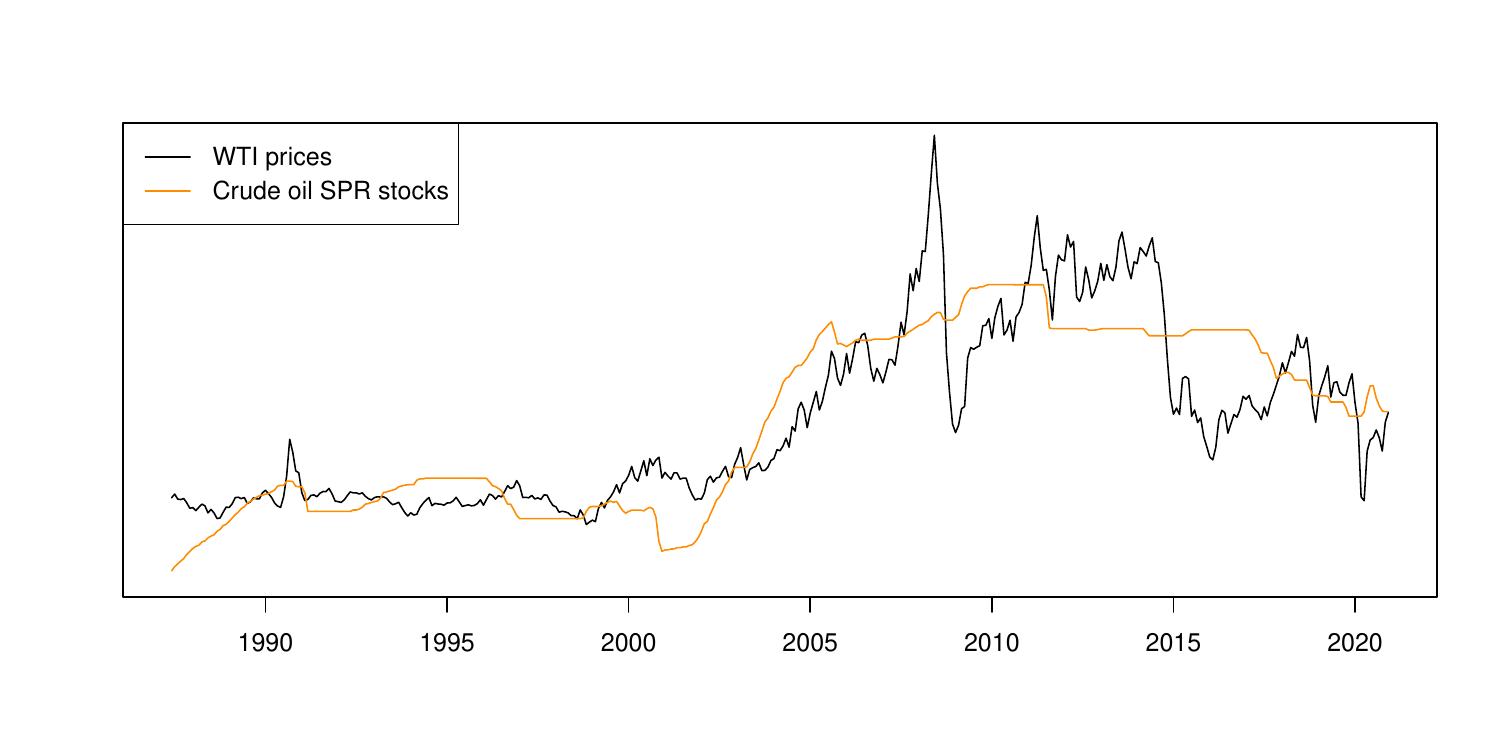}
    \caption{Raw WTI prices and US crude oil SPR stocks}
    \label{fig:SPR}
\end{figure}

There is an extensive literature on modeling oil prices using economic variables. As an example, \citeA{kilian2014role} construct a structural VAR model for the real price of oil, making use of stationary transformations of economic variables, namely the real economic activity index constructed in \citeA{kilian2009not} as well as inventories and production of crude oil. In this analysis we however do not construct a structural model for the price of oil, but instead we investigate ways of detrending prices without altering the inherent dynamics of the process. As such, we suggest employing the US crude oil Strategic Petroleum Reserve (hereafter SPR) levels. These reserves were established primarily to reduce the impact of disruptions in supplies of petroleum stocks \cite{kilian2020does}. This variable therefore incorporates not only expectations regarding the economic activity but also regarding the production of crude oil. US SPR stock is depicted against WTI crude oil prices in Figure \ref{fig:SPR}.  \\

SPR is significantly less volatile than total crude oil stocks as it is a last resort reserve and is not often made use of as it requires approval of the US President.\footnote{Limited release can be allowed by the Secretary of Energy for crude oil loans to non-governmental entities, as is described by the Energy department of the US.} This characteristic of the series makes it a good candidate for the smooth trend we intend to extract from oil price series. We hence detrend prices (both nominal and real) by taking the residuals from a standard OLS regression of prices on crude oil SPR levels.\footnote{As shown in Figure \ref{fig:oilprices}, WTI and Brent price series seem to follow a similar trend; we therefore also employ US SPR stocks to detrend Brent prices. We find statistical support for cointegration between prices (both nominal and real prices of WTI and Brent) and US SPR levels. Hence the remaining cycles are, as intended, stationary.}

\subsection{In-sample analysis}
\begin{figure}[h!]
    \centering
    \includegraphics[width=\linewidth,trim={0cm 1cm 0cm 0cm},clip]{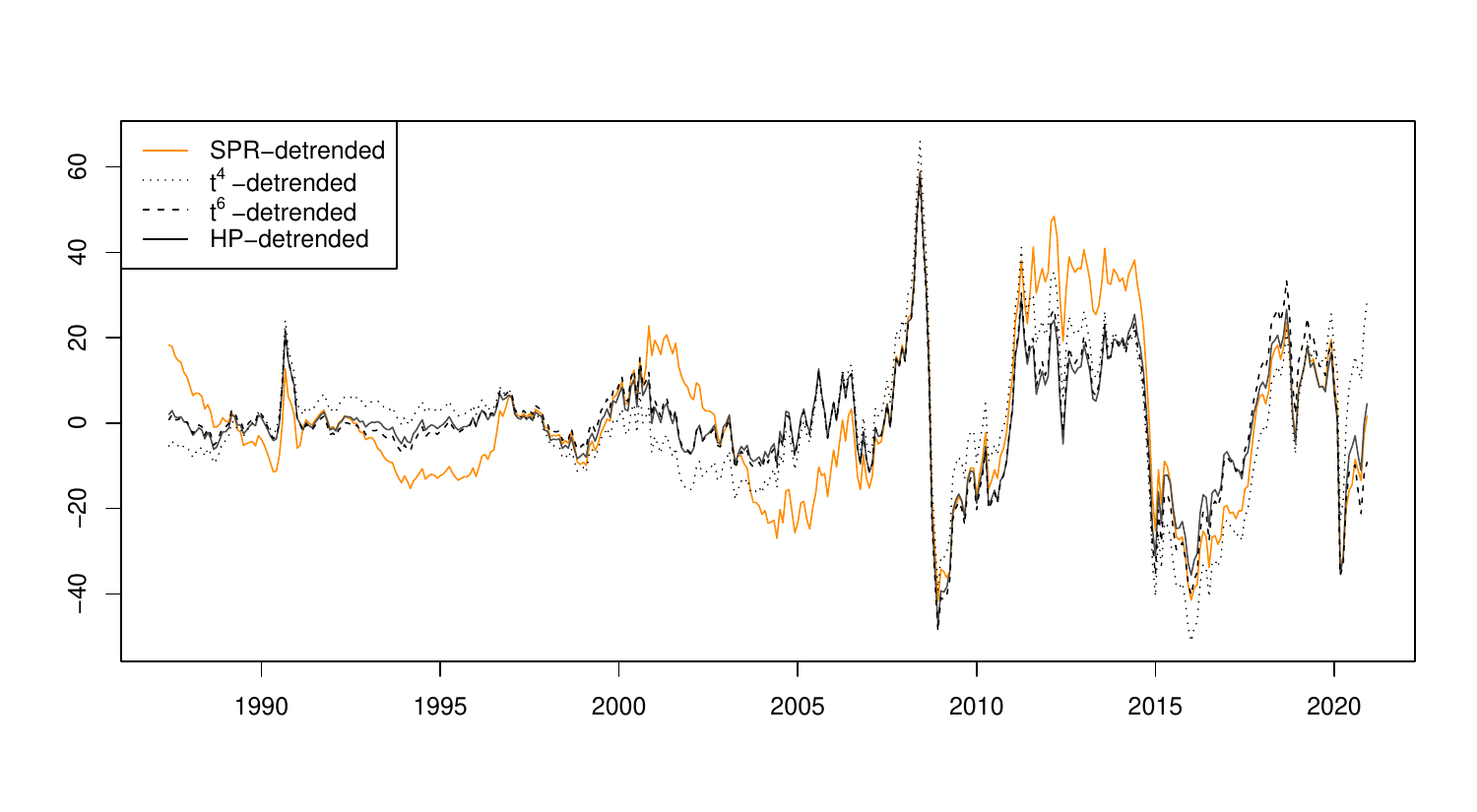}
    \caption{Detrended Brent prices}
    \label{fig:detrended_series}
\end{figure}

To save space, Figure \ref{fig:detrended_series} only depicts the detrended Brent series,\footnote{Data and results for the prices-adjusted series that are not presented here are available upon request.} after the polynomial trends, the HP filter and SPR levels were used to detrend the whole sample. The \textit{SPR}-detrended series consists of the residuals obtained from a standard OLS regression of the prices on the SPR levels. We can see that the \textit{HP}-detrended series (black solid line) and the $t^6$-detrended series (dashed line) are very much alike over the majority of the sample. The polynomial trend of order 4 (dotted line) however seems to induce some more variations than the other two mechanical detrending. This is especially visible at the beginning and at the end of the sample, stemming from the lack of flexibility of such trend due to its lower order. This latter detrending method suggests that the end of 2020 is as extreme as the period between 2010 and 2014, during which prices were in fact twice as large. We can see that overall \textit{SPR}-detrended series follows a similar pattern than the others but shows slightly more persistent dynamics. This could stem from the fact that the SPR series, while being rather smooth, still displays more dynamics than the 3 other trends considered here. Hence, it could slightly alter the dynamics in the remaining cycle. \textit{SPR}-detrended series has a correlation of 0.84 with both \textit{HP}- and $t^6$-detrended series. Note that until the end of the 1980s, there was a persistent increase in SPR due to the creation and initial filling of the reserves which started in 1977, explaining the induced downward trend at the beginning of the detrended sample.  \\

We estimate \textit{MAR} models with Student's \textit{t}-distributed errors and set the maximum pseudo lag length in the first stage on the estimation to 4. All resulting models are \textit{MAR}(1,1) and are reported in Table \ref{tab:MAR_wti_brent}. We report the lag and lead coefficients as well as the degrees of freedom of the distribution and their respective standard errors in parentheses. Models estimated on series that were detrended with a polynomial trend of order 6 and with the HP filter are the most similar, as suggested by Figure \ref{fig:detrended_series}. Models estimated after detrending with the polynomial trend of order 4 slightly deviate from the two others and always have a larger lead coefficient, hence indicating more persistence in the explosive episodes. Recall that Section \ref{sec:Montecarlo_MAR} suggests that underestimating the trend order in mixed causal-noncausal models induces on average an overestimation of the noncausal coefficient. All series are mostly forward looking, as the lead coefficients are at least 0.8 while the lag coefficients are at most 0.31.\footnote{The bi-modality of the coefficient distribution in the estimation can lead, in the optimization of the likelihood function, to a local maximum \cite{frederique2019mixed}. This phenomenon is subject to initial values and can induce a switch between the lag and lead coefficients. This was however thoroughly checked in the analysis.} We can see that, as expected, \textit{SPR}-detrended series are slightly more persistent in their noncausal dynamics with a lead coefficient up to 0.1 larger than other detrending methods and also slightly larger degrees of freedom induced by more persistent extreme events. The identification of the dynamics is overall consistent across series and their transformation. Note that adjusting the series for inflation leads to larger estimated degrees of freedom for the Student's \textit{t} distribution but overall to similar dynamics. \\

\begin{table}[h!]\small
    \caption{Estimated \textit{MAR} models}
	\centering
		\resizebox{\textwidth}{!}{%
	\begin{threeparttable}
	\setlength\tabcolsep{3.5pt} 
    	\begin{tabular}{lccccccccccccccc}
            \hline \hline 
        \multirow{3}{*}{Series} & \multicolumn{15}{c}{\textit{MAR}(1,1) estimations per detrending method} \\ \cline{2-16}
        & \multicolumn{3}{c}{$t^4$} && \multicolumn{3}{c}{$t^6$} && \multicolumn{3}{c}{$HP$} && \multicolumn{3}{c}{$SPR$}   \\\cline{2-4} \cline{6-8} \cline{10-12} \cline{14-16}
        & $\phi$ & $\psi$ & $t(\gamma)$ && $\phi$ & $\psi$ & $t(\gamma)$ && $\phi$ & $\psi$ & $t(\gamma)$ && $\phi$ & $\psi$ & $t(\gamma)$ \\ \hline
&\\
        {WTI}        
            &0.25 &0.88 &2.25
            &\quad \quad&0.29 &0.82 &1.93
            &\quad \quad&0.29 &0.80 &1.85
            &\quad \quad&0.24 &0.90 &2.60 \\
            &\textit{(0.03)} & 	\textit{(0.01)} & 	\textit{(0.36)}
            &&\textit{(0.03)} & 	\textit{(0.02)} & 	\textit{(0.29)}
            &&\textit{(0.03)} & 	\textit{(0.02)} & 	\textit{(0.28)}
            &\quad \quad&\textit{(0.03)} &\textit{(0.01)} &\textit{(0.37)} \\
&\\
        {$\text{WTI}_{\text{real}}$ \quad }    
            &0.22 &0.88 &3.05
            &&0.26 &0.83 &2.75
            &&0.26 &0.81 &2.63
            &\quad \quad&0.22 &0.91 &3.44 \\
            &\textit{(0.04)} & 	\textit{(0.02)} & 	\textit{(0.49)}
            &&\textit{(0.04)} & 	\textit{(0.02)} & 	\textit{(0.50)}
            &&\textit{(0.04)} & 	\textit{(0.02)} & 	\textit{(0.49)}
            &\quad \quad&\textit{(0.04)} &\textit{(0.02)} &\textit{(0.69)}\\

\hline
         &\\
        {Brent}
            &0.31 &0.89 &1.93
            &&0.31 &0.86 &1.82
            &&0.31 &0.83 &1.83   
            &\quad \quad&0.31 &0.92 &2.18 \\
            &\textit{(0.03)} & 	\textit{(0.01)} & 	\textit{(0.27)}
            &&\textit{(0.03)} & 	\textit{(0.02)} & 	\textit{(0.33)}
            &&\textit{(0.03)} & 	\textit{(0.02)} & 	\textit{(0.31)}
            &\quad \quad&\textit{(0.03)} &\textit{(0.01)} &\textit{(0.34)} \\
&\\
        {$\text{Brent}_{\text{real}}$}
            &0.26 &0.90 &2.59
            &&0.27 &0.86 &2.48
            &&0.27 &0.84 &2.45
            &\quad \quad&0.25 &0.92 &2.90 \\
            &\textit{(0.04)} & 	\textit{(0.02)} & 	\textit{(0.55)}
            &&\textit{(0.04)} & 	\textit{(0.02)} & 	\textit{(0.56)}
            &&\textit{(0.04)} & 	\textit{(0.02)} & 	\textit{(0.57)} 
            &\quad \quad&\textit{(0.04)} &\textit{(0.02)} &\textit{(0.53)}\\
\hline
        \end{tabular}
        \begin{tablenotes}
    	    \item \scriptsize  Notes: The models are obtained with a maximum pseudo lag order of 4 and for each series the model identified was an \textit{MAR}(1,1). $\phi$ is the lag coefficient, $\psi$ is the lead coefficient and $\gamma$ the degrees of freedom of the Student's \textit{t} distribution. The polynomial trend are trends up to the order indicated and the HP filtering is performed with a penalization parameter $\lambda=129\,600$. In parentheses are reported the standard error of the coefficients estimated obtained with the MARX package \cite{MARX}.
    	\end{tablenotes}
    \end{threeparttable}%
       }	
    \label{tab:MAR_wti_brent}
\end{table}

Lacking closed-form expressions for the predictive densities, we use the two data-driven approaches mentioned in Section \ref{sec:Discussion}. We employ the simulations-based approach of \citeA{lanne2012optimal}, which only depends on the model estimated and the last observed point and compare, it approximate the density by use of simulations. We compare this method with the sample-based approach of \citeA{filtering}, which uses past values in the forecasting step to approximate the conditional density. Table \ref{tab:1step_prediction_expost} shows the one-month ahead probabilities that the series will decrease (hence be lower than its last observed value) and the probabilities that the series will drop by more than 1 standard deviation (the standard deviations are calculated empirically over the whole sample). Forecasts are performed for January, February, March and April 2020 and results from the two prediction methods are reported for each of the detrended nominal series. We focus on the nominal series as they are the prices people observe and because the estimated models for real series are noticeably similar.\footnote{Results for price-adjusted series can be found in Appendix C in the online material, probabilities slightly vary however the patterns described in the results for nominal series are identical.} While we advocate the use of predictive densities to get the best picture of potential future prices, we choose 2 arbitrary probabilities to present for a matter of comparison and to save space. Nonetheless, the probabilities for any event can be computed from the methods used here, and they could for instance be employed in the construction of risk measures. \\

\begin{table}[ht]\small
    \caption{One-step ahead probabilities}
	\centering
	\resizebox{\textwidth}{!}{%
	\begin{threeparttable}
    	\begin{tabular}{l@{\extracolsep{0pt}}cccccccccccc}
            \hline \hline 
        \multirow{2}{*}{Series} & Detrended  & \multicolumn{2}{c}{Jan.} && \multicolumn{2}{c}{Feb.} && \multicolumn{2}{c}{Mar.} && \multicolumn{2}{c}{Apr.} \\ \cline{3-4} \cline{6-7} \cline{9-10} \cline{12-13}
        & with & samp. & sims. && samp. & sims. && samp. & sims. && samp. & sims. \\ \hline
        &\\
        \multirow{11}{*}{$WTI$}  && \multicolumn{11}{c}{Probability of a decrease} \\ \cline{3-13}

                                &$t^4$  &.444 & .423 && .784 & .762 && .722 & .681 && .828 & .825 \\
                                &$t^6$  &.414 & .437 && .873 & .851 && .726 & .748 && .583 & .705 \\
                                &$HP$   &.411 & .422 && .869 & .836 && .701 & .730 && .544 & .675 \\
                                &$SPR$  &.432 & .440 && .808 & .768 && .691 & .687 && .738 & .781  \\
                                &\\
                                && \multicolumn{11}{c}{Probability of a decrease $>$ 1 s.d.} \\\cline{3-13}
        
                                &$t^4$  &.052 & .044 && .016 & .017 && .006 & .008 && .018 & .015 \\
                                &$t^6$  &.041 & .042 && .007 & .011 && .004 & .005 && .177 & .322 \\
                                &$HP$   &.047 & .045 && .007 & .012 && .005 & .006 && .227 & .399 \\
                                &$SPR$  &.012 & .010 && .005 & .005 && .002 & .003 && .005 & .014  \\ \hline
                                &\\
        \multirow{11}{*}{$Brent$}  && \multicolumn{11}{c}{Probability of a decrease} \\ \cline{3-13}
        
                                &$t^4$  &.379 & .346 && .806 & .800 && .718 & .696 && .879 & .852 \\
                                &$t^6$  &.398 & .400 && .886 & .864 && .792 & .768 && .569 & .770 \\
                                &$HP$   &.397 & .396 && .880 & .853 && .757 & .745 && .500 & .720 \\
                                &$SPR$  &.386 & .390 && .861 & .824 && .786 & .731 && .678 & .860  \\
                                &\\
                                && \multicolumn{11}{c}{Probability of a decrease $>$ 1 s.d.} \\\cline{3-13}
                                &$t^4$  &.044 & .035 && .016 & .016 && .007 & .008 && .071 & .106 \\
                                &$t^6$  &.034 & .034 && .004 & .011 && .001 & .005 && .308 & .552 \\
                                &$HP$   &.037 & .038 && .008 & .012 && .003 & .006 && .350 & .591 \\
                                &$SPR$  &.012 & .010 && .003 & .006 && .000 & .003 && .025 & .065  \\
            
         \hline
        \end{tabular}
        \begin{tablenotes}
            \item \scriptsize For the simulations-based approach (sims.) the truncation parameter $M=100$ and 1\,000\,000 simulations were used. Standard deviations (s.d.) are calculated over the detrended samples and are around 15 for all nominal series.
    	\end{tablenotes}
    \end{threeparttable}%
    }
    \label{tab:1step_prediction_expost}
\end{table}

At the end of December 2019 oil prices were around \$60 per barrel, they had been fluctuating around this price over the last three years. All detrending methods yield values for December that are above the $90^{th}$ percentile of the samples, suggesting high but not extreme levels. At that point in time, no international alerts regarding the risk of a pandemic had been made yet. Probabilities that prices will drop in January are roughly 0.4 for all series and for both forecasting methods. However, probabilities that prices will drop by more than 1 standard deviation are at most 0.052. This confirms that crude oil prices are in a period of volatile and rather high prices, but it does not suggest a bubble behavior with a potential large drop. This can also be seen by the difference between the sample-based and simulations-based predictions. \citeA{voisin2019forecasting} show that discrepancies between the two approaches mostly arise during extreme episodes. Here, they do not differ by more than 3.3\% for the probabilities of a decrease, and by no more than 0.9\% for the probabilities of a sharper decrease. \\

At the end of January 2020, international alerts regarding the spread of the novel coronavirus had been made, which induced an unforeseeable drop in prices. Yet, the $t^4$-detrended series only fell by half a standard deviation and the other two by 75\% (resp. 80\%) of a standard deviation for the Brent (resp. WTI) series. Values remained however above median values. Forecasts based on both methods suggest a continuity in the decrease for February with probabilities ranging from 0.76 to 0.88, yet, they indicate almost zero probability that the drop will be substantial (more than a standard deviation). They hence suggest a return to median values, meaning a return to fundamental prices. Both prediction methods again provide results diverging by no more than 3.3\%. By the end of February 2020, mass gatherings started to be forbidden and the first advice for the quarantine of individuals to contain the spread of the virus had be made. The increasing worldwide pressure hence kept pushing prices down. Yet, no decrease in the detrended series was larger than 60\% of a standard deviation, which was once again in line with the predictions. The series reached their median levels, forecasts for March suggested that series would remain stable around those values, yet favoring a further slight decrease as prices had been declining for the last three consecutive periods. Probabilities of a sharp drop decreased even more towards zero and both prediction methods yielded again similar probabilities. \\

In March 2020 the worldwide situation worsened significantly and the World Health Organization declared COVID-19 a global pandemic. Many countries imposed strict movement restrictions within and across borders, and curfews and lock-downs were implemented. This sudden drop in crude oil demand led to a considerable fall in prices, WTI prices fell by 55\% and Brent prices by 71\%. Values of the detrended series fell by more than 2 standard deviations and reached the $2^{nd}$ and $3^{rd}$ percentile for $HP$-, $SPR$- and $t^6$-detrending. This indicates a negatively explosive episode, and therefore a negative bubble below fundamental prices. The $t^4$-detrending values correspond to at least the $10^{th}$ percentile, suggesting a less extreme episode, compared to the previous behavior of the series. Until this point both predicting methods yielded similar probabilities. However, the discrepancy between the probabilities now attain 0.24 difference, where the simulations-based probabilities of a decrease are always larger than the sample-based probabilities. \citeA{voisin2019forecasting} show that the discrepancies between the sample- and simulations-based approaches widen during explosive episodes. This is why probabilities for $t^4$-detrending series are still very similar across the forecasting methods as opposed to the other detrending methods. They also show that the larger the lead coefficient, the more the sample-results tend to yield larger probabilities of a turning point than the ones computed with simulations. This stems from the fact that the series had attained a few times this point before (in 2008 and in 2015) and turned back towards median value. It is therefore, based on the learning mechanism of the sample-based approach, less likely that the series will keep on decreasing. It is important to notice that even though prices dropped significantly, probabilities that they will keep on decreasing are lower than before for $HP$- and $t^6$-detrended series as well as for $SPR$-detrended Brent. However, compared to previous forecasts, probabilities now suggest that if the series actually kept on decreasing, it could likely be by more than 1 standard deviation as it has now entered an explosive episode. $SPR$-detrended WTI series has a larger probabilities of decrease than for the previous month, however, as can be notice, the probabilities of the sharper decrease for both $SPR$-detrended series are much closer to 0 than with other detrending. This stems from the larger degrees of freedom as well as larger lead coefficient and slightly lower lag coefficient.\\

\begin{figure}[h!]
    \centering
    \begin{subfigure}{0.48\textwidth}
      \centering
      \caption{\hspace{0.25cm}\footnotesize January 2020}
      \vspace{-0.2cm}
      \includegraphics[width=0.9\linewidth,trim={0 0 0cm 0cm},clip]{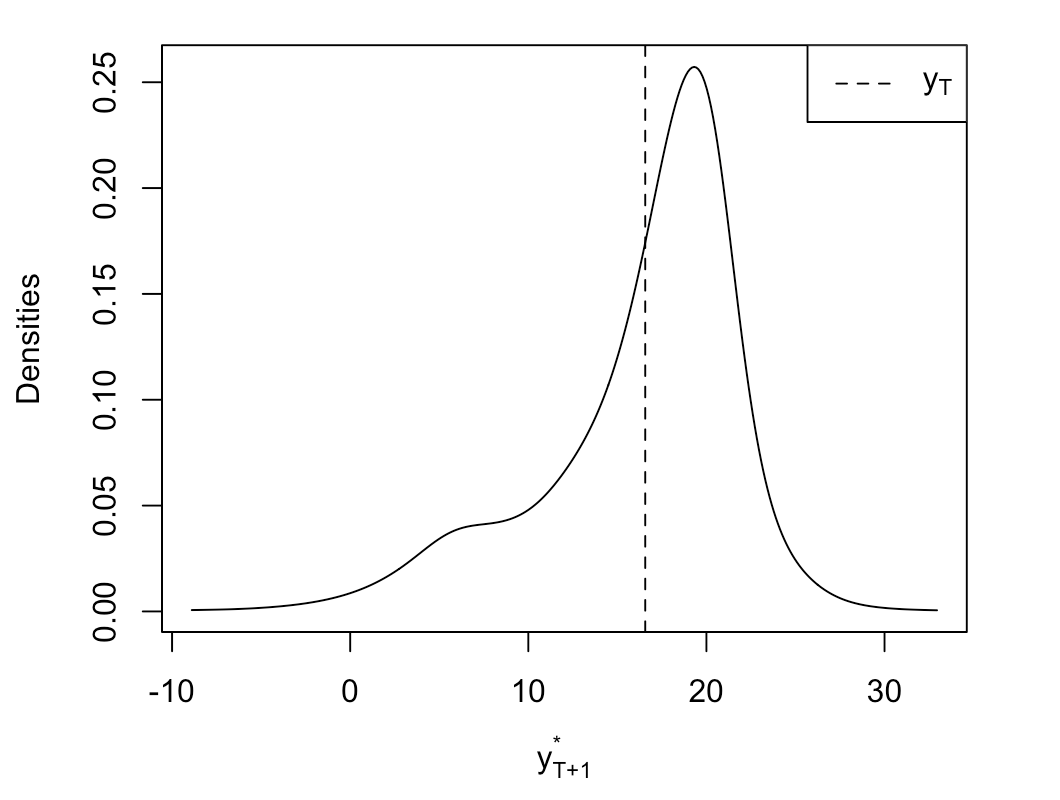}
    \end{subfigure}%
    \hspace{0.08cm}
    \begin{subfigure}{0.48\textwidth}
      \centering
        \caption{\hspace{0.3cm}\footnotesize February 2020}
        \vspace{-0.2cm}
      \includegraphics[width=0.9\linewidth,trim={0cm 0 0cm 0cm},clip]{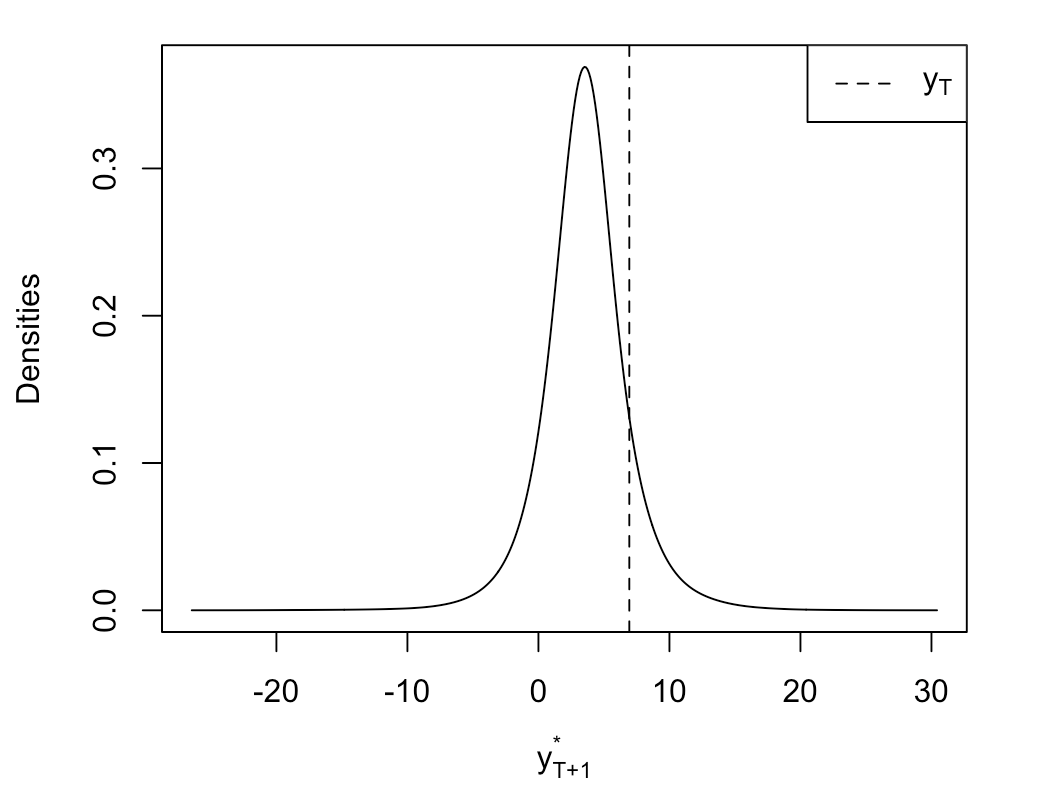}
    \end{subfigure}
    \newline
    \begin{subfigure}{0.48\textwidth}
      \centering
        \caption{\hspace{0.25cm}\footnotesize March 2020}
        \vspace{-0.2cm}
      \includegraphics[width=0.9\linewidth,trim={0 0 0cm 0cm},clip]{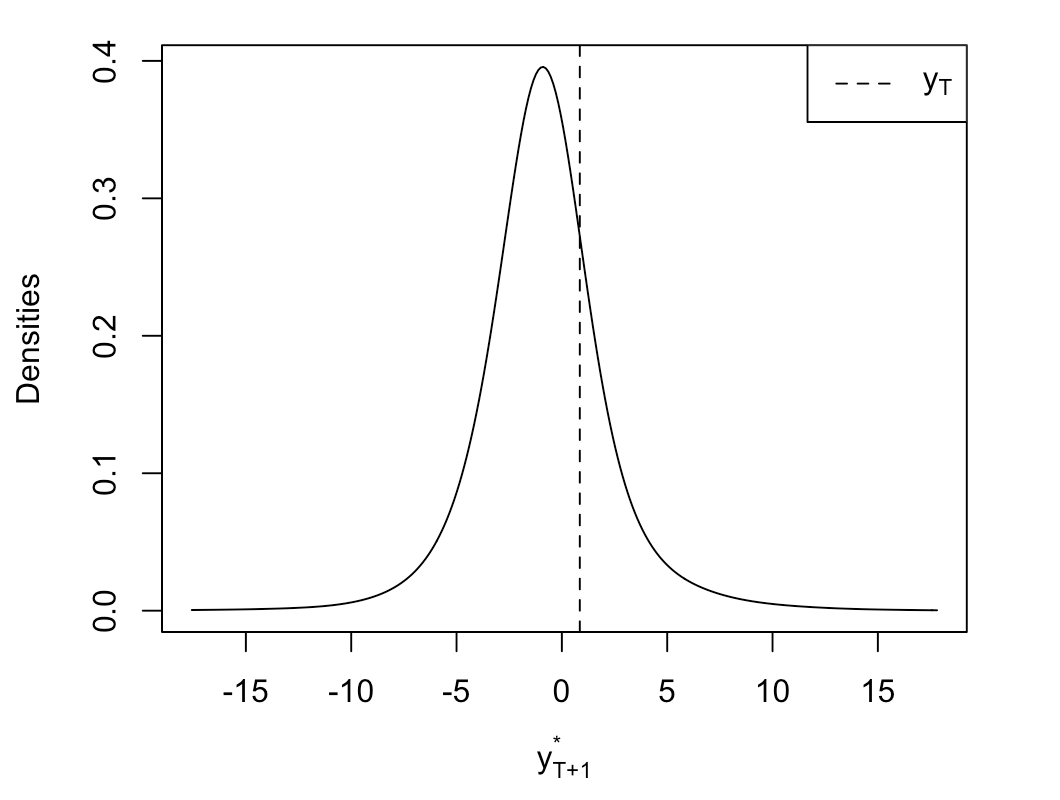}
    \end{subfigure}%
    \hspace{0.08cm}
    \begin{subfigure}{0.48\textwidth}
      \centering
        \caption{\hspace{0.3cm}\footnotesize April 2020}
        \vspace{-0.2cm}
      \includegraphics[width=0.9\linewidth,trim={0cm 0 0cm 0cm},clip]{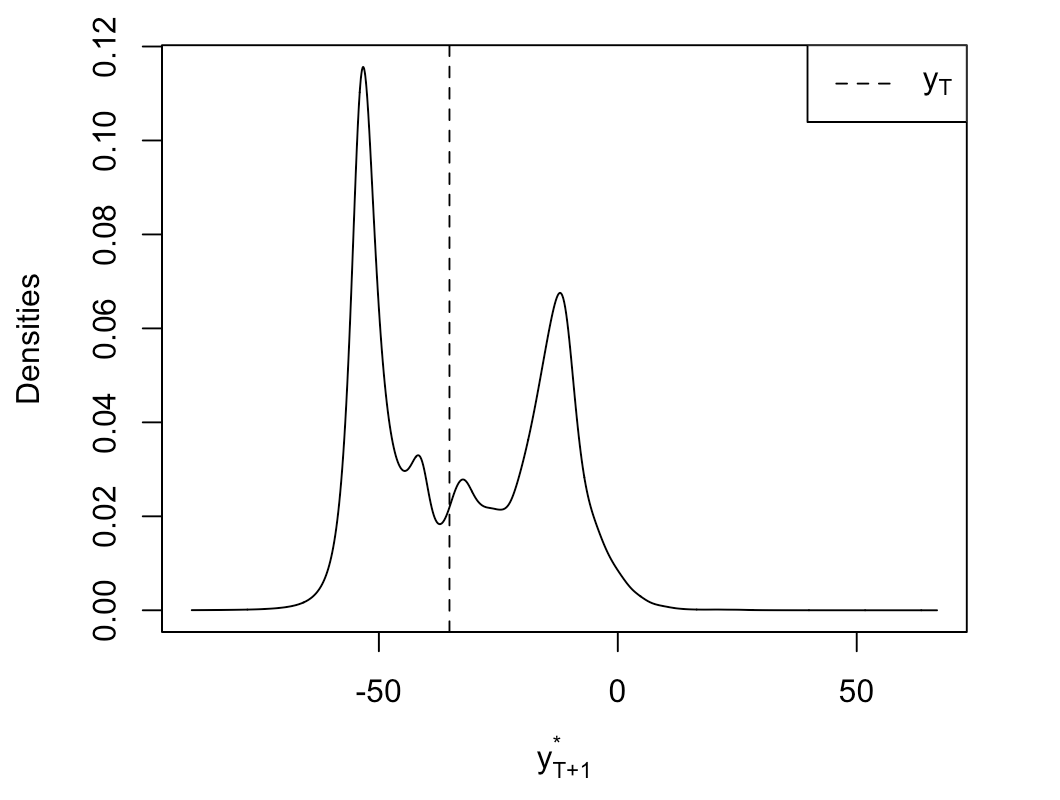}
    \end{subfigure}
    \vspace{-0.3cm}
    \caption{One-step ahead predictive densities of \textit{HP}-detrended Brent prices obtained with the sample-based prediction method.}
    \label{fig:density_brent}
\end{figure}

Figure \ref{fig:density_brent} illustrates the evolution of the predictive densities of the \textit{HP}-detrended Brent series over the time span. On the $x-axis$ are the predictions and on the $y-axis$ their corresponding probability density. The vertical dashed line corresponds to the last value, that is, in graph (a), the vertical line is the detrended value of Brent prices for December 2019. We can clearly observe the bi-modality of the distribution when the series deviates from median values, as shown for the forecasts of January and April, which exacerbate during the explosive negative episode. The range and shape of the density also explains the discrepancies between probabilities of a decrease and probabilities of a decrease of more than 1 standard deviation.\footnote{Results for all other series, available upon request, follow a similar pattern.}  \\

\begin{figure}[h]
    \centering
    \begin{subfigure}{0.46\textwidth}
      \centering
      \caption{\hspace{0.25cm}\footnotesize $HP$-detrended - \textit{AR}(2)}
      \includegraphics[width=0.9\linewidth,trim={0 0cm 0cm 2cm},clip]{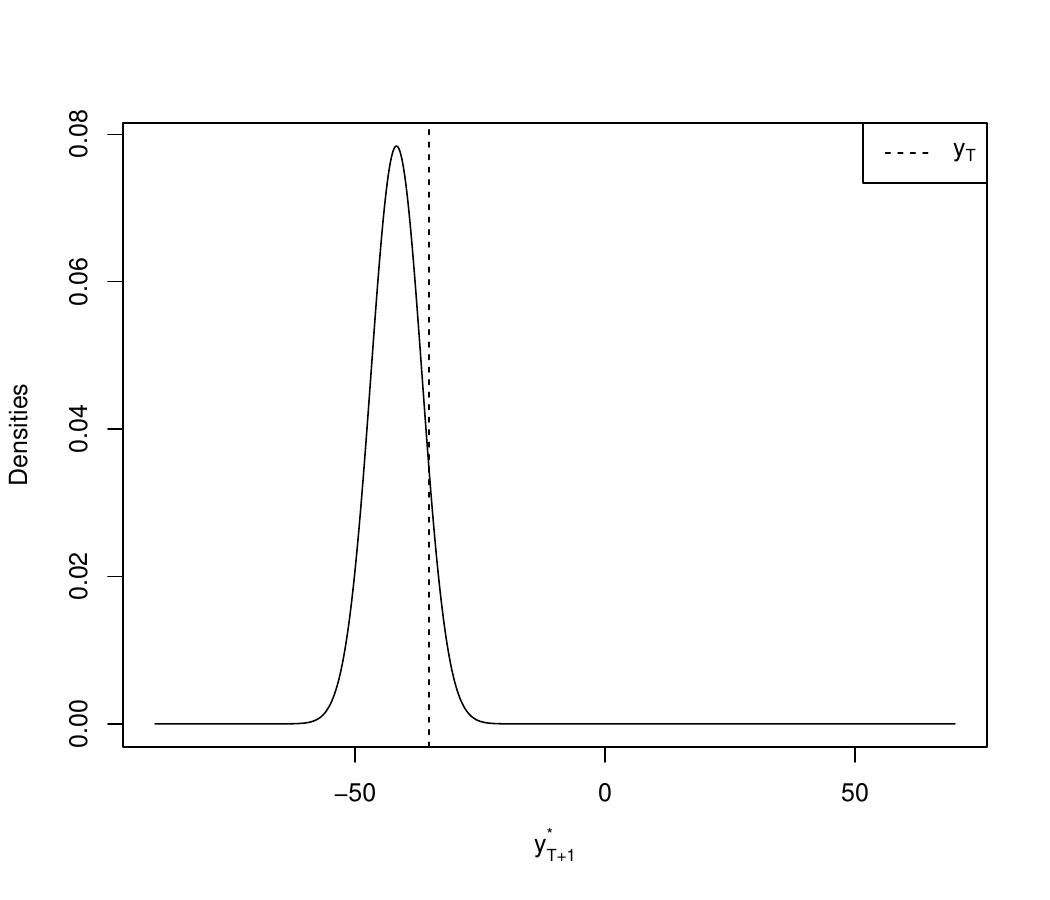}
    \end{subfigure}%
    \hspace{0.1cm}
    \begin{subfigure}{0.46\textwidth}
      \centering
        \caption{\hspace{0.3cm}\footnotesize $SPR$-detrended - \textit{MAR}(1,1)}
      \includegraphics[width=0.9\linewidth,trim={0cm 0 0cm 2cm},clip]{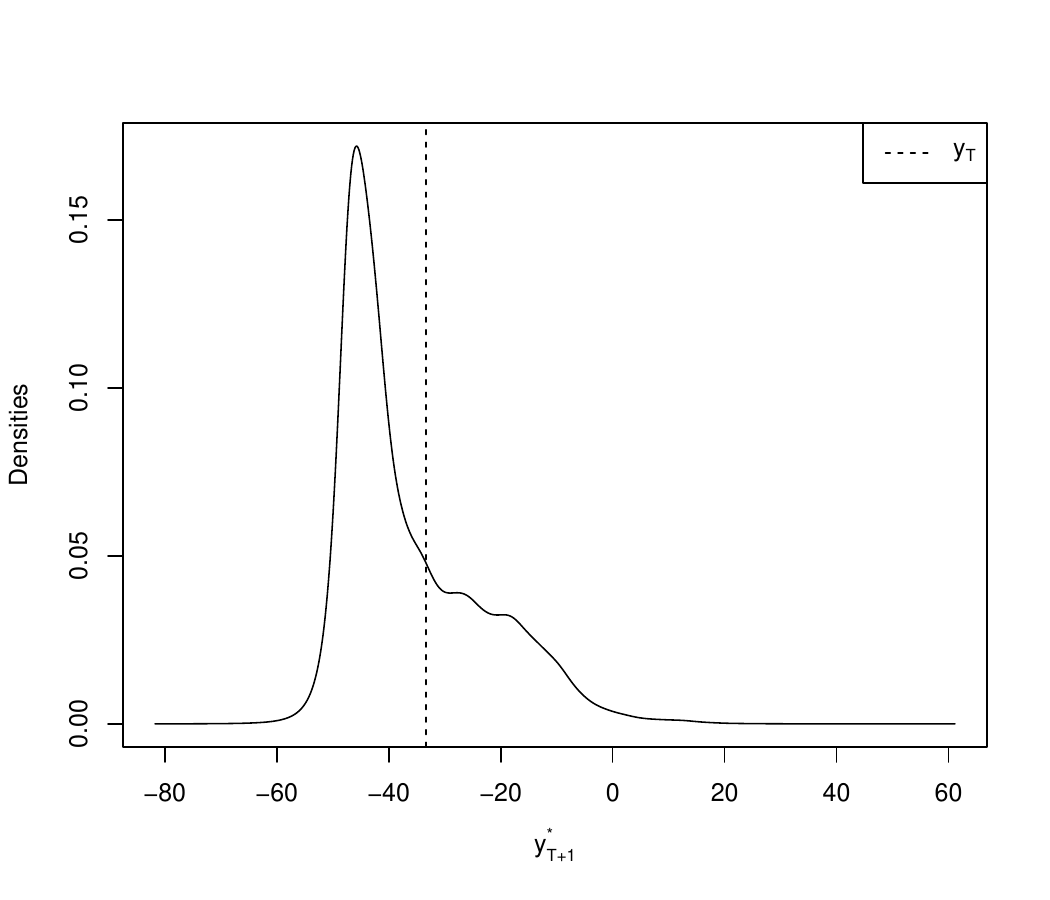}
    \end{subfigure}
        \vspace{-0.3cm}
    \caption{One-step ahead predictive densities of \textit{HP}-detrended Brent prices obtained with the sample-based prediction method.}
    \label{fig:other_density_brent}
\end{figure}

To illustrate the valuable information provided by the predictive densities of \textit{MAR} models, graph (a) of Figure \ref{fig:other_density_brent} depicts the predictive density for April 2020 using a Gaussian \textit{AR}(2) model instead of an \textit{MAR}(1,1) on $HP$-detrended Brent prices. The predictive density is obtained using the closed-form of the conditional normal distribution. We can see that the mode of the density corresponds to a further decrease, but it now lacks the bi-modality and therefore does not suggest a return to central values as does the \textit{MAR} predictive density shown on graph (d) of Figure \ref{fig:density_brent}. As such, once the series enters a locally -- here negative -- explosive episode, the \textit{AR}(2) only predicts a continuing decrease of the prices. Graph (b) of Figure \ref{fig:other_density_brent} displays the sample-based predictive density of the $SPR$-detrended Brent series. We can see that the larger lead coefficient implies a lower rate of decrease (this can be seen as the distance between the two modes), but it indicates larger probabilities of a further decrease as large lead coefficients imply longer lasting explosive episodes. This is why the right mode, which corresponds to a return to central values has a much lower weight on the density. \\

The \textit{MAR} models employed here are univariate, hence no exogenous information is incorporated, as opposed to MARX models (see \citeA{hecq2020mixed} and \citeA{brazilpaper}). Disregarding exogenous variables facilitates forecasting but can sometimes lead to consequential lack of information. For instance, it is expected that crude oil prices should be lower-bounded as they cannot decrease indefinitely and become increasingly negative. Simulations-based probabilities cannot not take that into account as they are only based on the model estimated. Sample-based probabilities however, since prices have never become negative (or at least not long enough to be visible on monthly series), will tend to limit the probabilities that it will happen in the future, even without incorporating additional information within the model, based on its learning mechanism. \\

Overall, \textit{HP}- and $t^6$-detrending provide similar results both for estimation and predictions. $SPR$-detrending, as mentioned earlier yields slightly different dynamics which might stem from the dynamics that are inherent to the stock variable itself. Detrending with $t^4$ yields slightly different results for the estimation but which in turn yields quite different results for predictions. We saw in Figure \ref{fig:detrended_series} that detrending with a trend polynomial of order 4 induced different dynamics in the remaining cycle than the other mechanical detrending. This also corroborates the results found in Section \ref{sec:Montecarlo_MAR} about the risks of underestimating the order of a polynomial trend on the dynamics of the series. We can also see in Figure \ref{fig:detrended_series} that the main differences between all detrending methods appear at the end of the sample. $HP$ and $SPR$ detrending are almost identical while $t^6$ provides slightly lower values. On the other hand, $t^4$-detrending yields significantly larger value than the others for the end of the sample.  \\

\subsection{Real-time analysis}
To illustrate the difficulties and the limitations of detrending and forecasting in real time, we compare the results obtained in real time to the ones obtained in-sample for Brent prices with $t^4$, $t^6$ and $HP$ detrending. We did not include $SPR$ detrending in this Section as we are interested in detrending methods that are affected by sample expansion and while with $SPR$ detrending we still need to re-estimate the model at each point, the trend itself does not change. Table \ref{tab:MAR_brent_realtime} shows the estimated \textit{MAR} models for the expanding samples after each detrending. We can see that the expansion of the sample, even with the inclusion of the large drop of March 2020 did not affect the identification of the model nor the dynamics. Lead and lag coefficients vary by no more than 0.03. The estimated degrees of freedom of the Student's \textit{t} distribution are rather stable until the data point of March is included, which induced decrease between 0.07 and 0.1 for all series, getting therefore closer to the parameter estimated ex-post. This stability in the estimation of the models suggest that probabilities should not significantly differ either. \\

\begin{table}[h!]\small
    \caption{Estimated \textit{MAR} models on different Brent prices samples}
	\centering
	\begin{threeparttable}
	\resizebox{\textwidth}{!}{%
	\setlength\tabcolsep{3.5pt} 
    	\begin{tabular}{lccccccccccc}
            \hline \hline 
        \multirow{3}{*}{Sample} & \multicolumn{11}{c}{\textit{MAR}(1,1) estimations per detrending method} \\ \cline{2-12}
        & \multicolumn{3}{c}{$t^4$} && \multicolumn{3}{c}{$t^6$} && \multicolumn{3}{c}{$HP$}   \\\cline{2-4} \cline{6-8} \cline{10-12}
        & $\phi$ & $\psi$ & $t(\gamma)$ && $\phi$ & $\psi$ & $t(\gamma)$ && $\phi$ & $\psi$ & $t(\gamma)$ \\ \hline
&\\
        {In-sample}        
                            &0.31    &0.89      &1.93
            &\quad \quad    &0.31    &0.86      &1.82
            &\quad \quad    &0.31    &0.83      &1.83\\
                &\textit{(0.03)}    &\textit{(0.01)}    &\textit{(0.27)}
            &   &\textit{(0.03)}    &\textit{(0.02)}    &\textit{(0.33)}
            &   &\textit{(0.03)}    &\textit{(0.02)}    &\textit{(0.31)}\\
\hline
&\\ 

        {$\rightarrow$ Dec}
                            &0.30    &0.89      &2.06
            &\quad \quad    &0.29    &0.86      &1.98
            &\quad \quad    &0.29    &0.84      &1.97\\
                &\textit{(0.03)}    &\textit{(0.01)}    &\textit{(0.27)}
            &   &\textit{(0.03)}    &\textit{(0.02)}    &\textit{(0.33)}
            &   &\textit{(0.03)}    &\textit{(0.02)}    &\textit{(0.31)}\\

        {$\rightarrow$ Jan}
                            &0.30    &0.89      &2.05
            &\quad \quad    &0.29    &0.86      &1.97
            &\quad \quad    &0.28    &0.84      &1.97\\
                &\textit{(0.03)}    &\textit{(0.01)}    &\textit{(0.31)}
            &   &\textit{(0.03)}    &\textit{(0.02)}    &\textit{(0.33)}
            &   &\textit{(0.03)}    &\textit{(0.02)}    &\textit{(0.31)}\\

        {$\rightarrow$ Feb}
                            &0.30    &0.89      &2.07
            &\quad \quad    &0.31    &0.85      &1.97
            &\quad \quad    &0.30    &0.83      &1.97\\
                &\textit{(0.03)}    &\textit{(0.01)}    &\textit{(0.31)}
            &   &\textit{(0.03)}    &\textit{(0.02)}    &\textit{(0.32)}
            &   &\textit{(0.03)}    &\textit{(0.02)}    &\textit{(0.31)}\\

        {$\rightarrow$ Mar}
                            &0.30    &0.89      &1.97
            &\quad \quad    &0.31    &0.85      &1.89
            &\quad \quad    &0.30    &0.83      &1.90\\
                &\textit{(0.03)}    &\textit{(0.02)}    &\textit{(0.36)}
            &   &\textit{(0.03)}    &\textit{(0.02)}    &\textit{(0.36)}
            &   &\textit{(0.03)}    &\textit{(0.02)}    &\textit{(0.33)}\\

\hline
        \end{tabular}%
       }
        \begin{tablenotes}
    	    \item \scriptsize  Notes: See Table \ref{tab:MAR_wti_brent}.
    	\end{tablenotes}
    \end{threeparttable}	
    \label{tab:MAR_brent_realtime}
\end{table}

\begin{figure}[h!]
    \centering
    \begin{subfigure}{\textwidth}
      \centering
      \caption{\hspace{0.5cm}\footnotesize $t^4$-detrended Brent}
      \vspace{-0.2cm}
      \includegraphics[width=0.85\linewidth,trim={0 0 0cm 0cm},clip]{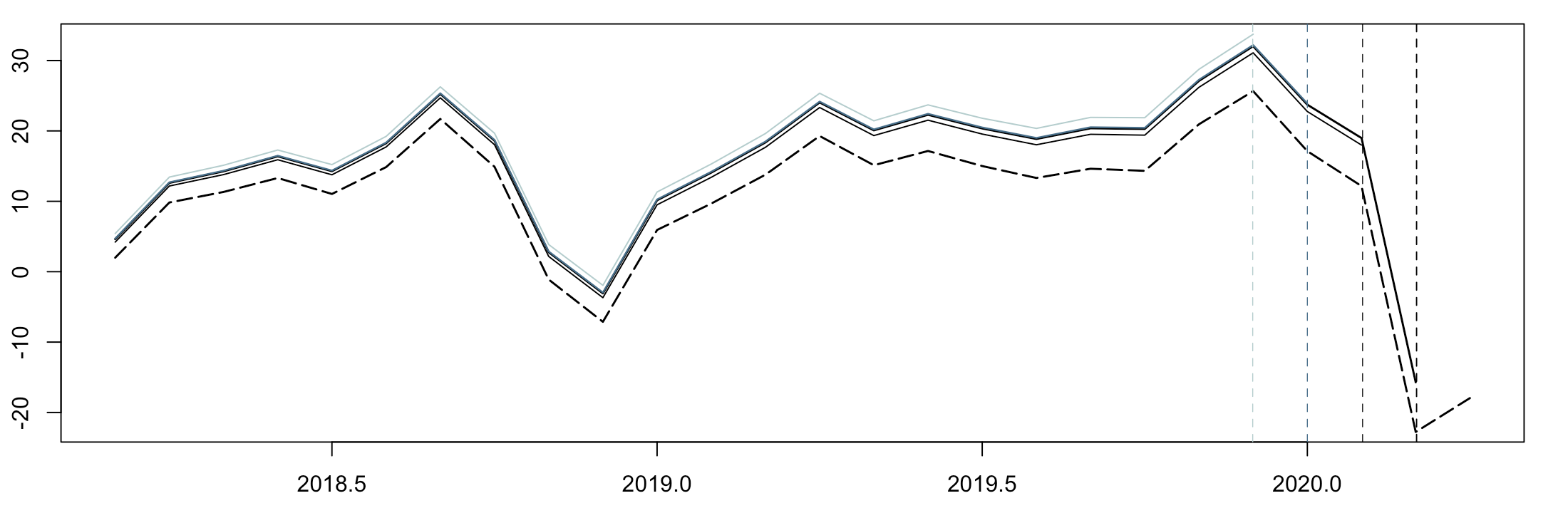}
    \end{subfigure}
    \newline
    \begin{subfigure}{\textwidth}
      \centering
        \caption{\hspace{0.5cm}\footnotesize $t^6$-detrended Brent}
        \vspace{-0.2cm}
      \includegraphics[width=0.85\linewidth,trim={0 0 0cm 0cm},clip]{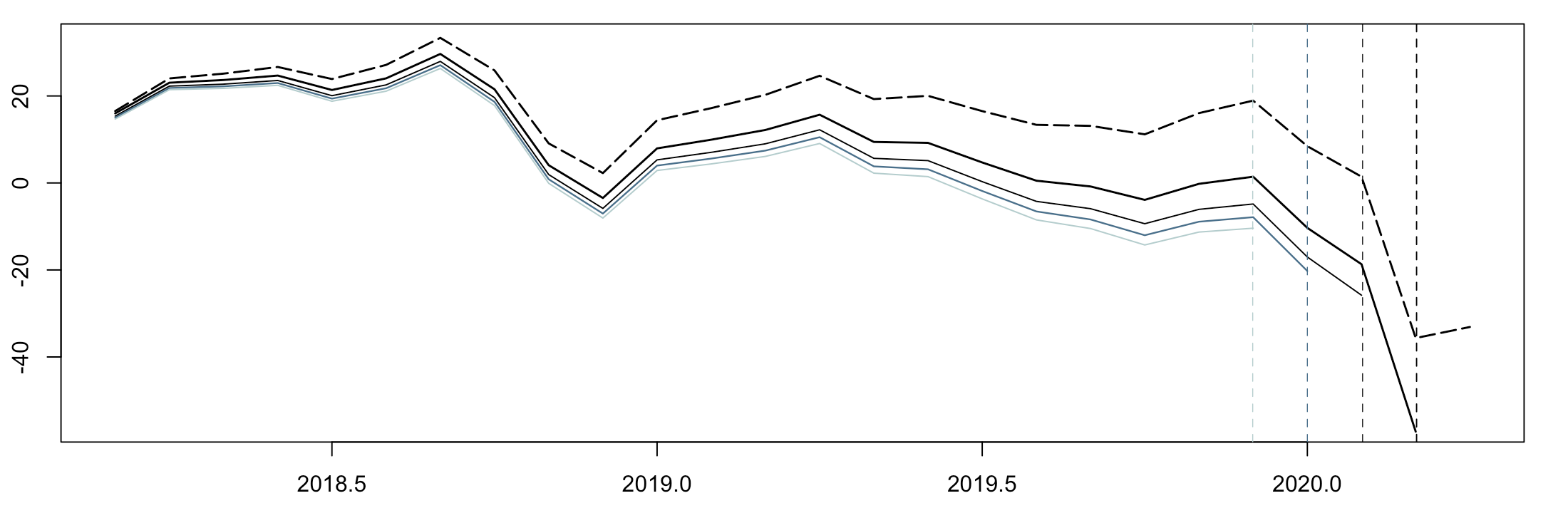}
    \end{subfigure}
    \newline
    \begin{subfigure}{\textwidth}
      \centering
        \caption{\hspace{0.5cm}\footnotesize\ $HP$-detrended Brent}
        \vspace{-0.2cm}
      \includegraphics[width=0.85\linewidth,trim={0 0 0cm 0cm},clip]{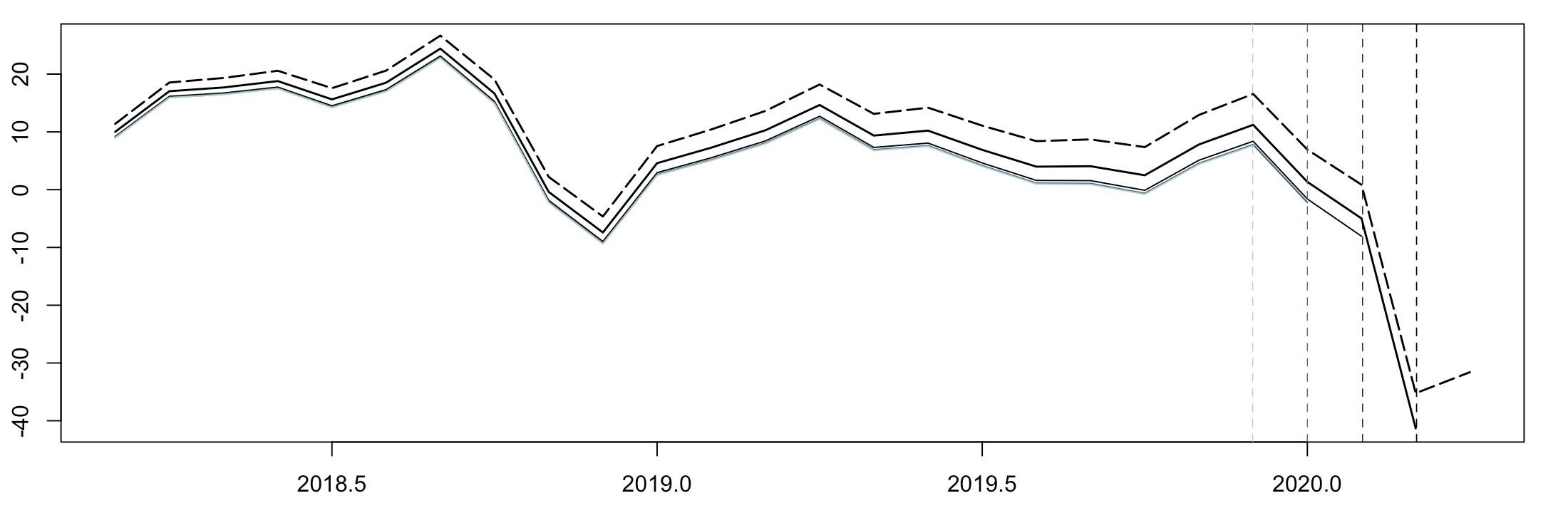}
    \end{subfigure}
    \vspace{-0.3cm}
    \caption{In-sample (dashed curve) vs real-time detrending of Brent prices}
    \label{fig:real_vs_expost_detending}
\end{figure}

To investigate the sensitivity of the detrending methods to the addition of new data points, Figure \ref{fig:real_vs_expost_detending} shows how the detrended series vary based on the stopping point of the sample. The dashed line corresponds to the ex-post detrended series, hence when all data points until December 2020 are included. Then, the expanding samples are depicted from the light blue curve (sample stopping in December 2019) to the black curve (sample until March 2020). While detrending with $t^4$ induced the most spurious dynamics over the sample, it seems, as well as the \textit{HP} filter, to be less affected by the addition of the new points than the $t^6$-detrending. In graph (a), we can see that the 4 detrended series are almost identical, even once the point for March is added. In graph (c), corresponding to the \textit{HP}-detrended series, we can see that the 3 first detrended series are almost identical but that the inclusion of March creates a slight shift in the detrended series. In this case also, the inclusion of even later points will induce further shifts of the estimated trend. However, for the polynomial trend of order 6, as depicted in graph (b), we can see that the inclusion of each point creates a noticeable shift in the estimated trend. From this, we expect the $t^6$-detrended series to be the ones for which the probabilities differ the most from the in-sample probabilities. Indeed, even if the estimated model is almost identical, the substantial discrepancies between the real-time and ex-post detrended series may impact probabilities, especially during (mildly) explosive episodes.\\

\begin{figure}[h!]
    \centering
    \begin{subfigure}{\textwidth}
      \centering
      \includegraphics[width=\linewidth,trim={0 0 0cm 0cm},clip]{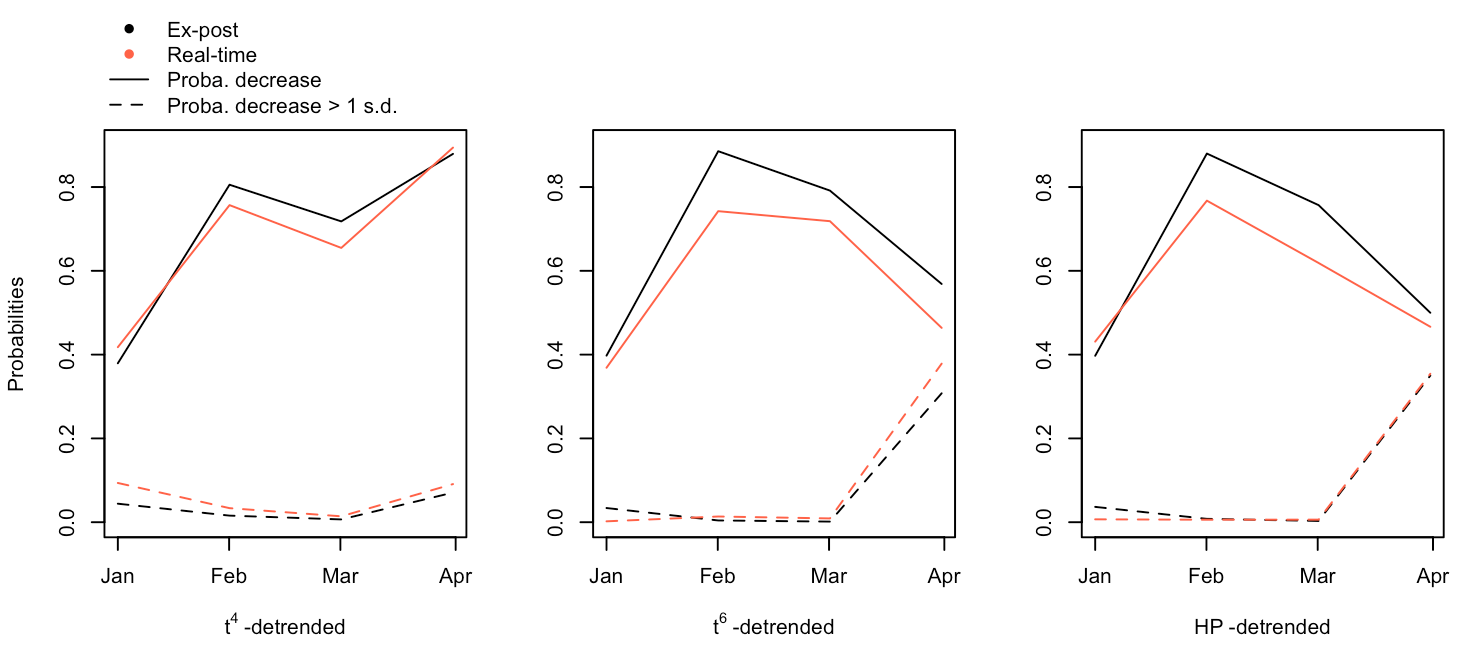}
    \caption{\hspace{0.5cm}\footnotesize Sample-based probabilities}
    \end{subfigure}
    \vspace{0.5cm}
    \newline
    \begin{subfigure}{\textwidth}
      \centering
      \includegraphics[width=\linewidth,trim={0 0 0cm 0cm},clip]{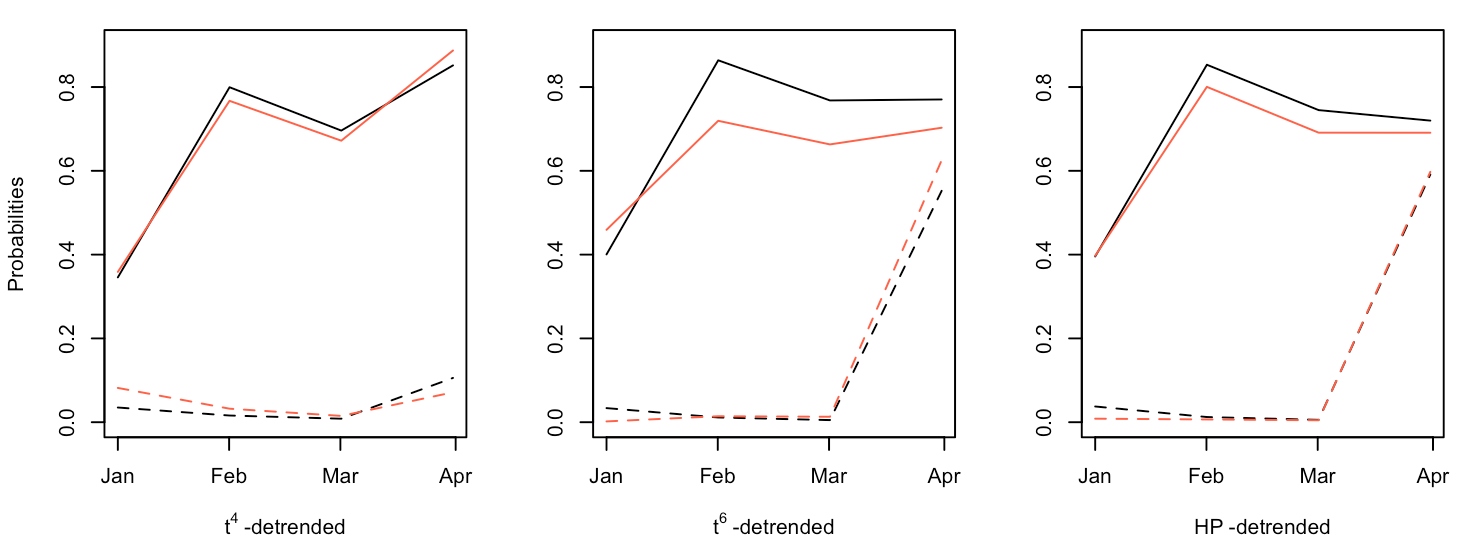}
      \caption{\hspace{0.5cm}\footnotesize Simulations-based probabilities}
    \end{subfigure}
    \caption{Evolution of in-sample (black solid and dashed lines) and real-time (orange solid and dashed lines) probabilities over time}
    \label{fig:evolution_real_vs_expost_proba}
\end{figure}

Figure \ref{fig:evolution_real_vs_expost_proba} depicts the evolution of the one-month ahead probabilities with expanding window. In black are the in-sample probabilities and in orange the real-time probabilities. For the real-time analysis, the trend and the model is re-estimated at each point. The full lines are the probabilities of a decrease and the dashed lines are the probabilities of a decrease of more than 1 standard deviation. Graph (a) (resp. (b)) represents the sample-based (resp. simulations-based) probabilities. As expected, the simulations-based probabilities are the least affected by the re-estimation of the model at each point, since we did not observe significant alteration in the estimations. However, as shown in Figure \ref{fig:real_vs_expost_detending}, it is indeed the $t^6$-detrending that is the most sensitive to the expansion of the sample. Furthermore, we can see that mostly the probabilities of a decrease are affected, as the probabilities of a drop of more than 1 standard deviation are not significantly deviating from the in-sample probabilities. Overall, this indicates that real-time forecasting would have indicated on average lower probabilities of a decrease, at each point and for both approaches. Yet, it would have indicated equal, if not slightly higher, probabilities for the larger drop. Hence, probabilities of more extreme events, namely the tails of the predictive densities, seem to be the least affected by alteration of the trend. \\ 

Overall, it seems that the \textit{HP}-filter is the least sensitive to the change of sample size within this analysis. Results with $t^4$-detrending also emphasizes the risks of underestimating the order of the trend. Moreover, while the \textit{HP}-filter and the polynomial trend of order 6 perform similarly in this analysis, assuming the order of a trend polynomial requires additional understanding regarding the deviations of the series from its fundamental trend. Deterministic trends appear also to be more sensitive to the addition of points in a real-time exercise than the HP filter. Furthermore, while simulations-based probabilities are not characterized by the learning mechanism of the sample-based approach, they are less affected by expanding samples, as long as the model estimated remains consistent. However, as mentioned earlier, with a model that lacks exogenous information, the sample-based approach relying more on past behavior can potentially offset the shortcomings.

%% file: Appendix.tex
\newpage

\newpage
\section*{Appendix A - Prediction methods for \textit{MAR} processes}\label{sec:Predictions}
The focus of this paper is on predicting probabilities of turning points, for example the probabilities of a crash or of entering a positive or a negative bubble. For such inquiries, density forecasts are therefore more adequate than point forecasts. However, the anticipative aspect of \textit{MAR} models complicates their use for predictions. An \textit{MAR}(\textit{r},\textit{s}) model can also be expressed as a causal \textit{AR} model, 
\begin{equation}\label{eq:u_y}
    \Phi(L)y_t=u_t,
\end{equation}
where $u_t$ is the forward-looking component of the error term. Hence, it can itself be expressed in a purely noncausal \textit{AR} process,
\begin{equation}\label{eq:u_eps}
    \Psi(L^{-1})u_t=\varepsilon_t.
\end{equation}
If the model is correctly identified and the parameters consistently estimated, it is therefore sufficient to forecast the purely noncausal process $u_t$ to forecast the variable of interest $y_t$. However, only a few specifications admit a closed-form conditional density  (see for instance \citeNP{gourieroux2013explosive} for the \textit{MAR}(0,1) Cauchy-distributed process). The assumption of other fat-tail distributions, such as Student's \textit{t}, can lead to the absence of closed-form expressions for the conditional moments and densities. Two approximations methods have been developed to estimate these predictive densities, for any distribution, also allowing for a larger lead order. The first method, based on simulations, was developed by \citeA{lanne2012optimal}. The second approach uses the information carried by the sample and was developed by \citeA{filtering}. For a detailed description and guidance in using those approximations methods, see \citeA{voisin2019forecasting}. We focus on processes with a unique lead as this is what we identify on the WTI and Brent series in Section \ref{sec:Empirics}.\\

\subsection*{Simulations-based approach}
The purely noncausal component of the errors, $u$, assumed with one lead, can be expressed as an infinite sum of future error terms in its MA representation. \citeA{lanne2012optimal} base their methodology on the fact that there exists an integer \textit{M} large enough so that any future point of the noncausal component can be approximated by the following finite sum,
\begin{equation}
    u^*_{T+h}\approx\sum_{i=0}^{M-h}\psi^i\varepsilon^*_{T+h+i},
    \label{u.in.eps}
\end{equation}
for any forecast horizon $h\geq 1$, and where $\psi$ is the lead coefficient of the purely noncausal \textit{MAR}(0,1) process $u_t$. \\

Let $\varepsilon_{+}^{*(j)}=\Big(\varepsilon_{T+1}^{*(j)},\dots,\varepsilon_{T+M}^{*(j)}\Big)$, with $1\leq j \leq N$, be the \textit{j}-th simulated series of \textit{M} independent errors, randomly drawn from the chosen distribution of the process with estimated parameters (whose probability density function (hereafter \textit{pdf}) is denoted by $g$). We are interested in the conditional cumulative probabilities,
\begin{equation}
    \begin{split}
        \mathbb{P}\Big(y_{T+h}^*\leq x\big|\mathcal{F}_T\Big)&= \mathbb{E}_T \bigg[\mathbf{1}\big(y^*_{T+h}\leq  x\big)\bigg] \\
        &\approx \mathbb{E}_T\Bigg[\mathbf{1}\Bigg(\iota '\mathbf{\Phi}^h\mathbf{y}_T+\sum_{i=0}^{h-1}\iota '\mathbf{\Phi}^i\iota \sum_{j=0}^{M-h+i}\psi^j\varepsilon^*_{T+h-i+j}\leq x \Bigg)\Bigg],
    \end{split}
    \label{eq:sim.estimator_1}
\end{equation} 
where the indicator function $\mathbf{1}()$ is equal to 1 when the condition is met and 0 otherwise. The variable $y^*_{T+h}$ is replaced by an approximation using recursive substitution of its companion form with truncation parameter \textit{M} and the following stacked form,
 \[ \mathbf{y}_T = \left[ \begin{matrix} 
y_T \\ y_{T-1} \\ \vdots \\ y_{T-r+1}  \end{matrix}\right]\text{,} \quad \mathbf{\Phi}= \left[ \begin{matrix} 
\phi_1 & \phi_2 & \dots & \dots & \phi_r \\
1 & 0 & \dots & \dots & 0\\
0 & 1 & 0 & \dots & 0\\
\vdots & \ddots & \ddots & \ddots & \vdots \\
0 & \dots & 0 & 1 & 0
 \end{matrix} \right] (r\times r) \quad  \text{and} \quad \iota=\left[\begin{matrix}1\\0\\ \vdots \\ 0 \end{matrix}\right] (r\times 1).
\]

Given the information set known at time \textit{T}, the indicator function in \eqref{eq:sim.estimator_1} is only a function of the \textit{M} future errors, $\varepsilon_+^*$. Let us denote this indicator function by $q(\varepsilon_+^*)$. Assuming that the number of simulations \textit{N} and the truncation parameter \textit{M} are large enough, the conditional cumulative probabilities of \textit{MAR}(\textit{r},1) processes can be approximated as follows \cite{lanne2012optimal},

\begin{equation}
    \begin{split}
        \mathbb{P}\Big(y_{T+h}^*\leq x\big|\mathcal{F}_T\Big)&\approx\mathbb{E}_T\bigg[q\big(\varepsilon^*_+\big)\bigg]\\
        &\approx \frac{N^{-1}\sum_{j=1}^N q\Big(\varepsilon_{+}^{*(j)}\Big)g\Big(u_T-\sum_{i=1}^M\psi^i\varepsilon^{*(j)}_{T+i}\Big)}{N^{-1}\sum_{j=1}^Ng\Big(u_T-\sum_{i=1}^{M}\psi^i\varepsilon^{*(j)}_{T+i}\Big)}, 
    \end{split}
    \label{eq:sim.estimator_2}
\end{equation} 

By computing its value for all possible $x$ covering the range of potential values for $y_{T+h}^*$, we can obtain the whole conditional cummulative density function (hereafter \textit{cdf}) of $y_{T+h}^*$.\\ 

\citeA{voisin2019forecasting} show that with Cauchy-distributed errors, this approach is a good estimator of theoretical probabilities but are significantly sensitive to the number of simulations \textit{N} chosen. For Student's \textit{t} distributions however, results cannot be compared to theoretical ones, but as the number of simulations gets very large, the derived densities converge to a unique function. Moreover, analogously to theoretical probabilities, once the series has significantly departed from its central values and diverges, the probabilities of a crash at a given horizon tend to a constant.

\subsection*{Sample-based approach}
As an alternative to using simulations, \citeA{filtering} employ all past observed values of the noncausal process. The predictive density function of a purely noncausal process with one lead is approximated as follows, 
\begin{equation}
    \begin{split}
        &{l}\big(u^*_{T+1},\dots,u^*_{T+h}\big|{u}_{T}\big)\\ 
        & \approx {g}({u}_{T}-\psi u^*_{T+1})\dots {g}(u^*_{T+h-1}-\psi u^*_{T+h})
        \frac{\sum_{i=1}^{T}
        {g}(u^*_{T+h}-{\psi} u_i)}{\sum_{i=1}^{T} 
        {g}(u_{T}-{\psi} u_i)},
    \end{split}
    \label{eq:samplebasedestim}
\end{equation}
where $g$ is the \textit{pdf} of the assumed errors distribution.\\

With this method, the predicted probabilities are a combination of theoretical probabilities and probabilities induced by past events. Results are therefore case-specific and are based on a sort of learning mechanism \cite{voisin2019forecasting}. If this method is used when errors are Cauchy distributed, results can be compared to the theoretical predictive distribution to evaluate the influence of past behaviors on the obtained probabilities. However, if the errors follow a Student's \textit{t} distribution for instance, results cannot be compared to theoretical probabilities as no closed-form expressions exists. In such cases an approximation of theoretical results can be derived using the simulations-based approach presented above to gauge how much of the probabilities are induced by the underlying process and by past behaviors.  \\

For values around the median of the series, both methods yield identical results. Discrepancies widen as the level of the series increases. Additionally, the larger the lead coefficient, the more the sample-based method tend to overestimate probabilities of a crash \cite{voisin2019forecasting}. That is, for low lead coefficients, they on average yield very similar results, even for explosive episodes, while for large lead coefficients probabilities induced by the two methods can be considerably different. Overall, both methods depend on the whole sample since they both depend on the estimated coefficients. Hence a wrong detrending would affect both methods. Overestimating the lead coefficient for instance would imply lower probabilities of a crash. For Cauchy distributed processes, one-step ahead probabilities of a crash tend to $(1-\psi)$ during explosive episodes. Thus, identifying a model with a lead coefficient of 0.9 instead of 0.7 for instance would induce a 20\% difference in the \textit{theoretical} probabilities. The sample-based probabilities could be even more distorted based on past behaviors, or on the contrary past behaviors could potentially alleviate the impact on the wrong detrending, but this is case-specific. This is why it is important to investigate the effects of various detrending methods on model identification and on the estimation of the dynamics. Note that formulas for higher lead orders can be found in the respective articles of \citeA{lanne2012optimal} and \citeA{filtering}.

\section*{Appendix B - Impact of detrending on estimated coefficients}
We now investigate the persistence of the dynamics from the magnitude of the estimated coefficients. For instance, a lower lead coefficient will indicate shorter lived bubbles compared to the true generated process and thus increases the probabilities of a crash during an explosive episode. The same goes for larger degrees of freedom when the errors follow a Student's \textit{t} distribution: larger degrees of freedom correspond to thinner tails, and thus rarer extreme values and thus makes less probable long lasting explosive episodes. \\

\begin{figure}[H]
    \centering
    \begin{subfigure}{0.35\textwidth}
      \centering
      \caption*{\hspace{1.15cm}\footnotesize \textit{MAR}(0,1) + no trend}
      \vspace{-0.2cm}
      \includegraphics[width=\linewidth,trim={0 0 3.1cm 2cm},clip]{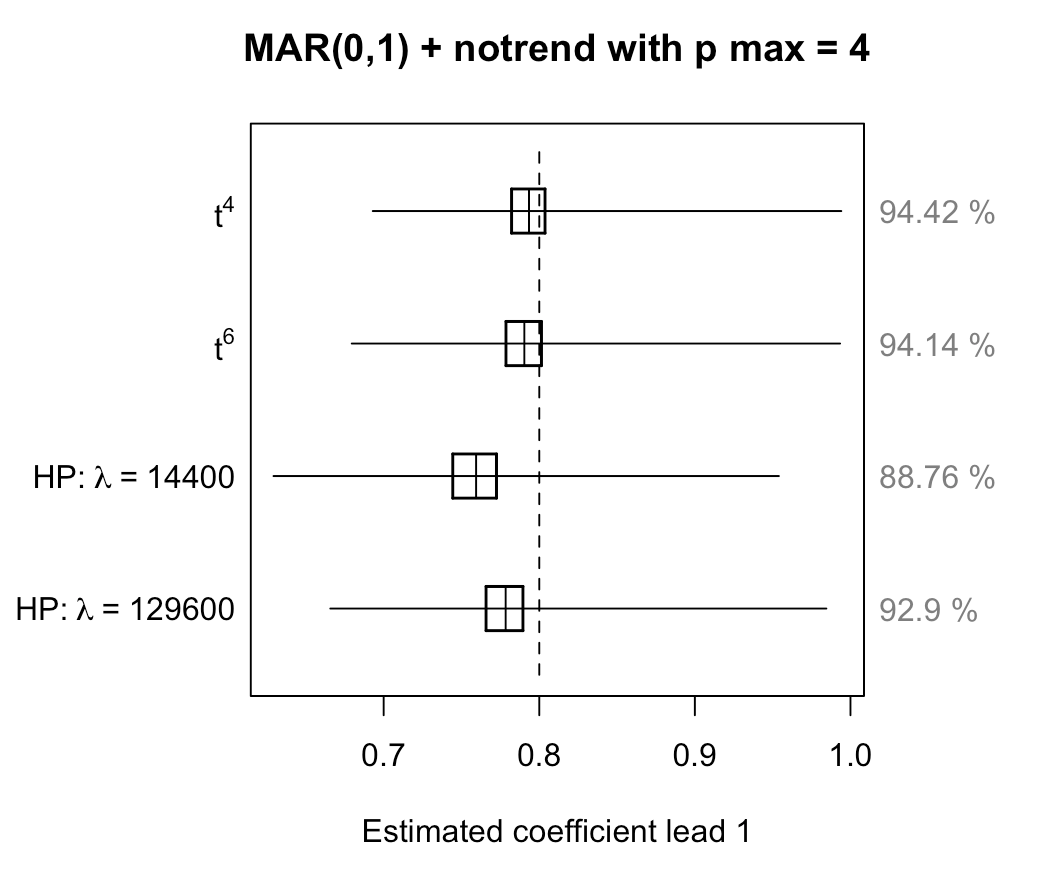}
    \end{subfigure}%
    \hspace{0.08cm}
    \begin{subfigure}{0.2525\textwidth}
      \centering
        \caption*{\hspace{1.4cm}\footnotesize\textit{MAR}(1,1)}
        \vspace{-0.2cm}
      \includegraphics[width=\linewidth,trim={4.3cm 0 3.1cm 2cm},clip]{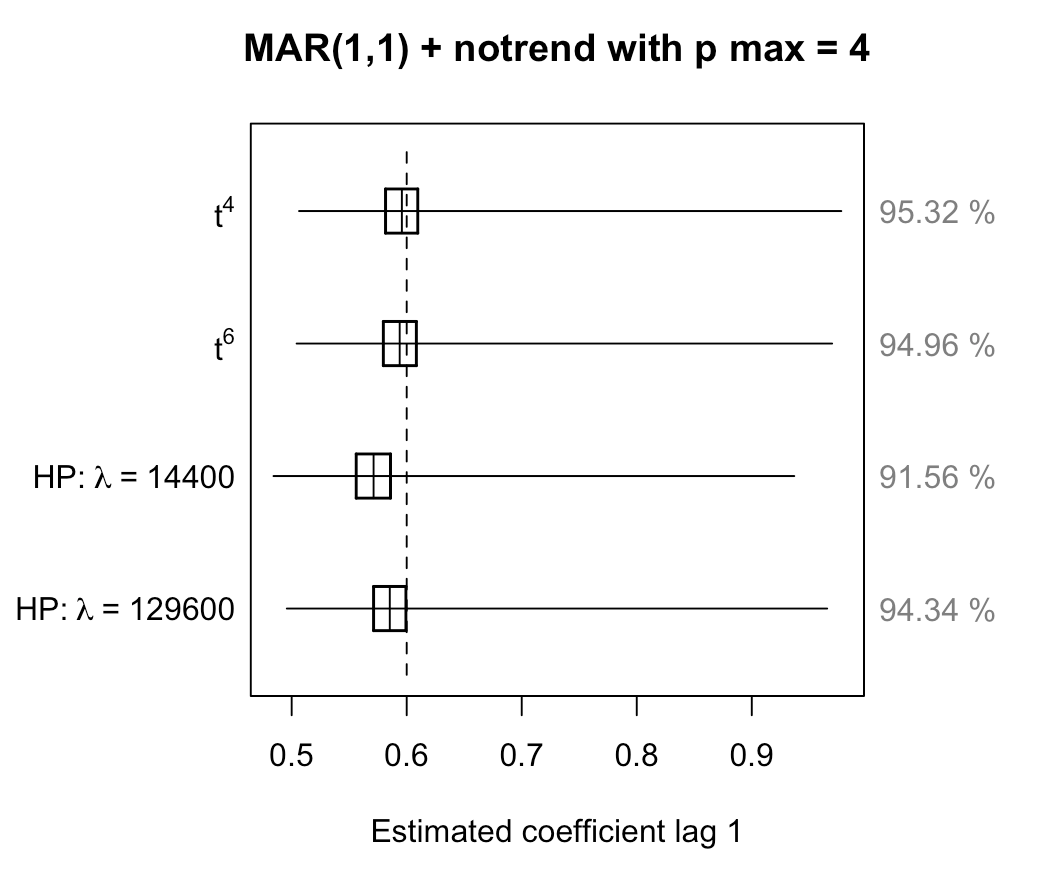}
    \end{subfigure}%
    \begin{subfigure}{0.2525\textwidth}
      \centering
        \caption*{\hspace{-1.7cm}\footnotesize+ no trend}
        \vspace{-0.2cm}
      \includegraphics[width=\linewidth,trim={4.3cm 0 3.1cm 2cm},clip]{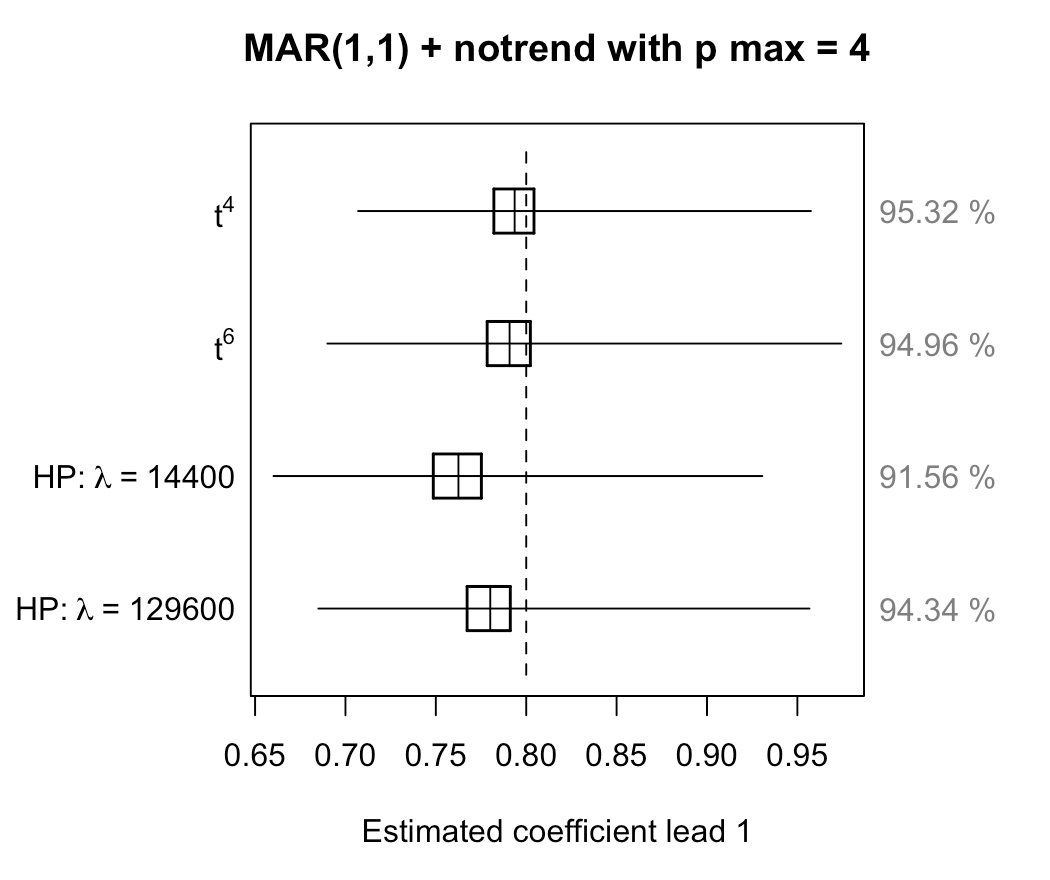}
    \end{subfigure}
    \newline
    \begin{subfigure}{0.35\textwidth}
      \centering
        \caption*{\hspace{1.4cm}\footnotesize\textit{MAR}(0,1) + $\tau^4$}
        \vspace{-0.2cm}
      \includegraphics[width=\linewidth,trim={0 0 3cm 2cm},clip]{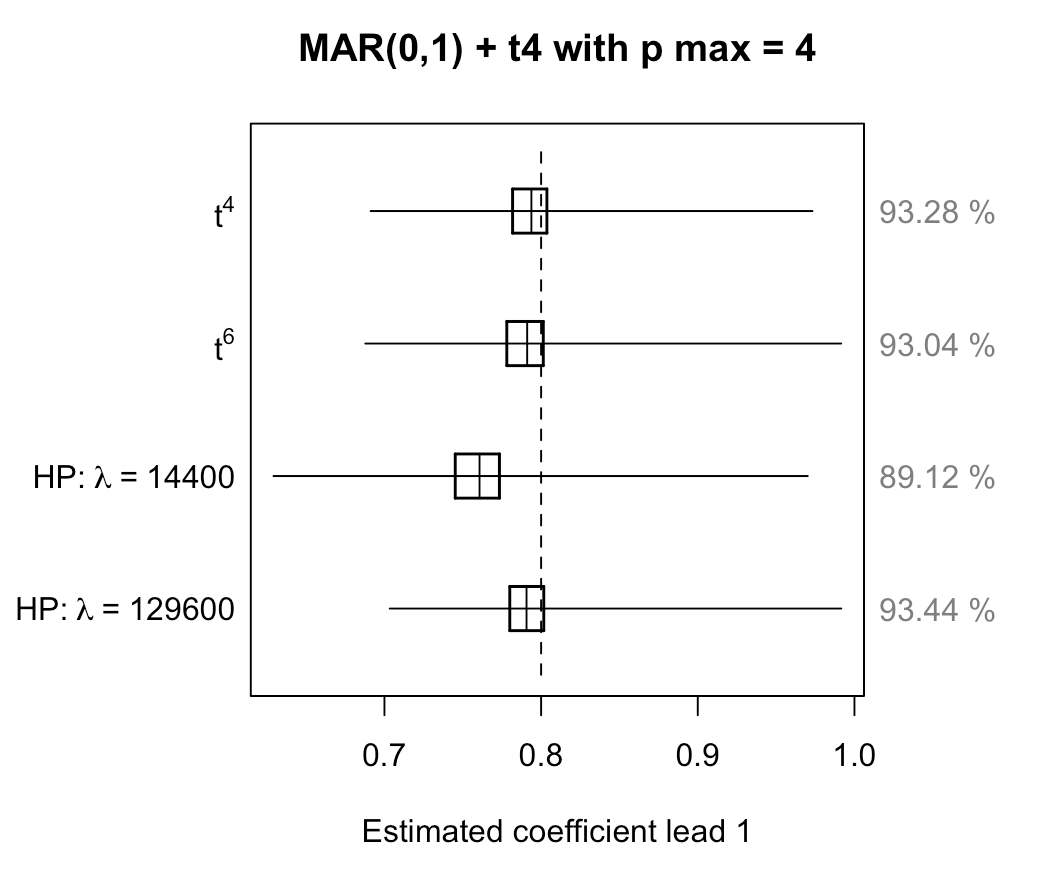}
    \end{subfigure}%
    \hspace{0.08cm}
    \begin{subfigure}{0.2525\textwidth}
      \centering
        \caption*{\hspace{1.9cm}\footnotesize\textit{MAR}(1,1)}
        \vspace{-0.2cm}
      \includegraphics[width=\linewidth,trim={4.3cm 0 3cm 2cm},clip]{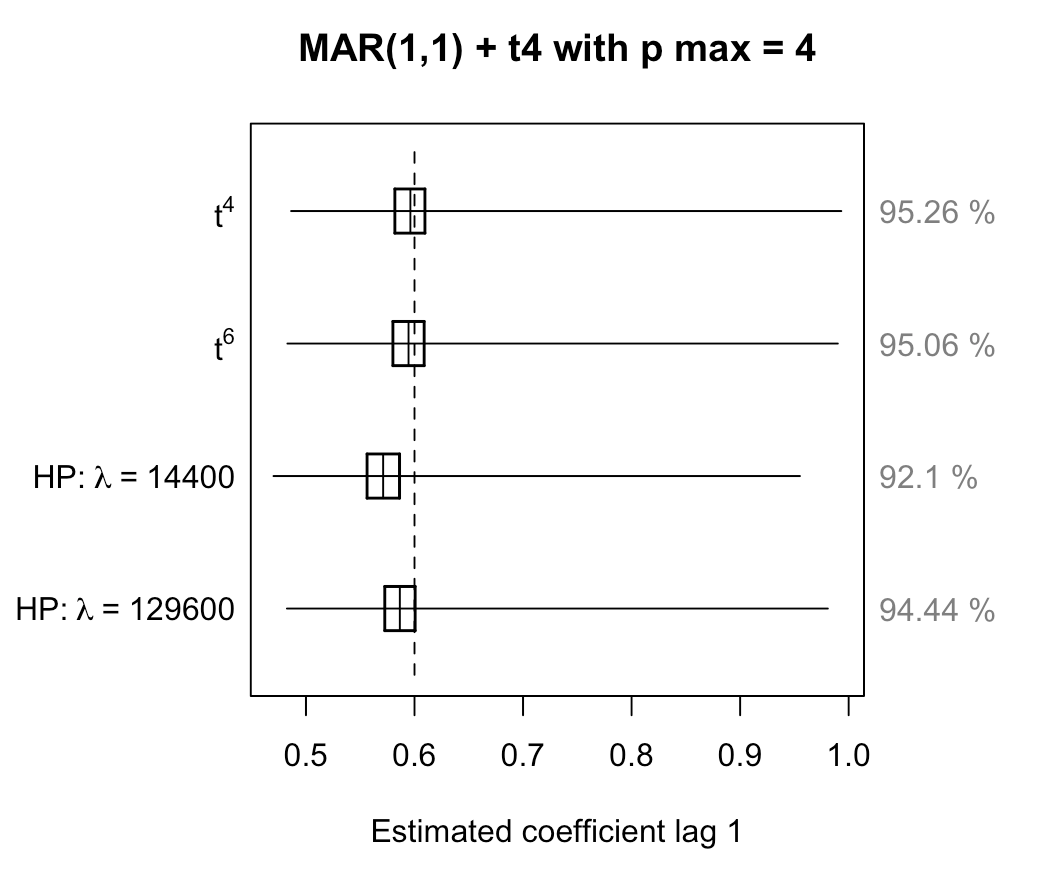}
    \end{subfigure}%
    \begin{subfigure}{0.2525\textwidth}
      \centering
        \caption*{\hspace{-2cm}\footnotesize+ $\tau^4$}
        \vspace{-0.2cm}
      \includegraphics[width=\linewidth,trim={4.3cm 0 3cm 2cm},clip]{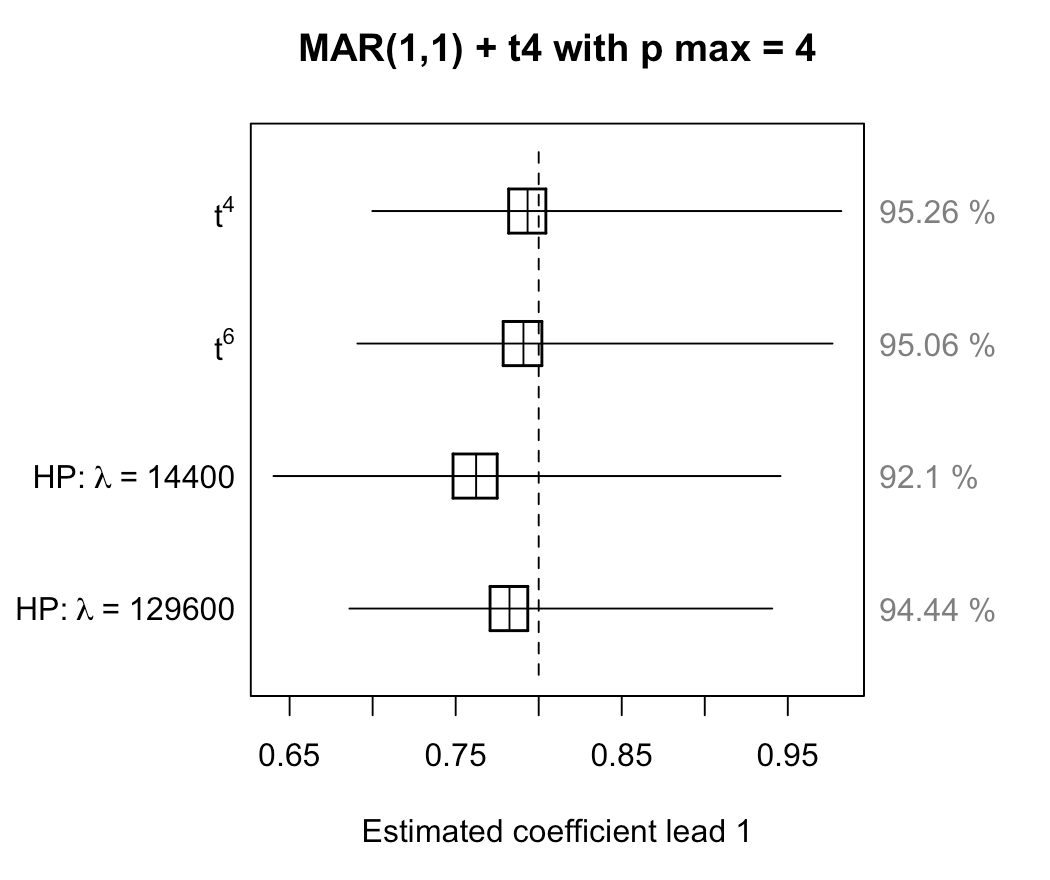}
    \end{subfigure}
    \newline
    \begin{subfigure}{0.35\textwidth}
      \centering
        \caption*{\hspace{1.6cm}\footnotesize\textit{MAR}(0,1) + $\tau^6$}
        \vspace{-0.2cm}
      \includegraphics[width=\linewidth,trim={0 0 3cm 2cm},clip]{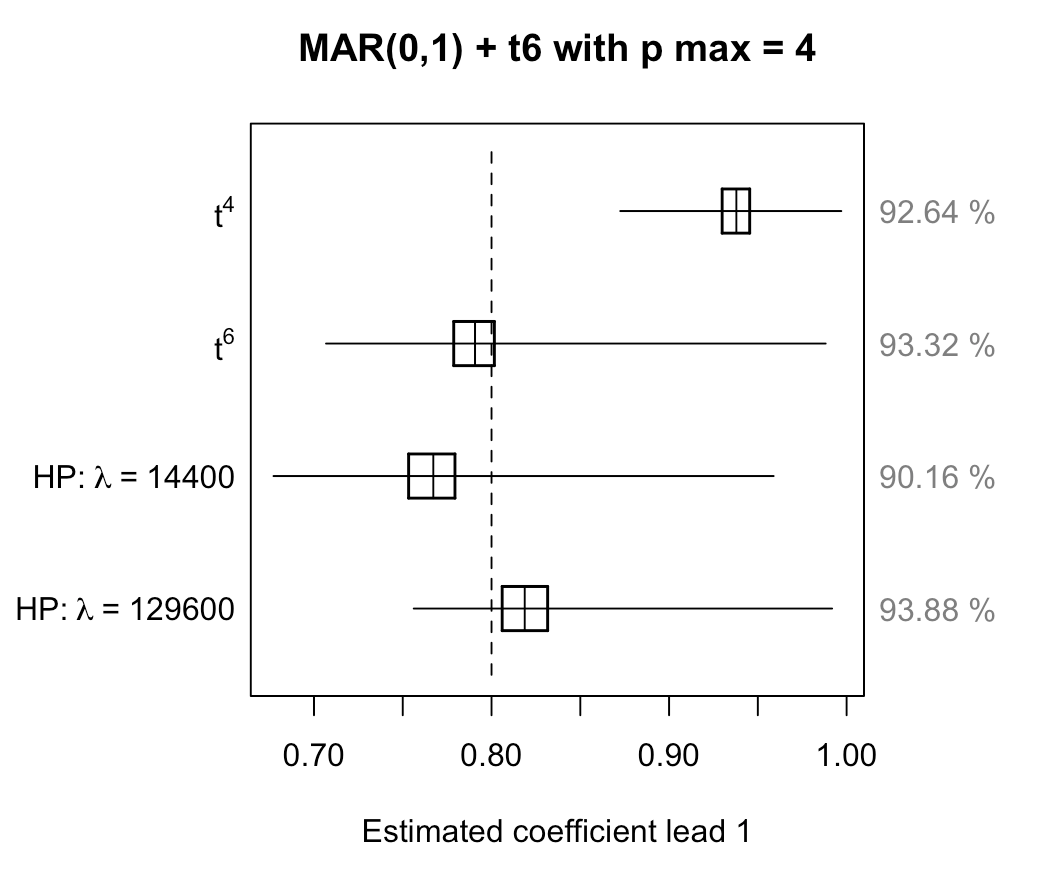}
    \end{subfigure}%
    \hspace{0.08cm}
    \begin{subfigure}{0.2525\textwidth}
      \centering
          \caption*{\hspace{1.9cm}\footnotesize\textit{MAR}(1,1)}
          \vspace{-0.2cm}
      \includegraphics[width=\linewidth,trim={4.3cm 0 3cm 2cm},clip]{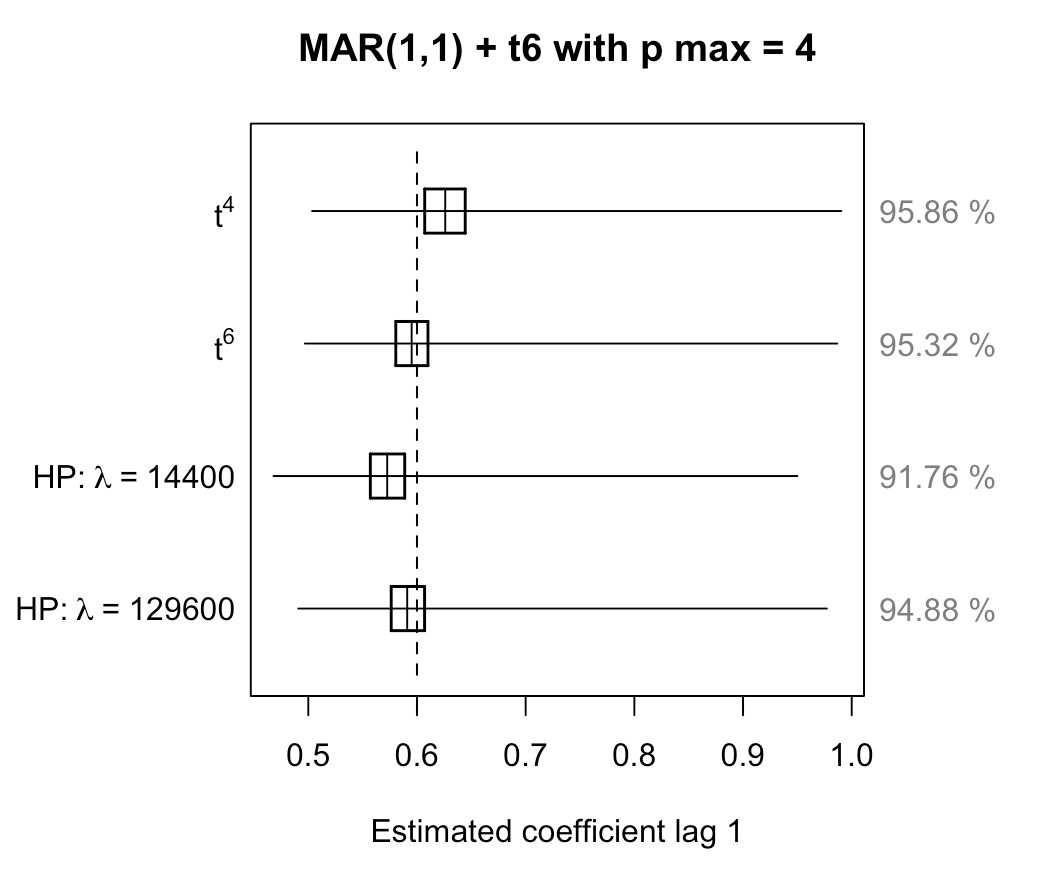}
    \end{subfigure}%
    \begin{subfigure}{0.2525\textwidth}
      \centering
        \caption*{\hspace{-2cm}\footnotesize + $\tau^6$}
        \vspace{-0.2cm}
      \includegraphics[width=\linewidth,trim={4.3cm 0 3cm 2cm},clip]{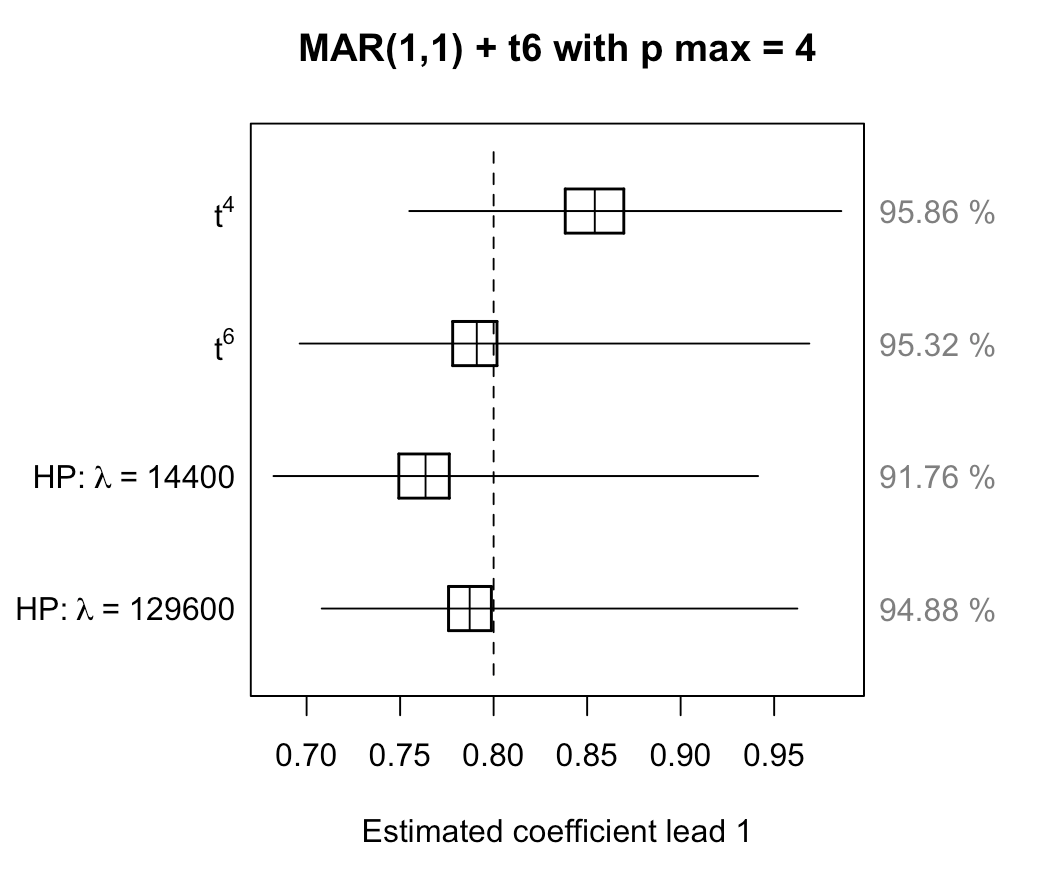}
    \end{subfigure}
    \newline
    \hspace*{-0.5cm}
    \begin{subfigure}{0.35\textwidth}
      \centering
        \caption*{\hspace{1.2cm}\footnotesize\textit{MAR}(0,1) + breaks}
        \vspace{-0.2cm}
      \includegraphics[width=\linewidth,trim={0 0 3cm 2cm},clip]{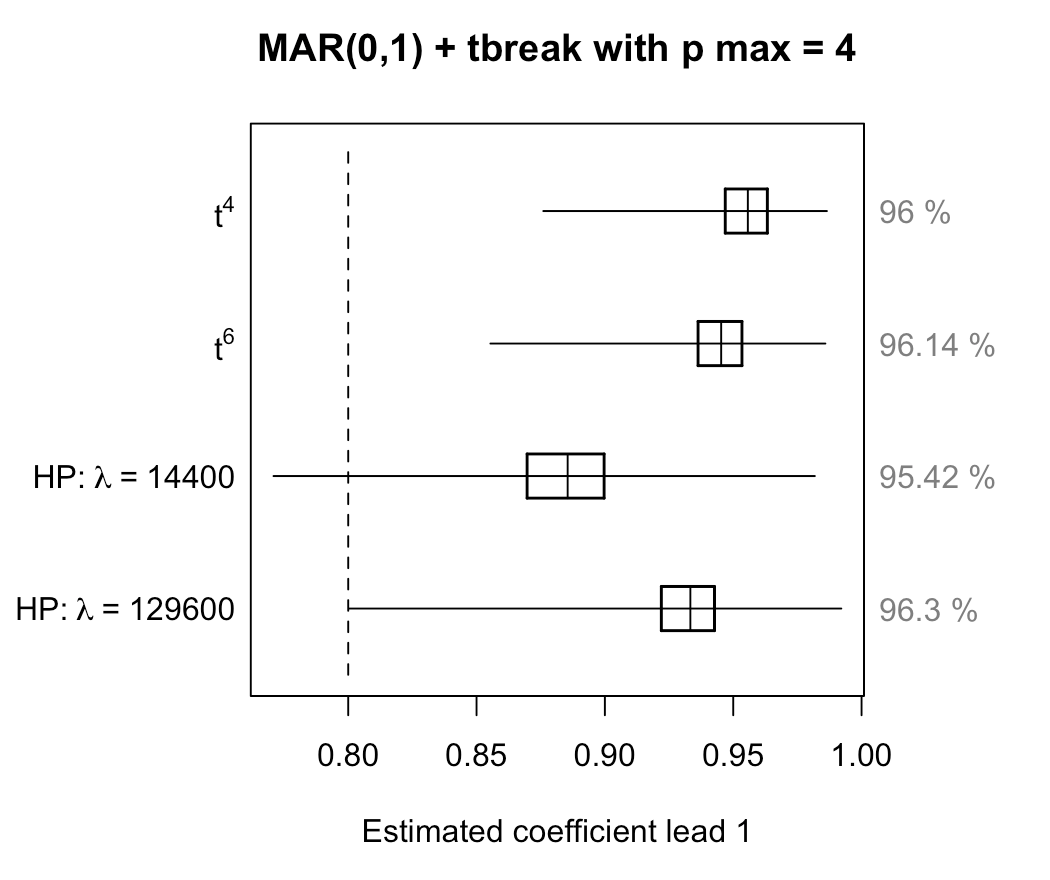}
    \end{subfigure}%
    \hspace{0.08cm}
    \begin{subfigure}{0.2525\textwidth}
      \centering
          \caption*{\hspace{1.5cm}\footnotesize\textit{MAR}(1,1)}
          \vspace{-0.2cm}
      \includegraphics[width=\linewidth,trim={4.3cm 0 3cm 2cm},clip]{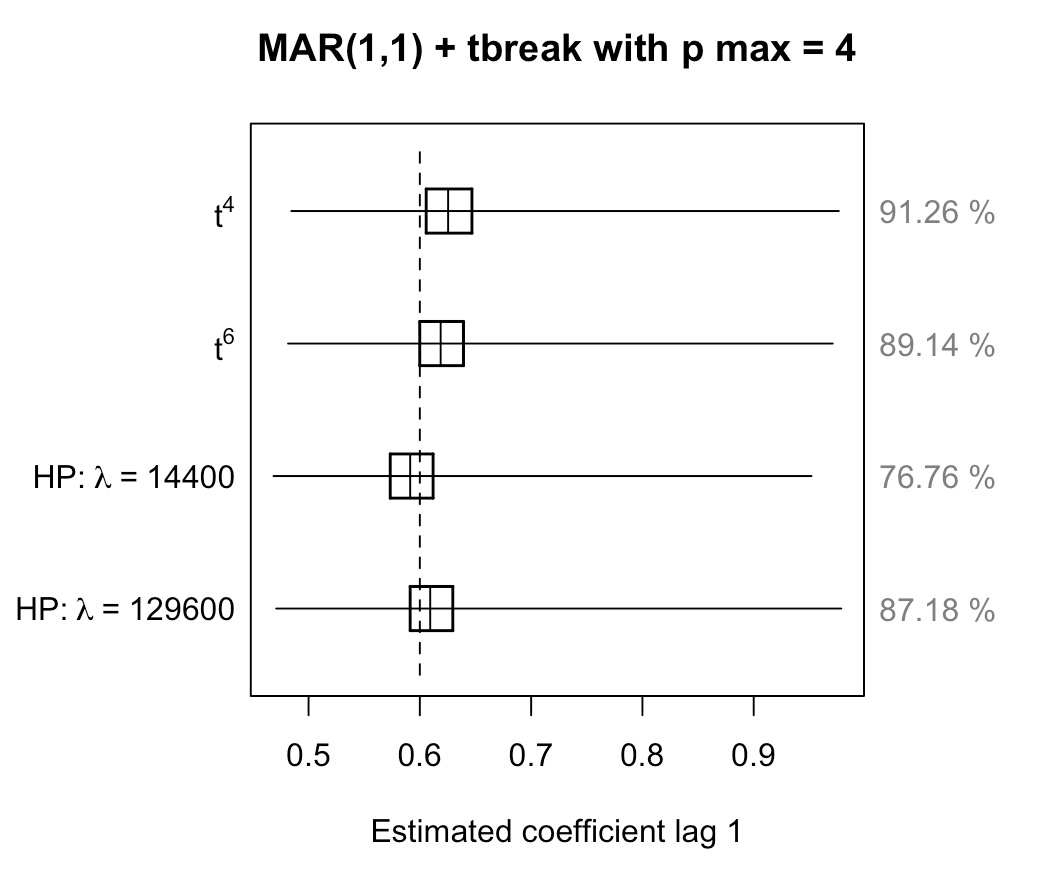}
    \end{subfigure}%
    \begin{subfigure}{0.2525\textwidth}
      \centering
        \caption*{\hspace{-1.9cm}\footnotesize+ breaks}
        \vspace{-0.2cm}
      \includegraphics[width=\linewidth,trim={4.3cm 0 3cm 2cm},clip]{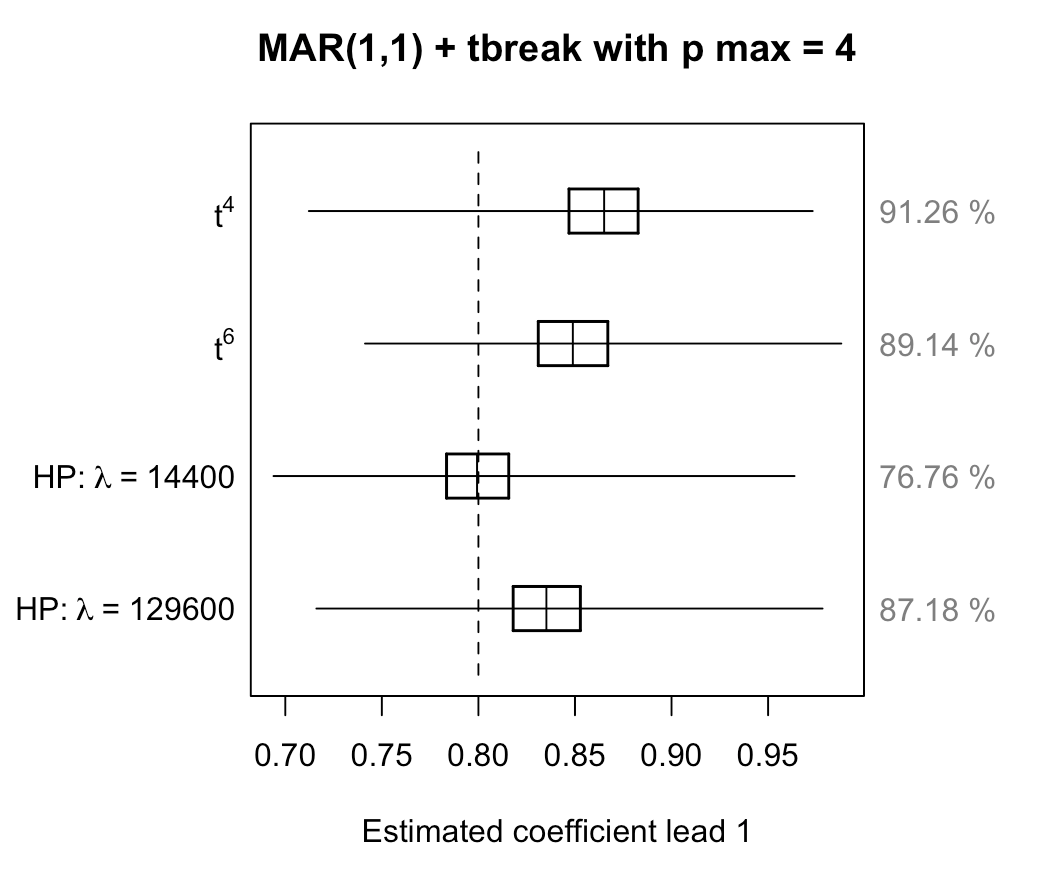}
    \end{subfigure}
    \vspace{-0.3cm}
    \caption{\small Distribution of estimated \textit{MAR} coefficients}
    \label{fig:distrib_coeffs}
\end{figure}

We investigate the distribution of the estimated coefficients given a correctly identified model. Frequencies of wrongly identified models per \textit{dgp} and detrending method are shown in the columns ‘wrong \textit{MAR}' of Table \ref{tab:estimated_MARs_pmax}. Hence, proportions of correctly identified models range between 76.76\% and 96.3\% of the 5\,000 replications, but are almost always above 90\%. Figure \ref{fig:distrib_coeffs} reports the box plots of estimated coefficients for the purely noncausal (left column) and mixed causal-noncausal (center and right columns, for the lag and lead coefficients respectively) processes after each of the four detrending approaches is applied. We indicate the true coefficients, 0.6 and 0.8 for the lag and lead respectively, by the vertical dotted line. The box plots indicate the minimum, maximum, the interquartile range and the median. The $HP_1$-filtered series (with $\lambda=14\,000$) are on average characterized by lower estimated lead and lag coefficients than the other detrended series. This is due to the low penalization of the filter, capturing too much of the dynamics, reducing the persistence of the true noncausal process. Furthermore, we can see that using polynomial trends does not affect estimations of the coefficients, on average, as long as the order of the trend estimated is at least that of the true trend. That is, underestimating the order of the trend leads to an alteration of the dynamics and in our case, to more persistent noncausal dynamics. The $HP_2$ filter performs similarly to $t^6$, but we can expect that if the true trend was a higher order, $HP_2$ would perform better. The constructed linear trend with breaks leads to much larger noncausal coefficients for all detrending methods. The second break in the trend mimics the crash of a bubble and the long expansion preceding it leads to the identification of the model with a larger lead coefficient, which corroborates the earlier findings. Importantly, lag coefficients are on average correctly identified (the distributions of the estimated degrees of freedom, available upon request, show that they are not significantly affected by the detrendings either). A wrong detrending therefore mostly affects the noncausal dynamics of the processes. \\

\section*{Appendix C - Price adjusted series}\label{sec:App.expost_real}
\appendix
\begin{figure}[h!]
    \centering
        \vspace{-0.5cm}

    \includegraphics[width=\linewidth]{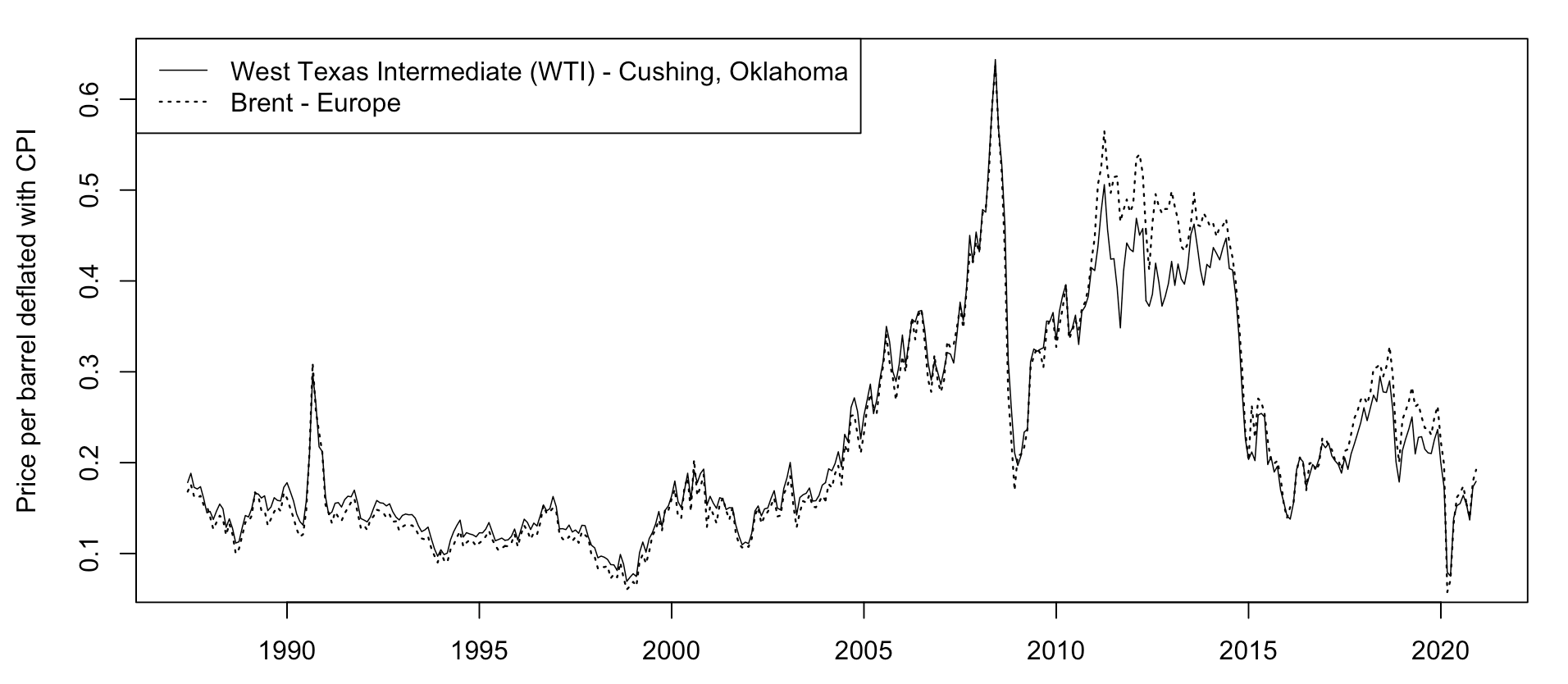}
    \vspace{-0.5cm}
    \caption{Monthly crude oil prices deflated with Consumer Price Index}
    \label{fig:real_oilprices}
\end{figure}

\begin{table}[h!]\small
    \caption{One-step ahead probabilities}
	\centering
		\resizebox{\textwidth}{!}{%
	\begin{threeparttable}
    	\begin{tabular}{l@{\extracolsep{0pt}}cccccccccccc}
            \hline \hline 
        \multirow{2}{*}{Series} & Detrended  & \multicolumn{2}{c}{Jan.} && \multicolumn{2}{c}{Feb.} && \multicolumn{2}{c}{Mar.} && \multicolumn{2}{c}{Apr.} \\ \cline{3-4} \cline{6-7} \cline{9-10} \cline{12-13}
        & with & samp. & sims. && samp. & sims. && samp. & sims. && samp. & sims. \\ \hline
        &\\
        \multirow{11}{*}{$WTI_{real}$}  && \multicolumn{11}{c}{Probability of a decrease} \\ \cline{3-13}

                                        &$t^4$  &.533 & .476 && .735 & .704 && .668 & .637 && .794 & .789 \\
                                        &$t^6$  &.511 & .503 && .775 & .753 && .646 & .662 && .696 & .656 \\
                                        &$HP$   &.501 & .492 && .765 & .738 && .645 & .642 && .660 & .626 \\
                                        &$SPR$  &.470 & .446 && .637 & .656 && .529 & .586 && .639 & .693  \\
                                        &\\
                                        && \multicolumn{11}{c}{Probability of a decrease $>$ 1 s.d.} \\\cline{3-13}
                             
                                        &$t^4$  &.034 & .030 && .014 & .016 && .008 & .008 && .010 & .007 \\
                                        &$t^6$  &.030 & .029 && .010 & .013 && .005 & .006 && .064 & .090 \\
                                        &$HP$   &.034 & .032 && .010 & .015 && .006 & .008 && .104 & .125 \\
                                        &$SPR$  &.002 & .002 && .002 & .002 && .001 & .001 && .002 & .007  \\
\hline
                                        &\\
        \multirow{11}{*}{$Brent_{real}$}  && \multicolumn{11}{c}{Probability of a decrease} \\ \cline{3-13}
                                        
                                        &$t^4$  &.461 & .428 && .739 & .714 && .666 & .639 && .856 & .850 \\
                                        &$t^6$  &.485 & .468 && .790 & .765 && .693 & .677 && .721 & .736 \\
                                        &$HP$   &.476 & .462 && .780 & .750 && .675 & .656 && .651 & .675 \\
                                        &$SPR$  &.526 & .440 && .715 & .705 && .606 & .627 && .643 & .797  \\
                                        &\\
                                        && \multicolumn{11}{c}{Probability of a decrease $>$ 1 s.d.} \\\cline{3-13}
                             
                                        &$t^4$  &.033 & .026 && .015 & .015 && .008 & .009 && .027 & .023 \\
                                        &$t^6$  &.027 & .026 && .009 & .013 && .000 & .007 && .134 & .215 \\
                                        &$HP$   &.030 & .029 && .011 & .015 && .000 & .008 && .210 & .294 \\
                                        &$SPR$  &.002 & .003 && .002 & .003 && .001 & .002 && .007 & .020  \\
        \hline
        \end{tabular}
        \begin{tablenotes}
            \item \scriptsize For the simulations-based approach (sims.) the truncation parameter $M=100$ and 1\,000\,000 simulations were used. Standard deviations (s.d.) are calculated over the detrended samples.
    	\end{tablenotes}
    \end{threeparttable}	%
        }
    \label{tab:1step_prediction_real}
\end{table}

%% file: 0-main.bbl
\begin{thebibliography}{}

\bibitem [\protect \citeauthoryear {%
Alquist%
\ \BBA {} Kilian%
}{%
Alquist%
\ \BBA {} Kilian%
}{%
{\protect \APACyear {2010}}%
}]{%
alquist2010we}
\APACinsertmetastar {%
alquist2010we}%
\begin{APACrefauthors}%
Alquist, R.%
\BCBT {}\ \BBA {} Kilian, L.%
\end{APACrefauthors}%
\unskip\
\newblock
\APACrefYearMonthDay{2010}{}{}.
\newblock
{\BBOQ}\APACrefatitle {What do we learn from the price of crude oil futures?}
  {What do we learn from the price of crude oil futures?}{\BBCQ}
\newblock
\APACjournalVolNumPages{Journal of Applied econometrics}{25}{4}{539--573}.
\PrintBackRefs{\CurrentBib}

\bibitem [\protect \citeauthoryear {%
Alquist%
, Kilian%
\BCBL {}\ \BBA {} Vigfusson%
}{%
Alquist%
\ \protect \BOthers {.}}{%
{\protect \APACyear {2013}}%
}]{%
alquist2013forecasting}
\APACinsertmetastar {%
alquist2013forecasting}%
\begin{APACrefauthors}%
Alquist, R.%
, Kilian, L.%
\BCBL {}\ \BBA {} Vigfusson, R\BPBI J.%
\end{APACrefauthors}%
\unskip\
\newblock
\APACrefYearMonthDay{2013}{}{}.
\newblock
{\BBOQ}\APACrefatitle {Forecasting the price of oil} {Forecasting the price of
  oil}.{\BBCQ}
\newblock
\BIn{} \APACrefbtitle {Handbook of economic forecasting} {Handbook of economic
  forecasting}\ (\BVOL~2, \BPGS\ 427--507).
\newblock
\APACaddressPublisher{}{Elsevier}.
\PrintBackRefs{\CurrentBib}

\bibitem [\protect \citeauthoryear {%
Backus%
\ \BBA {} Kehoe%
}{%
Backus%
\ \BBA {} Kehoe%
}{%
{\protect \APACyear {1992}}%
}]{%
backus1992international}
\APACinsertmetastar {%
backus1992international}%
\begin{APACrefauthors}%
Backus, D\BPBI K.%
\BCBT {}\ \BBA {} Kehoe, P\BPBI J.%
\end{APACrefauthors}%
\unskip\
\newblock
\APACrefYearMonthDay{1992}{}{}.
\newblock
{\BBOQ}\APACrefatitle {International evidence on the historical properties of
  business cycles} {International evidence on the historical properties of
  business cycles}.{\BBCQ}
\newblock
\APACjournalVolNumPages{The American Economic Review}{}{}{864--888}.
\PrintBackRefs{\CurrentBib}

\bibitem [\protect \citeauthoryear {%
Baumeister%
\ \BBA {} Kilian%
}{%
Baumeister%
\ \BBA {} Kilian%
}{%
{\protect \APACyear {2016}}%
}]{%
baumeister2016forty}
\APACinsertmetastar {%
baumeister2016forty}%
\begin{APACrefauthors}%
Baumeister, C.%
\BCBT {}\ \BBA {} Kilian, L.%
\end{APACrefauthors}%
\unskip\
\newblock
\APACrefYearMonthDay{2016}{}{}.
\newblock
{\BBOQ}\APACrefatitle {Forty years of oil price fluctuations: Why the price of
  oil may still surprise us} {Forty years of oil price fluctuations: Why the
  price of oil may still surprise us}.{\BBCQ}
\newblock
\APACjournalVolNumPages{Journal of Economic Perspectives}{30}{1}{139--60}.
\PrintBackRefs{\CurrentBib}

\bibitem [\protect \citeauthoryear {%
Bec%
, Bohn~Nielsen%
\BCBL {}\ \BBA {} Sa{\"i}di%
}{%
Bec%
\ \protect \BOthers {.}}{%
{\protect \APACyear {2019}}%
}]{%
frederique2019mixed}
\APACinsertmetastar {%
frederique2019mixed}%
\begin{APACrefauthors}%
Bec, F.%
, Bohn~Nielsen, H.%
\BCBL {}\ \BBA {} Sa{\"i}di, S.%
\end{APACrefauthors}%
\unskip\
\newblock
\APACrefYearMonthDay{2019}{}{}.
\newblock
\APACrefbtitle {Mixed Causal-Noncausal Autoregressions: Bimodality Issues in
  Estimation and Unit Root Testing} {Mixed causal-noncausal autoregressions:
  Bimodality issues in estimation and unit root testing}\
  \APACbVolEdTR{}{\BTR{}}.
\PrintBackRefs{\CurrentBib}

\bibitem [\protect \citeauthoryear {%
Bertelsen%
}{%
Bertelsen%
}{%
{\protect \APACyear {2019}}%
}]{%
bertelsen2019comparing}
\APACinsertmetastar {%
bertelsen2019comparing}%
\begin{APACrefauthors}%
Bertelsen, K\BPBI P.%
\end{APACrefauthors}%
\unskip\
\newblock
\APACrefYearMonthDay{2019}{}{}.
\newblock
\APACrefbtitle {Comparing Tests for Identification of Bubbles} {Comparing tests
  for identification of bubbles}\ \APACbVolEdTR{}{\BTR{}}.
\newblock
\APACaddressInstitution{}{Department of Economics and Business Economics,
  Aarhus University}.
\PrintBackRefs{\CurrentBib}

\bibitem [\protect \citeauthoryear {%
Breidt%
, Davis%
, Li%
\BCBL {}\ \BBA {} Rosenblatt%
}{%
Breidt%
\ \protect \BOthers {.}}{%
{\protect \APACyear {1991}}%
}]{%
breid1991maximum}
\APACinsertmetastar {%
breid1991maximum}%
\begin{APACrefauthors}%
Breidt, F\BPBI J.%
, Davis, R\BPBI A.%
, Li, K\BHBI S.%
\BCBL {}\ \BBA {} Rosenblatt, M.%
\end{APACrefauthors}%
\unskip\
\newblock
\APACrefYearMonthDay{1991}{}{}.
\newblock
{\BBOQ}\APACrefatitle {Maximum likelihood estimation for noncausal
  autoregressive processes} {Maximum likelihood estimation for noncausal
  autoregressive processes}.{\BBCQ}
\newblock
\APACjournalVolNumPages{Journal of Multivariate Analysis}{36}{2}{175--198}.
\PrintBackRefs{\CurrentBib}

\bibitem [\protect \citeauthoryear {%
Brooks%
, Prokopczuk%
\BCBL {}\ \BBA {} Wu%
}{%
Brooks%
\ \protect \BOthers {.}}{%
{\protect \APACyear {2015}}%
}]{%
brooks2015booms}
\APACinsertmetastar {%
brooks2015booms}%
\begin{APACrefauthors}%
Brooks, C.%
, Prokopczuk, M.%
\BCBL {}\ \BBA {} Wu, Y.%
\end{APACrefauthors}%
\unskip\
\newblock
\APACrefYearMonthDay{2015}{}{}.
\newblock
{\BBOQ}\APACrefatitle {Booms and busts in commodity markets: bubbles or
  fundamentals?} {Booms and busts in commodity markets: bubbles or
  fundamentals?}{\BBCQ}
\newblock
\APACjournalVolNumPages{Journal of Futures Markets}{35}{10}{916--938}.
\PrintBackRefs{\CurrentBib}

\bibitem [\protect \citeauthoryear {%
Campbell%
\ \BBA {} Shiller%
}{%
Campbell%
\ \BBA {} Shiller%
}{%
{\protect \APACyear {1987}}%
}]{%
campbell1987cointegration}
\APACinsertmetastar {%
campbell1987cointegration}%
\begin{APACrefauthors}%
Campbell, J\BPBI Y.%
\BCBT {}\ \BBA {} Shiller, R\BPBI J.%
\end{APACrefauthors}%
\unskip\
\newblock
\APACrefYearMonthDay{1987}{}{}.
\newblock
{\BBOQ}\APACrefatitle {Cointegration and tests of present value models}
  {Cointegration and tests of present value models}.{\BBCQ}
\newblock
\APACjournalVolNumPages{Journal of political economy}{95}{5}{1062--1088}.
\PrintBackRefs{\CurrentBib}

\bibitem [\protect \citeauthoryear {%
Canova%
}{%
Canova%
}{%
{\protect \APACyear {1998}}%
}]{%
canova1998detrending}
\APACinsertmetastar {%
canova1998detrending}%
\begin{APACrefauthors}%
Canova, F.%
\end{APACrefauthors}%
\unskip\
\newblock
\APACrefYearMonthDay{1998}{}{}.
\newblock
{\BBOQ}\APACrefatitle {Detrending and business cycle facts} {Detrending and
  business cycle facts}.{\BBCQ}
\newblock
\APACjournalVolNumPages{Journal of monetary economics}{41}{3}{475--512}.
\PrintBackRefs{\CurrentBib}

\bibitem [\protect \citeauthoryear {%
Cavaliere%
, Nielsen%
\BCBL {}\ \BBA {} Rahbek%
}{%
Cavaliere%
\ \protect \BOthers {.}}{%
{\protect \APACyear {2018}}%
}]{%
cavaliere2018bootstrapping}
\APACinsertmetastar {%
cavaliere2018bootstrapping}%
\begin{APACrefauthors}%
Cavaliere, G.%
, Nielsen, H\BPBI B.%
\BCBL {}\ \BBA {} Rahbek, A.%
\end{APACrefauthors}%
\unskip\
\newblock
\APACrefYearMonthDay{2018}{}{}.
\newblock
{\BBOQ}\APACrefatitle {Bootstrapping noncausal autoregressions: with
  Applications to explosive bubble modeling} {Bootstrapping noncausal
  autoregressions: with applications to explosive bubble modeling}.{\BBCQ}
\newblock
\APACjournalVolNumPages{Journal of Business \& Economic Statistics}{}{}{1--13}.
\PrintBackRefs{\CurrentBib}

\bibitem [\protect \citeauthoryear {%
Cubadda%
, Hecq%
\BCBL {}\ \BBA {} Telg%
}{%
Cubadda%
\ \protect \BOthers {.}}{%
{\protect \APACyear {2019}}%
}]{%
cubadda2019detecting}
\APACinsertmetastar {%
cubadda2019detecting}%
\begin{APACrefauthors}%
Cubadda, G.%
, Hecq, A.%
\BCBL {}\ \BBA {} Telg, S.%
\end{APACrefauthors}%
\unskip\
\newblock
\APACrefYearMonthDay{2019}{}{}.
\newblock
{\BBOQ}\APACrefatitle {Detecting Co-Movements in Non-Causal Time Series}
  {Detecting co-movements in non-causal time series}.{\BBCQ}
\newblock
\APACjournalVolNumPages{Oxford Bulletin of Economics and
  Statistics}{81}{3}{697--715}.
\PrintBackRefs{\CurrentBib}

\bibitem [\protect \citeauthoryear {%
Diba%
\ \BBA {} Grossman%
}{%
Diba%
\ \BBA {} Grossman%
}{%
{\protect \APACyear {1988}}%
}]{%
diba1988explosive}
\APACinsertmetastar {%
diba1988explosive}%
\begin{APACrefauthors}%
Diba, B\BPBI T.%
\BCBT {}\ \BBA {} Grossman, H\BPBI I.%
\end{APACrefauthors}%
\unskip\
\newblock
\APACrefYearMonthDay{1988}{}{}.
\newblock
{\BBOQ}\APACrefatitle {Explosive rational bubbles in stock prices?} {Explosive
  rational bubbles in stock prices?}{\BBCQ}
\newblock
\APACjournalVolNumPages{The American Economic Review}{78}{3}{520--530}.
\PrintBackRefs{\CurrentBib}

\bibitem [\protect \citeauthoryear {%
Fries%
}{%
Fries%
}{%
{\protect \APACyear {2021}}%
}]{%
fries2021conditional}
\APACinsertmetastar {%
fries2021conditional}%
\begin{APACrefauthors}%
Fries, S.%
\end{APACrefauthors}%
\unskip\
\newblock
\APACrefYearMonthDay{2021}{}{}.
\newblock
{\BBOQ}\APACrefatitle {Conditional moments of noncausal alpha-stable processes
  and the prediction of bubble crash odds} {Conditional moments of noncausal
  alpha-stable processes and the prediction of bubble crash odds}.{\BBCQ}
\newblock
\APACjournalVolNumPages{Journal of Business \& Economic Statistics}{}{}{}.
\PrintBackRefs{\CurrentBib}

\bibitem [\protect \citeauthoryear {%
Fries%
\ \BBA {} Zako{\"\i}an%
}{%
Fries%
\ \BBA {} Zako{\"\i}an%
}{%
{\protect \APACyear {2019}}%
}]{%
fries2019mixed}
\APACinsertmetastar {%
fries2019mixed}%
\begin{APACrefauthors}%
Fries, S.%
\BCBT {}\ \BBA {} Zako{\"\i}an, J\BHBI M.%
\end{APACrefauthors}%
\unskip\
\newblock
\APACrefYearMonthDay{2019}{}{}.
\newblock
{\BBOQ}\APACrefatitle {Mixed Causal-Noncausal {AR} Processes and the Modelling
  of Explosive Bubbles} {Mixed causal-noncausal {AR} processes and the
  modelling of explosive bubbles}.{\BBCQ}
\newblock
\APACjournalVolNumPages{Econometric Theory}{}{}{1–37}.
\PrintBackRefs{\CurrentBib}

\bibitem [\protect \citeauthoryear {%
Gourieroux%
\ \BBA {} Jasiak%
}{%
Gourieroux%
\ \BBA {} Jasiak%
}{%
{\protect \APACyear {2016}}%
}]{%
filtering}
\APACinsertmetastar {%
filtering}%
\begin{APACrefauthors}%
Gourieroux, C.%
\BCBT {}\ \BBA {} Jasiak, J.%
\end{APACrefauthors}%
\unskip\
\newblock
\APACrefYearMonthDay{2016}{}{}.
\newblock
{\BBOQ}\APACrefatitle {Filtering, prediction and simulation methods for
  noncausal processes} {Filtering, prediction and simulation methods for
  noncausal processes}.{\BBCQ}
\newblock
\APACjournalVolNumPages{Journal of Time Series Analysis}{37}{3}{405--430}.
\PrintBackRefs{\CurrentBib}

\bibitem [\protect \citeauthoryear {%
Gourieroux%
, Jasiak%
\BCBL {}\ \BBA {} Monfort%
}{%
Gourieroux%
\ \protect \BOthers {.}}{%
{\protect \APACyear {2020}}%
}]{%
gourieroux2020stationary}
\APACinsertmetastar {%
gourieroux2020stationary}%
\begin{APACrefauthors}%
Gourieroux, C.%
, Jasiak, J.%
\BCBL {}\ \BBA {} Monfort, A.%
\end{APACrefauthors}%
\unskip\
\newblock
\APACrefYearMonthDay{2020}{}{}.
\newblock
{\BBOQ}\APACrefatitle {Stationary bubble equilibria in rational expectation
  models} {Stationary bubble equilibria in rational expectation models}.{\BBCQ}
\newblock
\APACjournalVolNumPages{Journal of Econometrics}{218}{2}{714--735}.
\PrintBackRefs{\CurrentBib}

\bibitem [\protect \citeauthoryear {%
Gouri{\'e}roux%
\ \BBA {} Zako{\"\i}an%
}{%
Gouri{\'e}roux%
\ \BBA {} Zako{\"\i}an%
}{%
{\protect \APACyear {2013}}%
}]{%
gourieroux2013explosive}
\APACinsertmetastar {%
gourieroux2013explosive}%
\begin{APACrefauthors}%
Gouri{\'e}roux, C.%
\BCBT {}\ \BBA {} Zako{\"\i}an, J\BHBI M.%
\end{APACrefauthors}%
\unskip\
\newblock
\APACrefYearMonthDay{2013}{}{}.
\newblock
{\BBOQ}\APACrefatitle {Explosive Bubble Modelling by Noncausal Process}
  {Explosive bubble modelling by noncausal process}.{\BBCQ}
\newblock
\APACjournalVolNumPages{CREST. Paris, France: Centre de Recherche en Economie
  et Statistique}{}{}{}.
\PrintBackRefs{\CurrentBib}

\bibitem [\protect \citeauthoryear {%
Gouri{\'e}roux%
\ \BBA {} Zako{\"\i}an%
}{%
Gouri{\'e}roux%
\ \BBA {} Zako{\"\i}an%
}{%
{\protect \APACyear {2015}}%
}]{%
gourieroux2015uniqueness}
\APACinsertmetastar {%
gourieroux2015uniqueness}%
\begin{APACrefauthors}%
Gouri{\'e}roux, C.%
\BCBT {}\ \BBA {} Zako{\"\i}an, J\BHBI M.%
\end{APACrefauthors}%
\unskip\
\newblock
\APACrefYearMonthDay{2015}{}{}.
\newblock
{\BBOQ}\APACrefatitle {On Uniqueness of Moving Average Representations of
  Heavy-tailed Stationary Processes} {On uniqueness of moving average
  representations of heavy-tailed stationary processes}.{\BBCQ}
\newblock
\APACjournalVolNumPages{Journal of Time Series Analysis}{36}{6}{876--887}.
\PrintBackRefs{\CurrentBib}

\bibitem [\protect \citeauthoryear {%
Gouri{\'e}roux%
\ \BBA {} Zako{\"\i}an%
}{%
Gouri{\'e}roux%
\ \BBA {} Zako{\"\i}an%
}{%
{\protect \APACyear {2017}}%
}]{%
gourieroux2017local}
\APACinsertmetastar {%
gourieroux2017local}%
\begin{APACrefauthors}%
Gouri{\'e}roux, C.%
\BCBT {}\ \BBA {} Zako{\"\i}an, J\BHBI M.%
\end{APACrefauthors}%
\unskip\
\newblock
\APACrefYearMonthDay{2017}{}{}.
\newblock
{\BBOQ}\APACrefatitle {Local explosion modelling by non-causal process} {Local
  explosion modelling by non-causal process}.{\BBCQ}
\newblock
\APACjournalVolNumPages{Journal of the Royal Statistical Society: Series B
  (Statistical Methodology)}{79}{3}{737--756}.
\PrintBackRefs{\CurrentBib}

\bibitem [\protect \citeauthoryear {%
Hamilton%
}{%
Hamilton%
}{%
{\protect \APACyear {2018}}%
}]{%
hamilton2018you}
\APACinsertmetastar {%
hamilton2018you}%
\begin{APACrefauthors}%
Hamilton, J\BPBI D.%
\end{APACrefauthors}%
\unskip\
\newblock
\APACrefYearMonthDay{2018}{}{}.
\newblock
{\BBOQ}\APACrefatitle {Why you should never use the {H}odrick-{P}rescott
  filter} {Why you should never use the {H}odrick-{P}rescott filter}.{\BBCQ}
\newblock
\APACjournalVolNumPages{Review of Economics and Statistics}{100}{5}{831--843}.
\PrintBackRefs{\CurrentBib}

\bibitem [\protect \citeauthoryear {%
Hecq%
, Issler%
\BCBL {}\ \BBA {} Voisin%
}{%
Hecq%
\ \protect \BOthers {.}}{%
{\protect \APACyear {2021}}%
}]{%
brazilpaper}
\APACinsertmetastar {%
brazilpaper}%
\begin{APACrefauthors}%
Hecq, A.%
, Issler, J.%
\BCBL {}\ \BBA {} Voisin, E.%
\end{APACrefauthors}%
\unskip\
\newblock
\APACrefYearMonthDay{2021}{}{}.
\newblock
{\BBOQ}\APACrefatitle {That’s the limit! Evaluation of the {B}razilian
  inflation targeting system using mixed causal-noncausal models} {That’s the
  limit! evaluation of the {B}razilian inflation targeting system using mixed
  causal-noncausal models}.{\BBCQ}
\newblock

\PrintBackRefs{\CurrentBib}

\bibitem [\protect \citeauthoryear {%
Hecq%
, Issler%
\BCBL {}\ \BBA {} Telg%
}{%
Hecq%
\ \protect \BOthers {.}}{%
{\protect \APACyear {2020}}%
}]{%
hecq2020mixed}
\APACinsertmetastar {%
hecq2020mixed}%
\begin{APACrefauthors}%
Hecq, A.%
, Issler, J\BPBI V.%
\BCBL {}\ \BBA {} Telg, S.%
\end{APACrefauthors}%
\unskip\
\newblock
\APACrefYearMonthDay{2020}{}{}.
\newblock
{\BBOQ}\APACrefatitle {Mixed causal--noncausal autoregressions with exogenous
  regressors} {Mixed causal--noncausal autoregressions with exogenous
  regressors}.{\BBCQ}
\newblock
\APACjournalVolNumPages{Journal of Applied Econometrics}{35}{3}{328--343}.
\PrintBackRefs{\CurrentBib}

\bibitem [\protect \citeauthoryear {%
Hecq%
, Lieb%
\BCBL {}\ \BBA {} Telg%
}{%
Hecq%
\ \protect \BOthers {.}}{%
{\protect \APACyear {2017}}%
}]{%
MARX}
\APACinsertmetastar {%
MARX}%
\begin{APACrefauthors}%
Hecq, A.%
, Lieb, L.%
\BCBL {}\ \BBA {} Telg, S.%
\end{APACrefauthors}%
\unskip\
\newblock
\APACrefYearMonthDay{2017}{}{}.
\newblock
{\BBOQ}\APACrefatitle {Simulation, Estimation and Selection of Mixed
  Causal-Noncausal Autoregressive Models: The MARX Package} {Simulation,
  estimation and selection of mixed causal-noncausal autoregressive models: The
  marx package}.{\BBCQ}
\newblock

\PrintBackRefs{\CurrentBib}

\bibitem [\protect \citeauthoryear {%
Hecq%
\ \BBA {} Voisin%
}{%
Hecq%
\ \BBA {} Voisin%
}{%
{\protect \APACyear {2021}}%
}]{%
voisin2019forecasting}
\APACinsertmetastar {%
voisin2019forecasting}%
\begin{APACrefauthors}%
Hecq, A.%
\BCBT {}\ \BBA {} Voisin, E.%
\end{APACrefauthors}%
\unskip\
\newblock
\APACrefYearMonthDay{2021}{}{}.
\newblock
{\BBOQ}\APACrefatitle {Forecasting bubbles with mixed causal-noncausal
  autoregressive models} {Forecasting bubbles with mixed causal-noncausal
  autoregressive models}.{\BBCQ}
\newblock
\APACjournalVolNumPages{Econometrics and Statistics}{20}{}{29-45}.
\PrintBackRefs{\CurrentBib}

\bibitem [\protect \citeauthoryear {%
Hencic%
\ \BBA {} Gouri{\'e}roux%
}{%
Hencic%
\ \BBA {} Gouri{\'e}roux%
}{%
{\protect \APACyear {2015}}%
}]{%
hencic2015noncausal}
\APACinsertmetastar {%
hencic2015noncausal}%
\begin{APACrefauthors}%
Hencic, A.%
\BCBT {}\ \BBA {} Gouri{\'e}roux, C.%
\end{APACrefauthors}%
\unskip\
\newblock
\APACrefYearMonthDay{2015}{}{}.
\newblock
{\BBOQ}\APACrefatitle {Noncausal autoregressive model in application to
  bitcoin/{USD} exchange rates} {Noncausal autoregressive model in application
  to bitcoin/{USD} exchange rates}.{\BBCQ}
\newblock
\BIn{} \APACrefbtitle {Econometrics of risk} {Econometrics of risk}\ (\BPGS\
  17--40).
\newblock
\APACaddressPublisher{}{Springer}.
\PrintBackRefs{\CurrentBib}

\bibitem [\protect \citeauthoryear {%
Hodrick%
\ \BBA {} Prescott%
}{%
Hodrick%
\ \BBA {} Prescott%
}{%
{\protect \APACyear {1997}}%
}]{%
hodrick1997postwar}
\APACinsertmetastar {%
hodrick1997postwar}%
\begin{APACrefauthors}%
Hodrick, R\BPBI J.%
\BCBT {}\ \BBA {} Prescott, E\BPBI C.%
\end{APACrefauthors}%
\unskip\
\newblock
\APACrefYearMonthDay{1997}{}{}.
\newblock
{\BBOQ}\APACrefatitle {Postwar {US} business cycles: an empirical
  investigation} {Postwar {US} business cycles: an empirical
  investigation}.{\BBCQ}
\newblock
\APACjournalVolNumPages{Journal of Money, credit, and Banking}{}{}{1--16}.
\PrintBackRefs{\CurrentBib}

\bibitem [\protect \citeauthoryear {%
Homm%
\ \BBA {} Breitung%
}{%
Homm%
\ \BBA {} Breitung%
}{%
{\protect \APACyear {2012}}%
}]{%
homm2012testing}
\APACinsertmetastar {%
homm2012testing}%
\begin{APACrefauthors}%
Homm, U.%
\BCBT {}\ \BBA {} Breitung, J.%
\end{APACrefauthors}%
\unskip\
\newblock
\APACrefYearMonthDay{2012}{}{}.
\newblock
{\BBOQ}\APACrefatitle {Testing for speculative bubbles in stock markets: a
  comparison of alternative methods} {Testing for speculative bubbles in stock
  markets: a comparison of alternative methods}.{\BBCQ}
\newblock
\APACjournalVolNumPages{Journal of Financial Econometrics}{10}{1}{198--231}.
\PrintBackRefs{\CurrentBib}

\bibitem [\protect \citeauthoryear {%
Karapanagiotidis%
}{%
Karapanagiotidis%
}{%
{\protect \APACyear {2014}}%
}]{%
karapanagiotidis2014dynamic}
\APACinsertmetastar {%
karapanagiotidis2014dynamic}%
\begin{APACrefauthors}%
Karapanagiotidis, P.%
\end{APACrefauthors}%
\unskip\
\newblock
\APACrefYearMonthDay{2014}{}{}.
\newblock
{\BBOQ}\APACrefatitle {Dynamic modeling of commodity futures prices} {Dynamic
  modeling of commodity futures prices}.{\BBCQ}
\newblock

\PrintBackRefs{\CurrentBib}

\bibitem [\protect \citeauthoryear {%
Kilian%
}{%
Kilian%
}{%
{\protect \APACyear {2009}}%
}]{%
kilian2009not}
\APACinsertmetastar {%
kilian2009not}%
\begin{APACrefauthors}%
Kilian, L.%
\end{APACrefauthors}%
\unskip\
\newblock
\APACrefYearMonthDay{2009}{}{}.
\newblock
{\BBOQ}\APACrefatitle {Not all oil price shocks are alike: Disentangling demand
  and supply shocks in the crude oil market} {Not all oil price shocks are
  alike: Disentangling demand and supply shocks in the crude oil
  market}.{\BBCQ}
\newblock
\APACjournalVolNumPages{American Economic Review}{99}{3}{1053--69}.
\PrintBackRefs{\CurrentBib}

\bibitem [\protect \citeauthoryear {%
Kilian%
\ \BBA {} Murphy%
}{%
Kilian%
\ \BBA {} Murphy%
}{%
{\protect \APACyear {2014}}%
}]{%
kilian2014role}
\APACinsertmetastar {%
kilian2014role}%
\begin{APACrefauthors}%
Kilian, L.%
\BCBT {}\ \BBA {} Murphy, D\BPBI P.%
\end{APACrefauthors}%
\unskip\
\newblock
\APACrefYearMonthDay{2014}{}{}.
\newblock
{\BBOQ}\APACrefatitle {The role of inventories and speculative trading in the
  global market for crude oil} {The role of inventories and speculative trading
  in the global market for crude oil}.{\BBCQ}
\newblock
\APACjournalVolNumPages{Journal of Applied econometrics}{29}{3}{454--478}.
\PrintBackRefs{\CurrentBib}

\bibitem [\protect \citeauthoryear {%
Kilian%
\ \BBA {} Zhou%
}{%
Kilian%
\ \BBA {} Zhou%
}{%
{\protect \APACyear {2020}}%
{\protect \APACexlab {{\protect \BCnt {1}}}}}]{%
kilian2020does}
\APACinsertmetastar {%
kilian2020does}%
\begin{APACrefauthors}%
Kilian, L.%
\BCBT {}\ \BBA {} Zhou, X.%
\end{APACrefauthors}%
\unskip\
\newblock
\APACrefYearMonthDay{2020{\protect \BCnt {1}}}{}{}.
\newblock
{\BBOQ}\APACrefatitle {Does drawing down the {US} {S}trategic {P}etroleum
  {R}eserve help stabilize oil prices?} {Does drawing down the {US} {S}trategic
  {P}etroleum {R}eserve help stabilize oil prices?}{\BBCQ}
\newblock
\APACjournalVolNumPages{Journal of Applied Econometrics}{35}{6}{673--691}.
\PrintBackRefs{\CurrentBib}

\bibitem [\protect \citeauthoryear {%
Kilian%
\ \BBA {} Zhou%
}{%
Kilian%
\ \BBA {} Zhou%
}{%
{\protect \APACyear {2020}}%
{\protect \APACexlab {{\protect \BCnt {2}}}}}]{%
kilian2020econometrics}
\APACinsertmetastar {%
kilian2020econometrics}%
\begin{APACrefauthors}%
Kilian, L.%
\BCBT {}\ \BBA {} Zhou, X.%
\end{APACrefauthors}%
\unskip\
\newblock
\APACrefYearMonthDay{2020{\protect \BCnt {2}}}{}{}.
\newblock
\APACrefbtitle {{The Econometrics of Oil Market {VAR} Models}} {{The
  Econometrics of Oil Market {VAR} Models}}\ \APACbVolEdTR {}{CESifo Working
  Paper Series\ \BNUM\ 8153}.
\PrintBackRefs{\CurrentBib}

\bibitem [\protect \citeauthoryear {%
Lanne%
, Luoto%
\BCBL {}\ \BBA {} Saikkonen%
}{%
Lanne%
\ \protect \BOthers {.}}{%
{\protect \APACyear {2012}}%
}]{%
lanne2012optimal}
\APACinsertmetastar {%
lanne2012optimal}%
\begin{APACrefauthors}%
Lanne, M.%
, Luoto, J.%
\BCBL {}\ \BBA {} Saikkonen, P.%
\end{APACrefauthors}%
\unskip\
\newblock
\APACrefYearMonthDay{2012}{}{}.
\newblock
{\BBOQ}\APACrefatitle {Optimal forecasting of noncausal autoregressive time
  series} {Optimal forecasting of noncausal autoregressive time series}.{\BBCQ}
\newblock
\APACjournalVolNumPages{International Journal of Forecasting}{28}{3}{623--631}.
\PrintBackRefs{\CurrentBib}

\bibitem [\protect \citeauthoryear {%
Lanne%
\ \BBA {} Saikkonen%
}{%
Lanne%
\ \BBA {} Saikkonen%
}{%
{\protect \APACyear {2011}}%
}]{%
lanne2011noncausal}
\APACinsertmetastar {%
lanne2011noncausal}%
\begin{APACrefauthors}%
Lanne, M.%
\BCBT {}\ \BBA {} Saikkonen, P.%
\end{APACrefauthors}%
\unskip\
\newblock
\APACrefYearMonthDay{2011}{}{}.
\newblock
{\BBOQ}\APACrefatitle {Noncausal autoregressions for economic time series}
  {Noncausal autoregressions for economic time series}.{\BBCQ}
\newblock
\APACjournalVolNumPages{Journal of Time Series Econometrics}{3}{3}{}.
\PrintBackRefs{\CurrentBib}

\bibitem [\protect \citeauthoryear {%
Lof%
\ \BBA {} Nyberg%
}{%
Lof%
\ \BBA {} Nyberg%
}{%
{\protect \APACyear {2017}}%
}]{%
lof2017noncausality}
\APACinsertmetastar {%
lof2017noncausality}%
\begin{APACrefauthors}%
Lof, M.%
\BCBT {}\ \BBA {} Nyberg, H.%
\end{APACrefauthors}%
\unskip\
\newblock
\APACrefYearMonthDay{2017}{}{}.
\newblock
{\BBOQ}\APACrefatitle {Noncausality and the commodity currency hypothesis}
  {Noncausality and the commodity currency hypothesis}.{\BBCQ}
\newblock
\APACjournalVolNumPages{Energy Economics}{65}{}{424--433}.
\PrintBackRefs{\CurrentBib}

\bibitem [\protect \citeauthoryear {%
Phillips%
\ \BBA {} Shi%
}{%
Phillips%
\ \BBA {} Shi%
}{%
{\protect \APACyear {2019}}%
}]{%
phillips2019boosting}
\APACinsertmetastar {%
phillips2019boosting}%
\begin{APACrefauthors}%
Phillips, P\BPBI C.%
\BCBT {}\ \BBA {} Shi, Z.%
\end{APACrefauthors}%
\unskip\
\newblock
\APACrefYearMonthDay{2019}{}{}.
\newblock
{\BBOQ}\APACrefatitle {Boosting the {H}odrick-{P}rescott Filter} {Boosting the
  {H}odrick-{P}rescott filter}.{\BBCQ}
\newblock
\APACjournalVolNumPages{arXiv preprint arXiv:1905.00175}{}{}{}.
\PrintBackRefs{\CurrentBib}

\bibitem [\protect \citeauthoryear {%
Phillips%
, Wu%
\BCBL {}\ \BBA {} Yu%
}{%
Phillips%
\ \protect \BOthers {.}}{%
{\protect \APACyear {2011}}%
}]{%
phillips2011explosive}
\APACinsertmetastar {%
phillips2011explosive}%
\begin{APACrefauthors}%
Phillips, P\BPBI C.%
, Wu, Y.%
\BCBL {}\ \BBA {} Yu, J.%
\end{APACrefauthors}%
\unskip\
\newblock
\APACrefYearMonthDay{2011}{}{}.
\newblock
{\BBOQ}\APACrefatitle {Explosive behavior in the 1990s {N}asdaq: When did
  exuberance escalate asset values?} {Explosive behavior in the 1990s {N}asdaq:
  When did exuberance escalate asset values?}{\BBCQ}
\newblock
\APACjournalVolNumPages{International economic review}{52}{1}{201--226}.
\PrintBackRefs{\CurrentBib}

\bibitem [\protect \citeauthoryear {%
Pindyck%
}{%
Pindyck%
}{%
{\protect \APACyear {1993}}%
}]{%
pindyck1992present}
\APACinsertmetastar {%
pindyck1992present}%
\begin{APACrefauthors}%
Pindyck, R\BPBI S.%
\end{APACrefauthors}%
\unskip\
\newblock
\APACrefYearMonthDay{1993}{}{}.
\newblock
{\BBOQ}\APACrefatitle {The present value model of rational commodity pricing}
  {The present value model of rational commodity pricing}.{\BBCQ}
\newblock
\APACjournalVolNumPages{The Economic Journal}{103}{}{511--530}.
\PrintBackRefs{\CurrentBib}

\bibitem [\protect \citeauthoryear {%
Ravn%
\ \BBA {} Uhlig%
}{%
Ravn%
\ \BBA {} Uhlig%
}{%
{\protect \APACyear {2002}}%
}]{%
ravn2002adjusting}
\APACinsertmetastar {%
ravn2002adjusting}%
\begin{APACrefauthors}%
Ravn, M\BPBI O.%
\BCBT {}\ \BBA {} Uhlig, H.%
\end{APACrefauthors}%
\unskip\
\newblock
\APACrefYearMonthDay{2002}{}{}.
\newblock
{\BBOQ}\APACrefatitle {On adjusting the {H}odrick-{P}rescott filter for the
  frequency of observations} {On adjusting the {H}odrick-{P}rescott filter for
  the frequency of observations}.{\BBCQ}
\newblock
\APACjournalVolNumPages{Review of economics and statistics}{84}{2}{371--376}.
\PrintBackRefs{\CurrentBib}

\end{thebibliography}
